\documentclass[%
 aip,
 amsmath,amssymb,nofootinbib,
 reprint,%
]{revtex4-1}

\usepackage{graphicx}
\usepackage{dcolumn}
\usepackage{bm}

\usepackage[utf8]{inputenc}
\usepackage[T1]{fontenc}
\usepackage{mathptmx}
\usepackage{etoolbox}
\usepackage{graphicx}
\usepackage[utf8]{inputenc}
\usepackage{flushend}
\usepackage{dcolumn}
\usepackage{bm}
\usepackage{balance}
\usepackage{mathrsfs,epigraph,bm}

\usepackage{tikz}
\usetikzlibrary{mindmap,trees}

\usepackage{scalerel}
\usepackage{stackengine,wasysym}

\usepackage{tcolorbox}
\usepackage{tabularx}
\usepackage{array}
\usepackage{colortbl}
\tcbuselibrary{skins}

\usepackage{tcolorbox}
\definecolor{mycolor}{rgb}{0.122, 0.435, 0.698}
\makeatletter
\newcommand{\mybox}[1]{%
  \setbox0=\hbox{#1}%
  \setlength{\@tempdima}{\dimexpr\wd0+13pt}%
  \begin{tcolorbox}[colframe=mycolor,boxrule=0.5pt,arc=4pt,
      left=6pt,right=6pt,top=6pt,bottom=6pt,boxsep=0pt,width=0.98\linewidth]
    #1
  \end{tcolorbox}
}
\makeatother

\newtcbox{\boxbello}{on line,
  colframe=orange,colback=yellow!20!white,
  boxrule=0.5pt,arc=4pt,boxsep=0pt,left=6pt,right=6pt,top=6pt,bottom=6pt,width=0.98\linewidth}
\newcolumntype{Y}{>{\raggedleft\arraybackslash}X}

\tcbset{tab1/.style={fonttitle=\bfseries\large,fontupper=\normalsize\sffamily,
colback=yellow!10!white,colframe=red!75!black,colbacktitle=purple!40!white,
coltitle=black,center title,freelance,frame code={
\foreach \n in {north east,north west,south east,south west}
{\path [fill=purple!75!black] (interior.\n) circle (3mm); };},}}

\tcbset{tab2/.style={enhanced,fonttitle=\bfseries,fontupper=\normalsize\sffamily,
colback=yellow!10!white,colframe=red!50!black,colbacktitle=purple!40!white,
coltitle=black,center title}}
\usepackage[normalem]{ulem}

\usepackage[colorlinks = true,
            linkcolor = blue,
            urlcolor  = blue,
            citecolor =green,
            anchorcolor = blue]{hyperref}
\usepackage{verbatim}
\usepackage{color,ulem}
\usepackage[english]{babel}

\usepackage[utf8]{inputenc}
\input Starburst.fd
\newcommand*\initfamily{\usefont{U}{Starburst}{xl}{n}}\initfamily

\newcommand{\beq}{\begin{eqnarray}}
\newcommand{\eeq}{\end{eqnarray}}
\usepackage{amsmath}
\usepackage{tikz}
\usetikzlibrary{decorations.pathmorphing}
\usetikzlibrary{shapes.misc}
\tikzset{cross/.style={cross out, draw=black, minimum size=8*(#1-\pgflinewidth), inner sep=0pt, outer sep=0pt},
cross/.default={1pt}}
\usetikzlibrary{patterns,math}
\makeatletter
\def\@email#1#2{%
 \endgroup
 \patchcmd{\titleblock@produce}
  {\frontmatter@RRAPformat}
  {\frontmatter@RRAPformat{\produce@RRAP{*#1\href{mailto:#2}{#2}}}\frontmatter@RRAPformat}
  {}{}
}%
\makeatother
\begin{document}


\title[]{A fresh look at the vibrational and thermodynamic properties of liquids within the soft potential model}
\author{Haichen Xu}
\email{xhc1021@gmail.com}
\affiliation{University of Michigan-Shanghai Jiao Tong University Joint Institute, Shanghai Jiao Tong University, Shanghai 200240, China}
\affiliation{Wilczek Quantum Center, School of Physics and Astronomy, Shanghai Jiao Tong University, Shanghai 200240, China}
\affiliation{Shanghai Research Center for Quantum Sciences, Shanghai 201315, China}
\author{Matteo Baggioli}%
 \email{b.matteo@sjtu.edu.cn}
\affiliation{Wilczek Quantum Center, School of Physics and Astronomy, Shanghai Jiao Tong University, Shanghai 200240, China}
\affiliation{Shanghai Research Center for Quantum Sciences, Shanghai 201315, China}

\author{Tom Keyes}
\email{keyes@bu.edu}
\affiliation{Chemistry Department, Boston University, Boston, Massachusetts 02215, USA}

\date{\today}

\begin{abstract}
Contrary to the case of solids and gases, where Debye theory and kinetic theory offer a good description for most of the physical properties, a complete theoretical understanding of the vibrational and thermodynamic properties of liquids is still missing. Liquids exhibit a vibrational density of states (VDOS) which does not obey Debye law, and a heat capacity which decreases monotonically with temperature, rather than growing as in solids. Despite many attempts, a simple, complete and widely accepted theoretical framework able to formally derive the aforementioned properties has not been found yet. Here, we revisit one of the theoretical proposals, and in particular we re-analyze the properties of liquids within the soft-potential model, originally formulated for glasses. We confirm that, at least at a qualitative level, many characteristic properties of liquids can be rationalized within this model. We discuss the validity of several phenomenological expressions proposed in the literature for the density of unstable modes, and in particular for its temperature and frequency dependence. We discuss the role of negative curvature regions and unstable modes as fundamental ingredients to have a linear in frequency VDOS. Finally, we compute the heat capacity within the soft potential model for liquids and we show that it decreases with temperature, in agreement with experimental and simulation data.
\end{abstract}

\maketitle
\section{Introduction}
Together with solids and gases, liquids are one of the most common phases of matter in Nature, as $\approx 71\%$ of the Earth's surface is covered with them. Nevertheless, understanding the physical properties of liquids is not as easy as ``\emph{drinking a glass of water}''. Atoms in liquids are not distributed in an ordered or even periodic structure as in solids, and their displacements around the equilibrium configuration are not ``small". Even worse, the equilibrium configuration constantly changes in time since the atomic positions re-arrange quickly via for example diffusive processes. At the same time, the density in liquids is comparable to that of solids and much higher than the one of gases. Finally, molecular interactions are strong, the ``potential energy landscape"  is highly anharmonic, and the dynamics very collective and highly correlated.

\begin{figure}
    \centering
    \includegraphics[width=\linewidth]{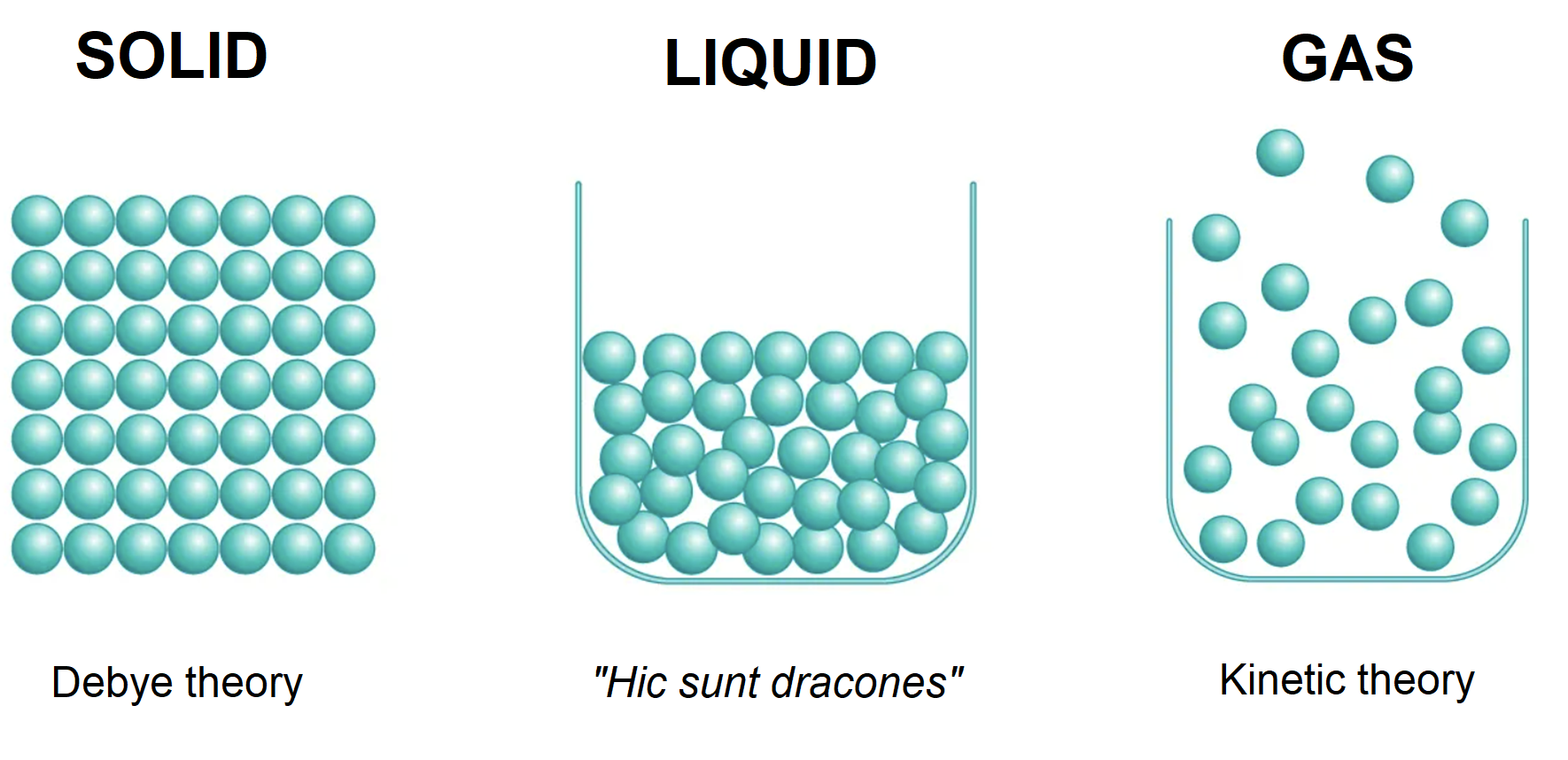}
    \caption{The physical properties of solids and gases are well described by Debye theory and kinetic theory respectively. The vibrational and thermodynamic properties of liquids are still subject of debate and a conclusive theoretical framework is missing.}
    \label{fig:0}
\end{figure}

In summary, liquids do not display any small parameter and they are therefore not amenable to any perturbative theoretical description \cite{trachenko_2023}. Because of the reasons listed above, standard theoretical frameworks such as Debye theory and kinetic theory, which are extremely successful for solids and gases respectively, are of no help in the case of liquids (see Fig.~\ref{fig:0}). Decades of effort have produced first-principles theories of liquid thermodynamics, mainly based upon the pair distribution function, but they are difficult to implement and properties are mainly obtained via computer simulation. Others are extrapolations either from the solid side or from the gas side, as we will see later. Constructing a successful theory of liquid dynamics is even more challenging.

Here, we are concerned with two important properties of liquids which determine their dynamics and thermodynamics: (I) the vibrational density of states (VDOS), $g(\omega)$, and (II) the heat capacity, $C(T)$.

The concept of VDOS in solids is well established. It counts the number of modes per unit frequency or equivalently per unit energy\footnote{As we will see, this distinction might become subtle in liquids.}:
\begin{equation}
    g(\omega)\equiv \frac{1}{V}\frac{dN(\omega)}{d\omega},
\end{equation}
where $V$ is the volume in real space and $N(\omega)$ the accumulated number of modes in the frequency range between $0$ and $\omega$. Given a concrete dispersion relation, $E(k)$ or $\omega(k)$, where $E$ is the energy, $\omega$ the frequency and $k$ the wave-vector, the VDOS in the continuum can be defined as:
\begin{equation}
    g(\omega)=\int_{V_k}\,\frac{d^dk}{(2\pi)^d}\,\delta \left( \omega-\omega(k)\right),
\end{equation}
where $V_k$ is the volume of the phase-space for the allowed wave-vector $k$ ($\mathbb{R}^d$ in the simplest cases).

In solids, the VDOS can be measured via neutron scattering experiments and computed numerically from molecular dynamics simulations as the Fourier transform of the normalized velocity auto-correlation function. In particular, defining the latter as:
\begin{equation}
    \text{VACF}(t)\equiv \frac{ \langle \vec{v}(t) \vec{v}(0)\rangle}{\langle| \vec{v}(0)|^2\rangle},
\end{equation}
where $\vec{v}$ is the velocity vector, then, the corresponding density of states is obtained from:
\begin{equation}
  \text{VACF}(t)=\int_{-\infty}^{\infty}\,\cos(\omega t)\,   g_{\text{VACF}}(\omega)\,d\omega\,.
\end{equation} 

Alternatively, using the harmonic approximation, one can compute a density of states from the distribution of the $3N$ frequencies, $\omega$, the square roots of the $3N$ eigenvalues of the Hessian matrix, defined as the second derivative of the total potential with respect to the individual particle coordinates.
We will call this VDOS, the focus of this paper, $g(\omega)$, with no subscript. Under the harmonic approximation, which is usually justified in low temperature solids by the fact that atoms are executing small displacements around a stable equilibrium configuration, the two definitions just presented are equivalent (see Fig.S3 in \cite{moon2022microscopic} for an explicit proof of this statement in crystalline silicon at $1$K).

What about liquids? One can still extract $g_{\text{VACF}}(\omega)$ from the VACF using MD simulations. However, its connection to actual vibrational modes is unclear. Zwanzig suggested that the dynamics is a sequence of intervals of harmonic vibrations around a local minimum of the many body potential \cite{doi:10.1063/1.446338}, denoted as Inherent Structures (IS) by Stillinger \cite{PhysRevA.28.2408,doi:10.1063/1.447498,stillinger2015energy,JCFHSfest}, interrupted by transitions to a new IS, and so on. The VACF is completely decorrelated by a transition. The resulting $g_{\text{VACF}}(\omega)$ can be considered to ``count" the harmonic modes at high $\omega$, but there is no simple connection to a well defined excitation at low $\omega$. This was confirmed with simulation by Seeley et al. \cite{doi:10.1063/1.457664}, who combined INM and Zwanzig's theory in a theory for the VACF. 

The computation of $g(\omega)$ from the Hessian matrix can proceed by extending the standard notion of normal modes in solids with the concept of ``instantaneous normal modes" (INM) \cite{doi:10.1021/jp963706h,doi:10.1021/ar00053a001} introduced by one of us. Use of a VDOS in liquids is not straightforward, but the Instantaneous Normal Mode (INM) VDOS describes diffusion very well \cite{doi:10.1021/jp963706h,doi:10.1063/1.468407,doi:10.1063/1.458192}, and it therefore represents a good candidate for such a role. Now, the question is how to use these finite-lifetime, anharmonic modes. 

Let us briefly explain this fundamental concept. Let us consider a collection of liquid particles in a potential $U$. At a given time $t$, the collective instantaneous configuration of all particles is indicated as $R_t$. For configurations $R$ near the instantaneous configuration, we can expand the potential:
\begin{equation}
    U(R)=U(R_t)-F \cdot \left(R-R_t\right)+\frac{1}{2}\left(R-R_t\right)\cdot H\cdot \left(R-R_t\right)+\dots
\end{equation}
where $F$ and $H$ are respectively the instantaneous force and the instantaneous Hessian matrix.
This description will hold for a time interval much less than a characteristic rearrangement time-scale $t^*$, $\Delta t \ll t^*$. To avoid clutter, we have neglected all tensorial and vectorial indices. 
 
 One can then diagonalize the $H$ matrix and obtain the eigenvalues $\lambda_\alpha\equiv \omega^2_\alpha$. In liquids, the eigenvalues can be either positive or negative, reflecting the fact that the potential energy landscape displays regions with both positive and negative curvature. We call the modes corresponding to positive eigenvalues the stable instantaneous normal modes. By contrast, the ones associated to negative eigenvalues are labelled as unstable instantaneous normal modes. The INM density of states can then obtained from the ensemble or time averaged distribution of frequencies obtained after averaging over many configurations $R_0$. In particular, upon normalizing to the total number of modes, $3N$, we obtain:
\begin{equation}
    g_{\text{INM}}(\omega)\equiv \Big\langle \frac{1}{3N}\sum_{\alpha=1}^{3N}\delta\left(\omega-\omega_\alpha(R_t)\right)\Big\rangle_{R_t} \,.
\end{equation}
Clearly, one can separate the total density of states into stable and unstable modes. We will indicate the two as $g_s(\omega)$ and $g_u(\omega)$ dropping the label INM. The unstable INM DOS is by convention plotted on the negative frequency axis, and for comparison of $g_s(\omega)$ and $g_u(\omega)$ we will treat both arguments as positive, hoping it does not cause confusion.

The INM framework is expected to describe the short-time liquid dynamics. Nevertheless, the INM analysis is surprisingly  relevant for the late-time dynamics, as the fraction of unstable modes, $f_u$ expresses the $T$-dependence of the diffusion constant $D$, very well \cite{doi:10.1063/1.458192,doi:10.1063/1.469947,doi:10.1063/1.468407,doi:10.1063/1.3701564,doi:10.1063/1.478211,doi:10.1063/1.475376,PhysRevE.65.026125}. Since this is not relevant for our discussion, we will not indulge into too many details, and we refer to the more complete reviews \cite{doi:10.1021/jp963706h,doi:10.1021/ar00053a001}.

Importantly, the low-frequency behavior for the density of states of liquids is substantially different from that of solids and does not exhibit any Debye-like scaling law, $g(\omega)\propto \omega^2$. On the contrary, the INM density of states of liquids displays a characteristic linear in frequency scaling at low frequency which exists in both its unstable and stable parts. In fact, at low frequency, $g_s(\omega)$ and $g_u(\omega)$ are identical, and vanish at $\omega = 0$. On the other hand, the density of states for liquids extracted from experiments \cite{stamper2022experimental}, and that obtained from the VACF, have a nonzero value at $\omega = 0$ determined by the  diffusion constant,
\begin{equation}
    g_{\text{VACF}}(\omega=0)=3D\,.
\end{equation}

Once such term is removed, also the experimental DOS displays a universal linear power-law $g(\omega)\propto \omega$ in the regime of low-frequency \cite{stamper2022experimental}. The observation of an approximate low-frequency linear scaling has been reported using neutron-scattering experiments in liquid Selenium by Phillips et al. as early as 1989 \cite{PhysRevLett.63.2381}, \footnote{We thank the anonymous Referee for bringing to our attention this important reference.} and discussed by Buchenau in the context of the SPM in \cite{BUCHENAU1993275}. It is a priori not clear whether the linear slope at low frequency in $g_{\textit{INM}}(\omega)$ coincides with that reported in experiments, or obtained using $g_{\text{VACF}}(\omega)$ (see \cite{jin2023dissecting} for a very recent study). In any case, no trace of any Debye contribution has been seen in the DOS of liquids at low frequency so far, even if often used in the phonon approaches to liquid dynamics.

This new linear scaling law, which is absent in solids, can therefore be thought as a qualitative difference between solids and liquids. 
It is natural to assume that it arises because of the emergence of unstable modes with imaginary frequency. Despite different theories having been proposed for this linear behavior and its slope \cite{doi:10.1063/1.457564,doi:10.1063/1.465563,doi:10.1063/1.467178,doi:10.1073/pnas.2022303118,doi:10.1073/pnas.2119288119,doi:10.1063/1.479810,PhysRevE.55.6917,doi:10.1063/1.463375}, a generic consensus has not been reached yet and the debate is still open. Nevertheless, from the point of view of simulations, the temperature and frequency dependence of the INM DOS has been studied in several systems (see for example \cite{doi:10.1063/1.468407,keyes1997instantaneous}), and general statements can be made.

A second quantity of interest is the heat capacity. The heat capacity of solids is well-described by Debye theory \cite{kittel2004introduction}. At low temperature, it grows as $C(T)\propto T^3$, which is an immediate consequence of the quadratic scaling of the density of state at low frequency and the quantum nature of phonons therein. At larger temperature, above the so-called Debye temperature, the system becomes classical and the heat capacity approaches the Dulong-Petit value, $C(T)=3N k_B$. 

In the opposite limit, at extremely large temperatures, within the gas phase, the potential energy is negligible and the heat capacity comes solely from the kinetic energy of the thermally activated free (or almost free) particles. Using the equipartition theorem, in that limit, each degree of freedom carries an amount of energy given by $(1/2) k_B T$. Thus, the heat capacity of monoatomic gases at constant volume is given by $C(T)=(3/2) N k_B$ which is smaller than that of solids in the classical limit, $C(T)=3 N k_B$.\footnote{In this discussion, we are neglecting internal degrees of freedom and anharmonic effects, which can be possibly included in a perturbative way (see for example \cite{TOGO20151}).} Were one to simply assume that $C(T)$ is a continuous and monotonic function, one would expect the heat capacity of liquids to decrease with temperature,
and this is exactly what happens (see for example the experimental data reported in \cite{bolmatov2012phonon}). However, there is nothing continuous about the liquid/vapor first order transition. Furthermore, the $(3/2) N k_B$ result for gases is obtained assuming no contribution from potential energy, and this is never the case in liquids. Hence, the validity of this simple argument is questionable.

In any event, solid and liquid behave very differently. The main question is not really why but rather how to rationalize and, even qualitatively, explain the decrease in the heat capacity of liquids with increasing temperature using a few simple but fundamental physical properties. Different theoretical explanations have been proposed for this phenomenon \cite{bolmatov2012phonon,bryk2015heat,PhysRevE.104.014103,madan1993unstable,doi:10.1080/00268979809483145,moon2022microscopic,PhysRevE.57.1717}. For the sake of brevity, we will not analyze in detail any of them but we limit ourselves to state that a final verdict on this question is yet to come.

For the temperature ranges usually studied, solids are in the quantum regime and liquids are not. This alone influences any comparison of their temperature dependences. For example, the Debye $T^3$ increase of the heat capacity is entirely due to quantized oscillators with $\omega>k_BT/\hbar$ giving no contribution, so more contribute with increasing $T$. We must take care to not compare quantum real systems theory with our classical liquid calculations.

Among the various theoretical ideas suggested to solve the problem of liquids, and in particular to describe their density of states and heat capacity, one of them is based on the attempt to borrow concepts developed in the field of amorphous systems \cite{doi:10.1142/q0371}. In particular, it has been proposed in \cite{PhysRevE.55.6917}, that the soft-potential model for glasses \cite{doi:10.1142/9781800612587_0008} (SPM), once opportunely modified, could describe part of the liquids dynamics, and give results in agreement with simulations.\footnote{See also \cite{moriel2023boson} for a recent connection between the linear VDOS of liquids and the reconstructed $\propto \omega^4$ VDOS in glasses.} Furthermore, another class of liquid theories is based on the potential energy landscape viewpoint \cite{stillinger2015energy}. The SPM may be viewed as providing a highly simplified landscape, allowing tests of ideas about how liquid properties depend on barrier heights and other landscape features.\\

In this work, we revisit and expand the theoretical analysis of such a model. In particular, we compute in detail the stable and unstable density of states of liquids from the soft potential model, building on the results of \cite{PhysRevE.55.6917}. Furthermore, we show that the soft potential model can qualitatively predict the correct liquid heat capacity and its dependence on temperature. Finally, in revisiting the computations of \cite{PhysRevE.55.6917}, we fix some minor mistakes therein which do not affect the main physical conclusions.

\section{Soft potential model}
The soft potential model (SPM) was originally introduced as an extension of the tunneling states model by the St. Petersburg group \cite{karpov1983theory,Parshin_1993}, and by Buchenau and collaborators \cite{PhysRevB.43.5039,PhysRevB.46.2798}, to explain the low-temperature anomalies in amorphous solids. The main idea is that in disordered systems quasi-localized soft modes coexist with vibrational modes (phonons) and they are described by a distribution of 1-dimensional anharmonic potentials of this form
\begin{equation}
    V(x)=W\left[D_1 x -D_2 x^2+ x^4\right],\label{pot}
\end{equation}
where $x$ is a dimensionless spatial coordinate, $W$ a characteristic energy scale and $D_1,D_2$ two random parameters. We notice that in the original version of the model, a cubic term was introduced and the linear term proportional to $D_1$ was neglected. As explicitly described in some of the early references, see for example \cite{PhysRevLett.70.182}, the cubic term can be set to vanish by appropriately choosing the origin of the generalized coordinate system, at the cost of introducing a linear contribution. In this sense, the two choices are equivalent, but, importantly, the corresponding distribution of the random parameters are not preserved by this map \cite{PhysRevB.43.5039}. In any case, in the rest of the manuscript, we will only consider the version of the SPM defined in Eq.\eqref{pot} (without the cubic term) which is the one mostly used in the modern literature. Furthermore, as we will argue extensively, the asymmetry of the potential, induced in our case by the linear term, does not play a major role in our discussion. Therefore, the distinction between the linear term or the cubic term as the source of such an asymmetry is only tangential to our work and its effects are very limited. Also, since we will be mostly interested in double well potentials, with the quadratic term $x^2$ being negative, we have defined the quadratic term in the potential Eq.\eqref{pot} with an overall minus sign. For an extensive review of the soft potential, its properties and definitions we refer to the chapter by U. Buchenau in \cite{doi:10.1142/9781800612587_0008}. Finally, depending on the value of the various parameters appearing in \eqref{pot}, the potential displays two (in general asymmetric) local minima separated by a barrier, or one global minimum together with a highly anharmonic shoulder. 

\begin{figure}[h]
    \centering
    \includegraphics[width=\linewidth]{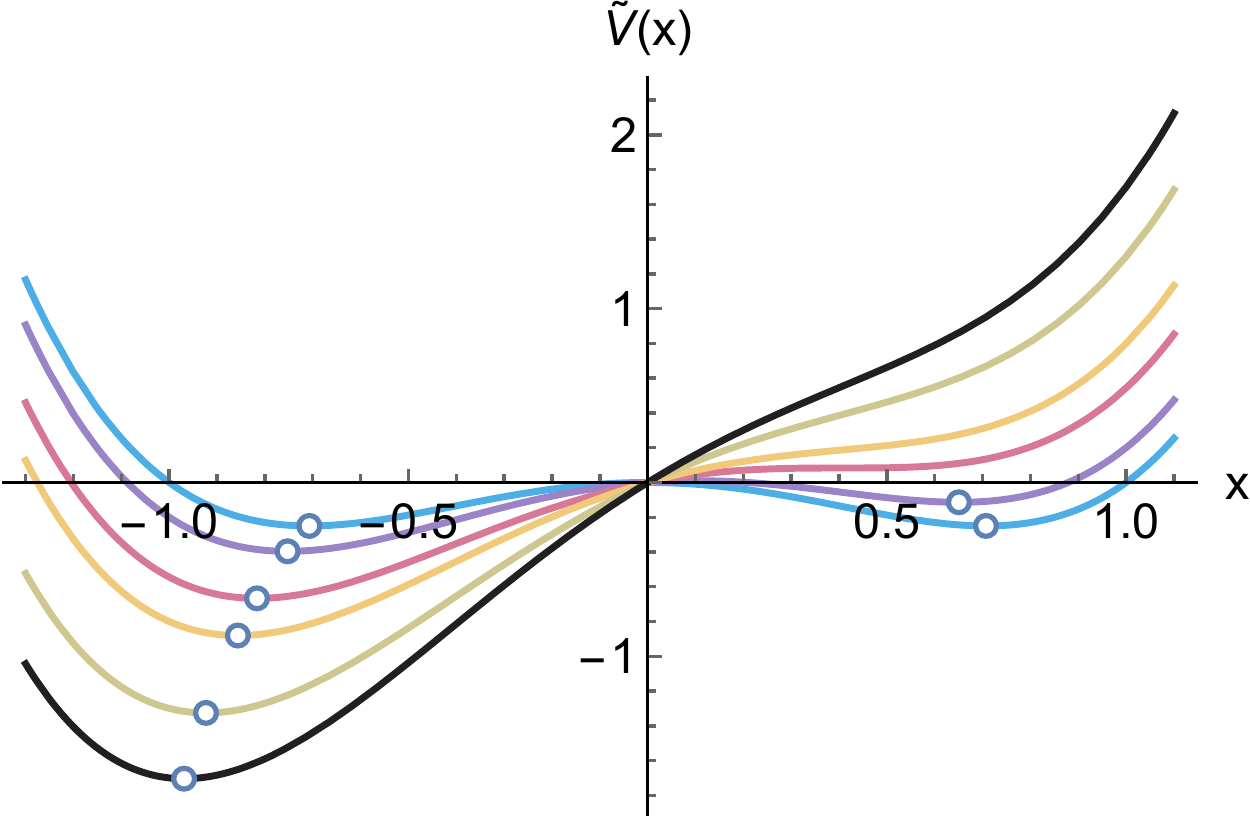}
    \caption{The reduced SPM potential $\Tilde{V}(x)\equiv V(x)/W$ for $D_2=1$ and $D_1=0,0.2,0.54,0.8,1.3,1.7$ (from light blue to black). The critical value for the presence of double wells corresponds to the pink curve. The open circles locate the position of the potential minima.}
    \label{fig:2}
\end{figure}

In general, the parameter $D_1$ controls the degree of asymmetry of the potential $V$, while the parameter $D_2$ controls how deep the well is. Large $D_2$ and small $D_1$ result in a double well potential. On the other hand, small $D_2$ and large $D_1$ give rise to a single well potential. More precisely, we have that
\begin{equation}
    \begin{aligned}
         (D_2/6)^3 &> (D_1/8)^2: \quad \text{double-well potential},\\
         (D_2/6)^3 &< (D_1/8)^2: \quad \text{single-well potential}\,.
    \end{aligned}
\end{equation}
In Fig.\ref{fig:2}, we display these features by dialing the parameter $D_1$.

Following the original discussion in \cite{PhysRevE.55.6917}, here, we assume that $D_2$ is a random variable with a uniform distribution up to an upper cutoff, while $D_1$ is Gaussian distributed:
\begin{align}
   & p_1(D_1)=0.231\left(\frac{T_g}{W}\right)^{3/4}\exp\left[-0.169\left(\frac{W}{T_g}\right)^{3/2}D_1^2\right],\nonumber\\
   & p_2(D_2)=p_2^{(0)}=\text{const. ,}\quad 0<D_2<\frac{\Omega^2}{4W}\,.\label{distr}
\end{align}
We have denoted with $T_g$ the glass transition temperature, and with $\Omega$ a frequency cutoff which determines the width of the $D_2$ distribution. A physical motivation for these choices can be found in \cite{PhysRevE.55.6917} and will not be repeated here.
Let us emphasize that all the discussion in this work will be totally classical, and we ignore the liquid/solid transition of classical liquids. Our simple treatment should be considered only above the glass transition temperature, $T>T_g$.\\

We also emphasize that, while we explore the full range of $D_1$ behavior above, the relevant values for the temperatures of interest to us are very small, corresponding to slightly asymetric double wells (See Fig~\ref{fignew} and associated discussion.)

Let us reiterate that several variants of the SPM are in the literature. Our model (and admittedly the same one used in \cite{PhysRevE.55.6917}) does not coincide exactly with the SPM for glasses, and it might be rather defined as an extension of it which is more suitable to describe liquids (see Appendix \ref{secfrom} for a discussion about the application of the SPM to glasses and liquids). Interestingly, U. Buchenau already considered in 1993 the possibility that modes corresponding to small positive or negative eigenvalues in liquids should be identical to the soft modes in the SPM \cite{BUCHENAU1994391,BUCHENAU1993275}, envisaging a connection between liquids dynamics and quasi-localized anharmonic soft modes.

Before proceeding, let us explain further the motivation behind the choice of the SPM model in the context of liquids. First, symmetric single wells are neglected as we do not explore the negative range of $D_2$, see \eqref{distr}. The reason for it is that we have pursued the viewpoint (Stillinger \cite{PhysRevA.28.2408}, Zwanzig \cite{doi:10.1063/1.446338}, Frenkel \cite{FrenkelBook46}) that the dynamics in low temperature liquids consist of solid-like vibrations around local minima of the potential interrupted by atomic-rearrangements in the form of jumps over energy barriers or saddle points in the potential. The simplest representation of this picture is a symmetric double well $V= -D_2 x^2+x^4$, with positive $D_2$. The cutoff in Eq.~\eqref{distr} sets the maximum barrier height.
We recognize that a symmetric single well potential of the form $V= x^2+x^4$ would describe many normal coordinates in a liquid, but not the diffusive reaction pathways.

Then, we add the $D_1 x$ term to include asymmetry, which has been invoked \cite{PhysRevLett.70.182} to explain the heat capacity of glasses and supercooled liquids. The term arises from thermal stress that is frozen in at the glass transition \cite{ferrari1987heat}, hence the form of $p(D_1)$. Notice that anharmonic single wells arise also in our potential for  large $D_1$. Nevertheless, given the distribution chosen, such a configuration has very low probability and plays a minor role, as will be explicitly demonstrated in the following sections. For this same reason, we do not think the choice of $D_1 x$ instead of $D_3 x^3$ as the source of asymmetry has significant effects.

Having selected and analyzed a variant of the SPM, applying it to liquids requires additional steps. The potential energy landscape for a real liquid consists of a very large number of local minima, supporting $3N$ independent modes, connected by barriers. In equilibrium, the particle dynamics contain two parts: oscillation around the local minima (as for solids), and jumps from one minimum to another local minimum. We aim to to provide an effective description of the corresponding low energy physics with 3N independent soft potentials. However, as just discussed, at least for the relevant values of $D_1$, we do not have single wells, so this idea does not include a simple, primarily single well, multidimensional basin. Note, however, that the individual wells of the double well potential are perfectly good deep single wells for large $D_2$, capable of supporting harmonic dynamics. Thus, the INM DOS in our SPM has stable and unstable branches, and the crucial fraction of unstable modes, $f_u$, behaves much like it does in liquids.

Nonetheless it must be admitted that the deep multidimensional basins in this picture are more like ``sombrero hats" with a barrier at the origin rather than simple harmonic minima. Probably a better procedure would be to model the basins with some soft modes, with fraction $f_{sp}$, and some harmonic modes, with fraction (1-$f_{sp})$. Then the total unstable fraction would be $f_{sp} f_u^{(SPM)}$, introducing at least one other parameter, and similar considerations apply to calculating other quantities. These ideas might be implemented by allowing an appropriate distribution of negative values of $D_2$. 

Without the help of simulations, or without performing a more quantitative comparison between the theory and simulation data, we are indeed unable to establish the number of atoms participating in the motion described by the SPM, \textit{i.e.}, $f_{sp}$. In summary, since we are neglecting the harmonic modes, our results for the fraction of unstable modes $f_u$ are for soft modes only, and those for the heat capacity are on a per-soft-mode basis. Conclusions about qualitative trends are unaffected. We leave for the near future a more comprehensive study which includes also the harmonic modes. 

\section{The density of states of liquids}
The first quantity of interest is the DOS for stable and unstable modes. As a disclaimer, let us notice that the DOS that we will compute is the one arising from all the modes described by the anharmonic potential in Eq.\eqref{pot}. At this point, it is not clear whether that is the totality of excitations in liquids. For simplicity, and for lack of any other means, we will consider that it is.
\subsection{Stable modes}
Let us start by considering stable modes for which $\omega^2>0$.
Let us define the unaveraged DOS, for fixed $D_1$ and $D_2$, as:
\begin{equation}
    \tilde g_s(\omega)= 2 ~\omega\, G_{s}(\omega^2),
\end{equation}
where:
\begin{equation}
    G_{s}(\omega^2)\equiv \Big\langle \delta \left(\frac{d^2 V}{dx^2}-\omega^2\right)\Big\rangle_x \,.\label{gen1}
\end{equation}
The average indicated with $x$ subscript is defined with respect to the Boltzmann factor as follows:
\begin{equation}
    \langle f(x) \rangle_x\equiv \frac{1}{N_x}\int_{-\infty}^{+\infty} f(x) e^{-V(x)/T}dx,
\end{equation}
where the normalization constant is given by
\begin{equation}
    N_x=\int_{-\infty}^{+\infty} e^{-V(x)/T}dx\,.
\end{equation}
Using standard mathematical identities, the delta function in \eqref{gen1} can be re-written as
\begin{equation}
    \delta \left(\frac{d^2 V}{dx^2}-\omega^2\right)=\frac{\delta( x-x_\omega^+)}{24\,W\,|x_\omega^+|}+ \frac{\delta( x-x_\omega^-)}{24\,W\,|x_\omega^-|},
\end{equation}
where $x^\pm_\omega$ are the roots of 
\begin{equation}
    \frac{d^2 V(x)}{dx^2}-\omega^2=0,
\end{equation}
which for the concrete potential in Eq.\eqref{pot} read
\begin{equation}
     x_\omega^\pm=\pm \frac{1}{\sqrt{12}}\,\sqrt{2 D_2+\omega^2/W}\,.
\end{equation}
Using standard mathematical properties of the delta function, the average in \eqref{gen1} can be performed analytically and gives
\begin{align}
    G_s(\omega^2)=&\left\langle \delta \left(\frac{d^2 V}{dx^2}-\omega^2\right)\right\rangle_x \nonumber\\=&\frac{1}{N_x}\left[\frac{e^{-V(x_\omega^+)/T}}{24\,W\,x_\omega^+}+ \frac{e^{-V(x_\omega^-)/T}}{24\,W\,x_\omega^+}\right],
\end{align}
and finally:
\begin{equation}
   \tilde  g_s(\omega)=\frac{1}{N_x} \frac{\omega}{12\,W}\left(\frac{e^{-V(x_\omega^+)/T}}{x_\omega^+}+ \frac{e^{-V(x_\omega^-)/T}}{x_\omega^+}\right),\label{fin1}
\end{equation}
which is a function of the parameters $D_1,D_2,W,T$ in addition to the frequency $\omega$. Here, we take the opportunity to correct a mistake in the original computations of \cite{PhysRevE.55.6917} (cfr. Eq.(3.10) therein) in which the $1/x_\omega^+$ factors are missing. As we will see, despite this mistake, all the main conclusions of \cite{PhysRevE.55.6917} remain.

Using the concrete expression for the potential $V$ in Eq.\eqref{pot}, we find our final expression
\begin{widetext}
\begin{equation}
\tilde g_s(\omega)=\frac{\omega}{N_x}\frac{\left(1+\exp \left[\frac{D_{1} W \sqrt{2 D_{2}+\omega^{2} W}}{\sqrt{3} T}\right]\right) \exp \left[\frac{4 W^{2}\left(5 D_{2}^{2}-6 D_{1} \sqrt{6 D_{2}+\frac{3 \omega^{2}}{W}}\right)+8 D_{2} \omega^{2} W-\omega^{4}}{144 T W}\right]}{2 W \sqrt{6 D_{2}+\frac{3 \omega^{2}}{W}}}\,.
\end{equation}
\end{widetext}
At low frequency, $\omega 
\ll1$, the previous expression reduces to

\begin{equation}
\tilde g_s(\omega) \approx \omega \,\frac{e^{\frac{W\left(5 D_{2}^{2}-6 \sqrt{6} D_{1} \sqrt{D_{2}}\right)}{36 T}}\left(1- e^{\frac{\sqrt{\frac{2}{3}} D_{1} \sqrt{D_{2}} W}{T}}\right)}{2  \sqrt{6D_{2}} W}\,+\dots
\end{equation}
where the ellipsis indicates higher corrections in the frequency $\omega$.

As anticipated, the resulting DOS of stable modes is linear in frequency at small frequencies contrary to the standard Debye behavior, $g(\omega)\sim \omega^2$. This is a direct consequence of the fact that $G_s(0)\neq 0$.\\
As a final step, we want to average over the random parameters in the potential $D_1,D_2$, using their distributions as defined in Eqs.\ref{distr}. The final density of states for stable INM is given by:
\begin{equation}
    g_s(\omega)=\int \int p_1(D_1)\,p_2(D_2)\,\tilde g_s(\omega)\,dD_1 dD_2\,.
\end{equation}

\subsection{Unstable modes}
We now switch to the computation of the density of states for unstable INM. In order to do that, let us consider the imaginary frequency $\nu$ defined as:
\begin{equation}
    \nu\equiv i \omega \,,
\end{equation}
which therefore corresponds to the square root of the absolute value of the negative eigenvalues, $\nu=\sqrt{|\lambda|}$.
The solutions of $ d^2 V / dx^2+\nu^2=0$ are now:
\begin{equation}\label{infle}
    x_\nu^\pm=\pm \frac{1}{\sqrt{12}}\,\sqrt{2 D_2-\nu^2/W}\,.
\end{equation}
\begin{figure}[t]
    \centering
\includegraphics[width=\linewidth]{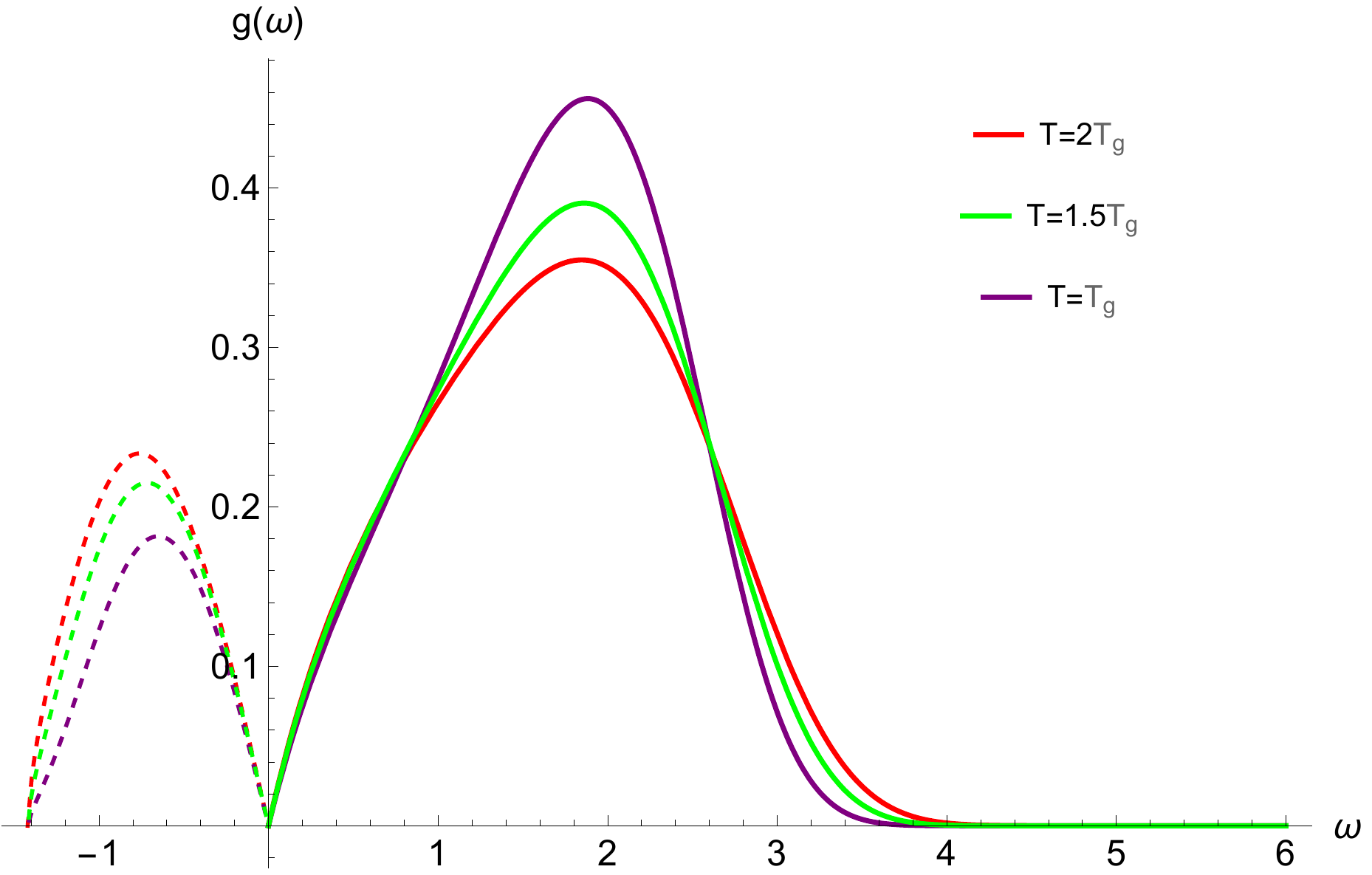}
    \caption{The density of states for stable and unstable modes as a function of the real (right axes) and imaginary (left part of the $x$-axes) frequency respectively. Here, $W=1$, $T_g=0.1$ and $\Omega=2$.}
    \label{fig3}
\end{figure}
The unaveraged DOS is then given by
\begin{align}
G_{u}(\nu^2)=&\frac{1}{N_x}\left[\frac{e^{-V(x_\nu^+)/T}}{24\,W\,x_\nu^+}+ \frac{e^{-V(x_\nu^-)/T}}{24\,W\,x_\nu^+}\right]\nonumber \\=&\frac{1}{24W N_x x_\nu^+ }\left[ e^{-V(x_\nu^+)/T} + e^{-V(-x_\nu^+)/T} \right]\,.\label{tt}
\end{align}
Using the same steps as in the previous section, together with the formula $\tilde g_u(\nu)=2 \nu G_u(\nu^2)$ we can obtain first the unaveraged DOS for unstable modes, and then its averaged version.
Notice that, differently from the case of stable modes, Eq.(\ref{infle}) now implies an upper bound for the imaginary frequencies given by
\begin{equation}\label{maxi}
    \nu_{max}=\sqrt{2D_2 W}.
\end{equation}
Finally, let us notice that the unaveraged unstable DOS obtained from Eq.\eqref{tt} displays a singularity at $\nu_{max}$, since $x_\nu^+$ approaches zero as $\nu \rightarrow \nu_{max}$. Interestingly, this singularity is resolved after averaging over the random parameters $D_1$ and $D_2$.
\subsection{Numerical results}
In Fig.\ref{fig3}, we show a benchmark example for the density of states of both stable and unstable INM. The DOS are normalized such that:
\begin{equation}
    \int g_s(\omega) d\omega+ \int g_u(\omega) d\omega\equiv N_s+N_u=1\,.
\end{equation}
In order to visualize both of them in the same plot, we used the common convention of plotting the stable ones on the positive frequency axes and the unstable ones on the negative frequency axes. We can observe that, as expected, both curves are linear in frequency for small frequencies independently of the values of the parameters. Moreover, at least at small frequency, $g_s(\omega)=g_u(\omega)$, as the curves are symmetric under reflection, $\omega \rightarrow - \omega$. Additionally, we see the both the slope of the linear regime and the shape of the curves strongly depend on the temperature $T$. Finally, in the distribution of stable modes, we see that the spectral weight is shifted towards higher frequencies by increasing the temperature and the distribution becomes less localized and more smoothed out.

\begin{figure}
    \centering
    \includegraphics[width=\linewidth]{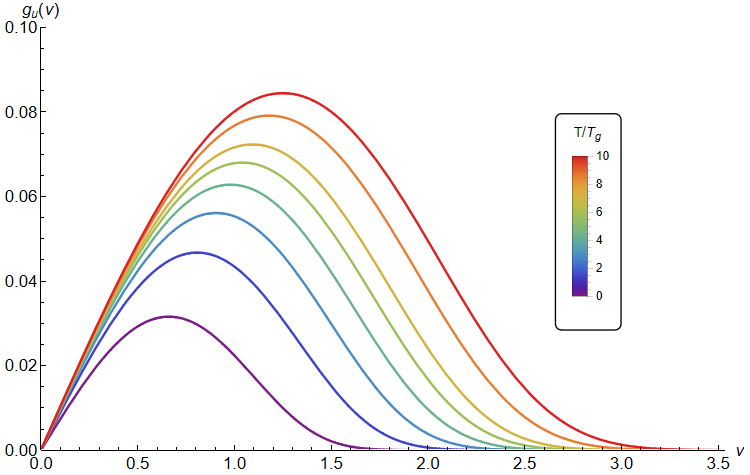}

    \vspace{0.5cm}
    
    \includegraphics[width=\linewidth]{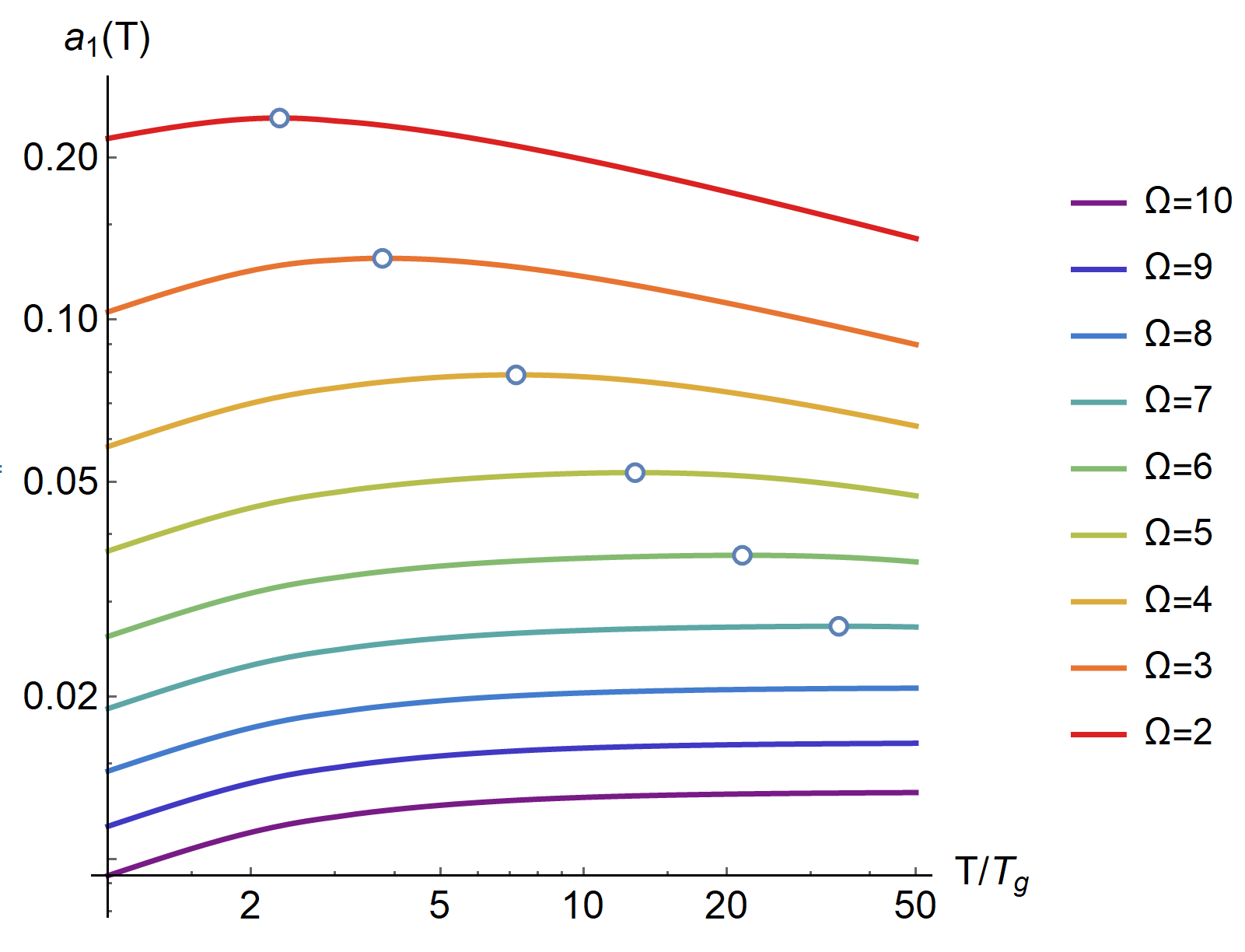}
    \caption{Top: The density of states for unstable modes $g_u(\nu)$ as a function of the imaginary frequency $\nu=i \omega$. Here, $W=1$, $T_g=0.1$ and $\Omega=5$. The maximum frequency is $\nu_{max}=5/ \sqrt{2} \approx 3.54$. Bottom: The low-frequency slope $a_1(T)$ as a function of the reduced temperature $T/T_g$ with $W=1,T_g=0.1$ for different values of the frequency cutoff $\Omega$. The empty circles indicate the position of the maxima.}
    \label{figdos}
\end{figure}

Our main interest is to characterize the density of unstable modes. Therefore, from now on, we will concentrate on the unstable part of the density of states and we will analyze in detail its features. In the top panel of Fig.\ref{figdos}, we show the behavior of the INM density of states of unstable modes as a function of the reduced temperature $T/T_g$. First, let us notice that the unstable DOS goes to zero at the maximum frequency $\nu_{\text{max}}$. Second, the spectral weight in the unstable DOS, and the position of its  maximum as a function of the imaginary frequency $\nu$, moves to higher frequencies by increasing the temperature. Its intensity grows as well with increasing $T$.

Finally, the linear slope at small frequencies, $g_u(\nu)\propto a_1 \nu$, depends on the temperature and in particular, at least within the soft potential model,  grows with it for this choice of parameters. To verify whether this is a general feature of the model or not, we have performed a more detailed analysis for different values of the ultraviolet (UV) cutoff $\Omega$. The results are shown in the bottom panel of Fig.\ref{figdos}. For large values of the cutoff $\Omega$, the linear slope $a_1(T)$ first grows with temperature and then reaches a plateau at very large values of $T\approx 10 T_g$. On the contrary, when the UV cutoff $\Omega$ becomes small, the slope exhibits a maximum as a function of $T$ which moves to lower temperatures by decreasing $\Omega$. Finally, the value of the linear coefficient at fixed temperature decreases monotonically with the UV cutoff $\Omega$.

\begin{figure}
    \centering
    \includegraphics[width=\linewidth]{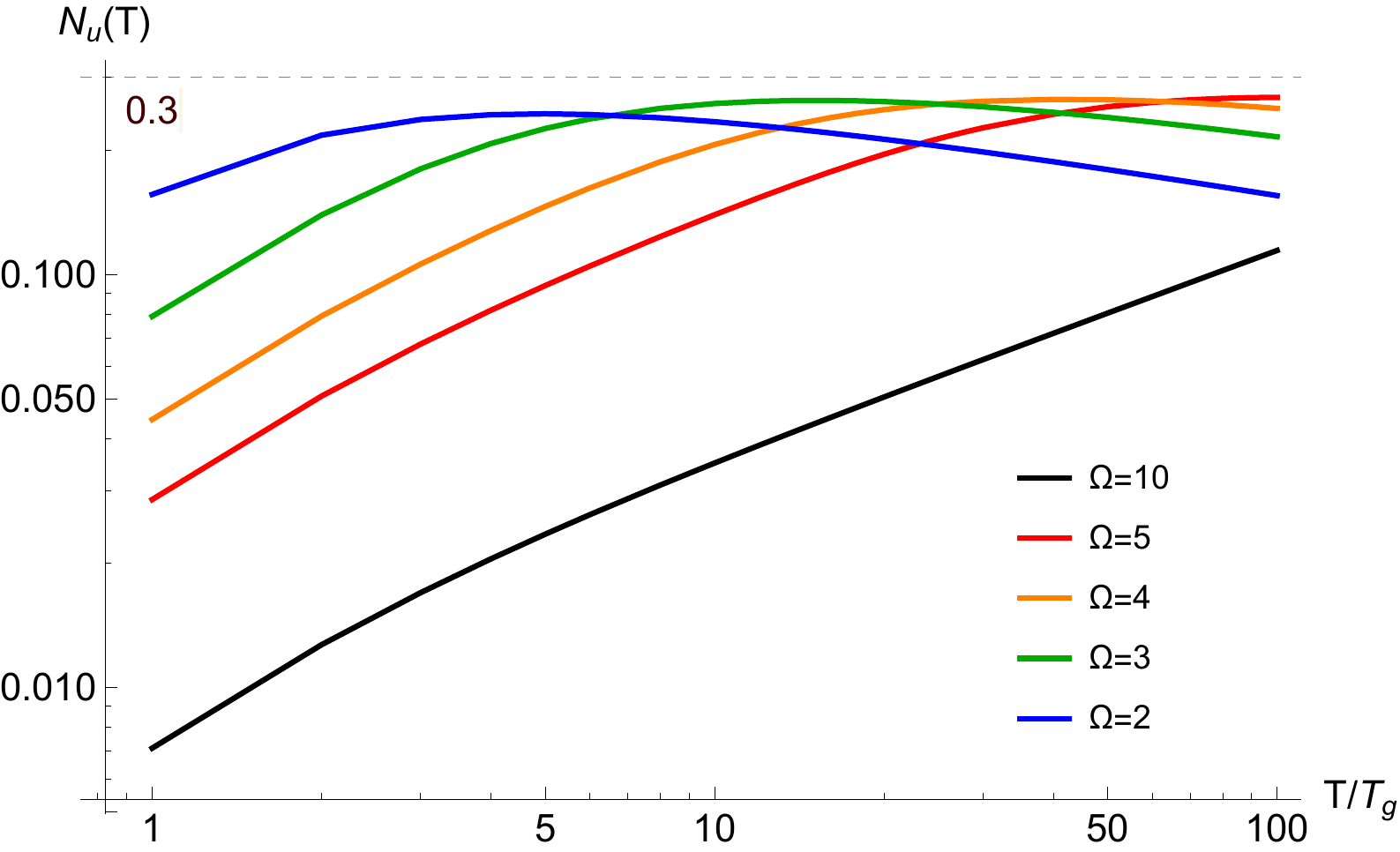}

    \vspace{0.5cm}
    
    \includegraphics[width=\linewidth]{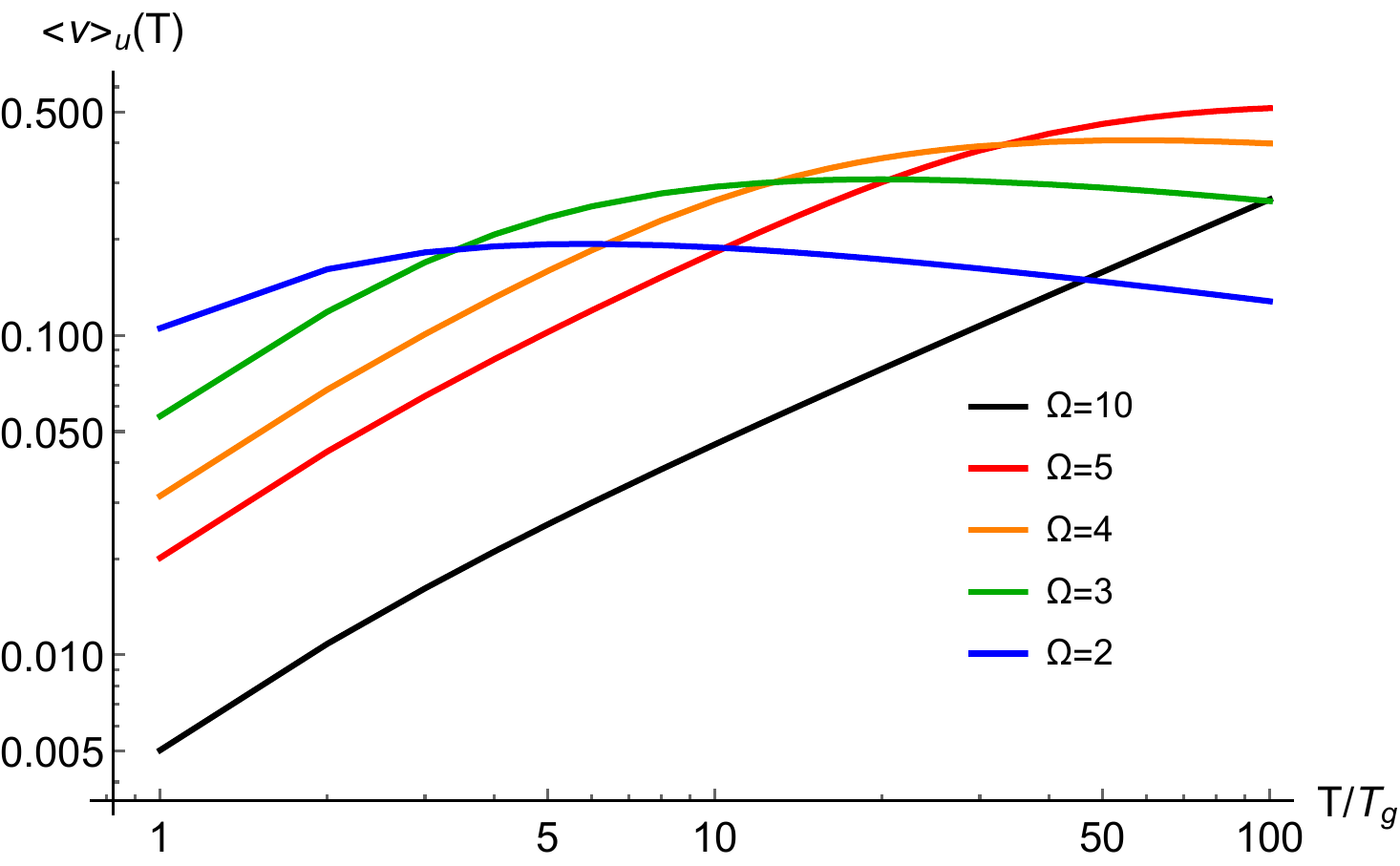}
    \caption{Top: The fraction of unstable modes $N_u$ from Eq.\eqref{numero} as a function of the reduced temperature $T/T_g$. Parameters are fixed to $W=1,T_g=0.1$. The horizontal dashed line indicates the reference value $N_u=0.3$. Bottom: The average frequency $\langle \nu\rangle_u$ from Eq.\eqref{ave} as a function of the reduced temperature $T/T_g$ for the same choice of parameters.}
    \label{fig:4}
\end{figure}

The number of unstable modes is:
\begin{equation}
    N_u\equiv \int_0^{\nu_{max}} g_u(\nu) d \nu\,,\quad \nu_{max}=\frac{\Omega}{\sqrt{2}}\,.\label{numero}
\end{equation}
Since we are studying the $1-$dimensional soft potential, \textit{i.e.}, $N_{tot}=1$, then $N_u$ in Eq.\eqref{numero} is actually the fraction of unstable modes, denoted $f_u$ in the literature. We will then use the two symbols interchangeably. The corresponding fraction of stable INMs can then be simply derived as:
\begin{equation}
    N_s=N_{tot}-N_u=1-f_u\,.
\end{equation}
Finally, we can also define the average imaginary frequency:
\begin{equation}
    \langle \nu \rangle_u\equiv \int_0^{\nu_{\max}} \nu~ g_u(\nu)\, d\nu \,.\label{ave}
\end{equation}
In Fig.\ref{fig:4}, we show the fraction of unstable INM and the average imaginary frequency as a function of the reduced temperature $T/T_g$ by dialing the UV cutoff $\Omega$. The behavior of the two quantities is qualitatively very similar and also analogous to that of the linear coefficient $a_1$, shown in the bottom panel of Fig.\ref{figdos}. For large UV cutoff, both quantities grow monotonically with $T$. On the contrary, by decreasing the UV cutoff, a maximum appears and it moves gradually to low temperature. In real liquids, the fraction of unstable modes grows monotonically with the temperature until reaching a high-$T$ plateau. Let us provide a heuristic argument to explain the difference between the model and real liquids.

In liquids, as T increases, the system explores higher and higher saddle/barrier regions, for more imaginary frequencies. However, in the SPM there is just one barrier region, and once T is high enough to explore it fully the only remaining new configurations are those further from the origin in either direction, and they have real frequencies. So adding more of them can reduce $f_u$. It is also reasonable that this effect is favored by small $\Omega$, as then the barriers are lower and their ``saturation" more easily attained. In other words, we can conclude that for small values of $\Omega$, the model is a good description for real liquids only in the low temperature region. By increasing the value of $\Omega$, this region extends to higher temperature and the regime of validity of the model becomes larger.

 In order to extract more information from the unstable DOS, we can use the fitting formula suggested in \cite{keyes1997instantaneous}, where the unstable DOS has been parametrized as
\begin{equation}\label{fitt}
    g_u(\nu)\,=\, 2 a_1(T)\,\nu\,\exp \left[-\left(\frac{a_2(T)\,\nu}{\sqrt{T}}\right)^{a_3(T)}\right].
\end{equation}

\begin{figure}
    \centering
    \includegraphics[width=\linewidth]{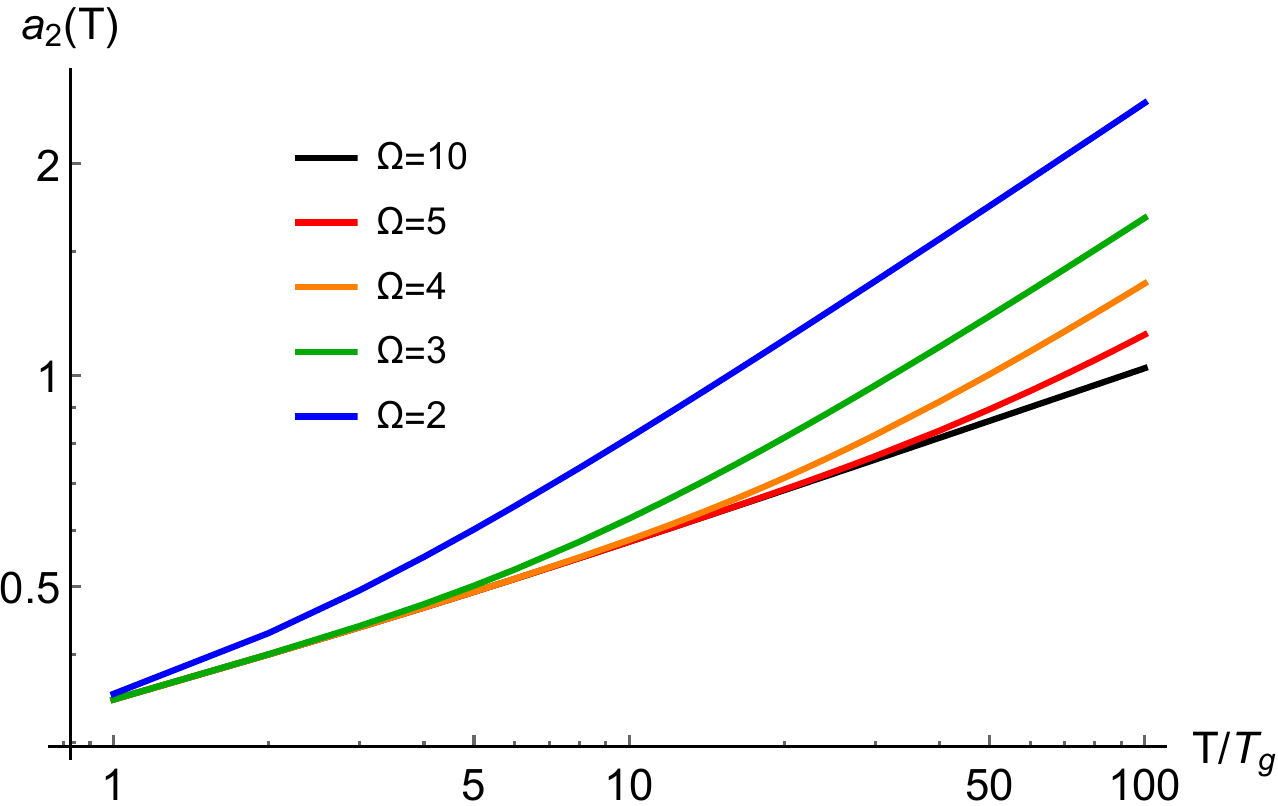}

    \vspace{0.5cm}
    
\includegraphics[width=\linewidth]{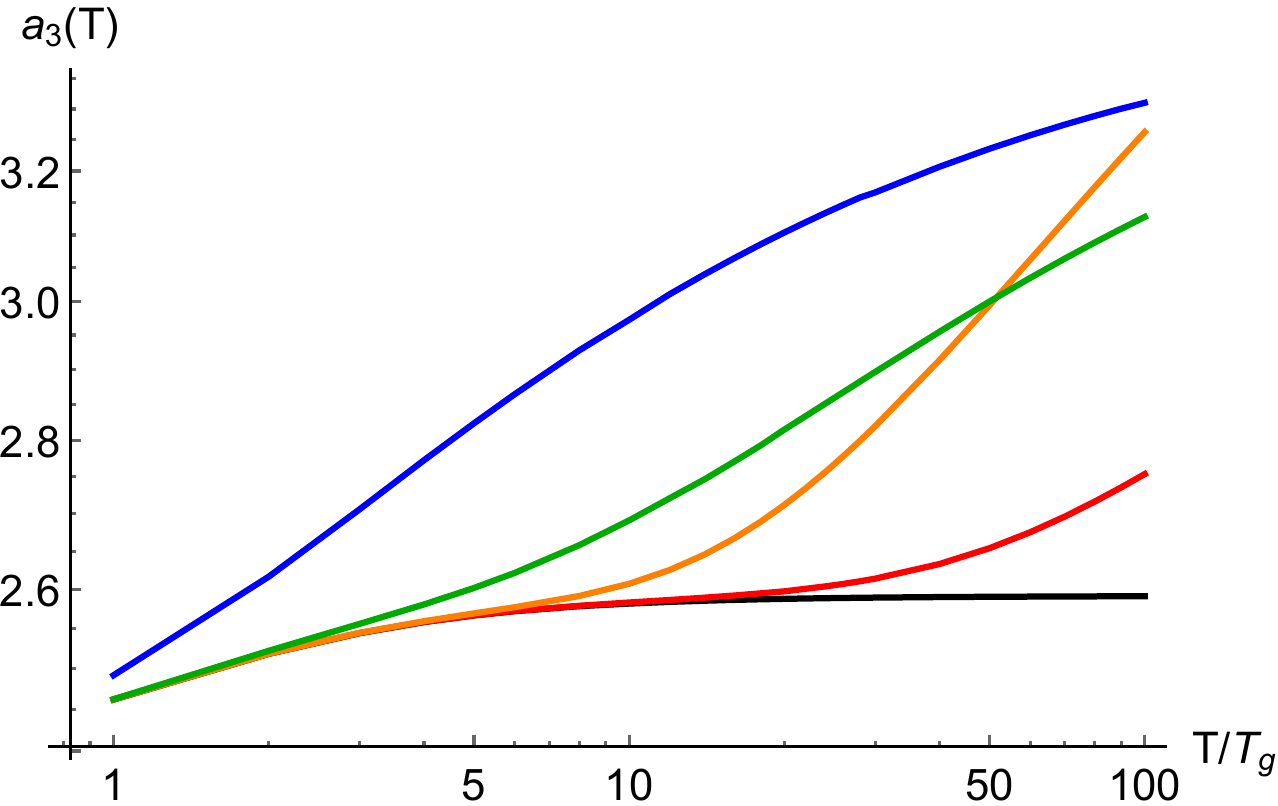}
    \caption{The temperature dependence of the $a_2,a_3$ parameters from the fitting function Eq.\eqref{fitt}. The parameters are fixed to $W=1,T_g=0.1$.}
    \label{acoff}
\end{figure}

We have verified numerically that Eq.\eqref{fitt} indeed fits reasonably well the numerical data within the soft potential model.
The coefficient $a_1(T)$ coincides with the linear slope at small imaginary frequency and it is already shown in the bottom panel of Fig.\ref{figdos}. The behavior of the other two parameters as a function of the temperature is shown in Fig.\ref{acoff}. The $a_2$ parameter grows monotonically with the temperature $T$ independently of the value of the UV cutoff $\Omega$. The behavior of the $a_3$ parameter is more interesting. At low temperature it grows monotonically independently of the value of the UV cutoff $\Omega$. After that, for large values of the UV cutoff, it approaches a constant value. On the contrary, for smaller values of $\Omega$, it keeps growing with temperature until up to $T\approx 100 T_g$.

Let us now try to understand the behavior of the coefficient, $a_3$. In the theory developed for liquids, the exponential factor expresses the probability of visiting a saddle barrier, representing $\langle \exp(-E(\nu)/T)\rangle \approx \exp(-\langle E(\nu) \rangle /T)$, where $E(\nu)$ is the height of a barrier with frequency $\nu$ above the associated local minimum. Some simple approximations are possible for the SPM with $D_1$=0. Consider the entire unstable region, not just the barrier top. The unstable frequency varies from zero at the inflection point to $(2D_2)^{1/2}$ at the barrier top. The smallest $D_2$ that contributes at a given $\nu$ is the one with the barrier top frequency equal to $\nu$. Thus soft potentials with $D_2$ between $\nu^2/2$ and the upper limit of the uniform distribution, $\Omega^2/4$, will contribute to $E(\nu)$. It is easy to find that
\begin{equation}
E(\nu) = \frac{D_2^2}{4} + \frac{(\nu^2 + 10 D_2)(\nu^2-2 D_2)}{144}.
\end{equation}

While the contribution of a given soft potential to $\langle E(\nu) \rangle$ is weighted by the Boltzmann factor, if $T$ is much higher than the highest barrier, $\Omega^4/64$, an unweighted average gives the high-$T$ limit, and an estimate of the exponent in the DOS. Averaging over the uniform $D_2$ distribution in the indicated range with normalization $(\Omega^2/4 - \nu^2/2)^{-1}$ yields the high-$T$ limit of $\langle E(\nu) \rangle$,
\begin{equation}
\langle E(\nu) \rangle = \frac{13 \nu^4 + 5 \nu^2 \Omega^2 + \Omega^4}{432}.
\end{equation}
If this expression appears in the exponent of the DOS, it is clear that the coefficient $a_3$ in Eq.~\ref{fitt} lies between $2$ and $4$, in agreement with Fig.~\ref{acoff} and with several computer simulations of liquids. Furthermore, increasing $\Omega$ will shift the power of $\nu$, and the value of $a_3$, at high $T$, away from $4$ and towards $2$, which is also found in Fig.~\ref{acoff}.

A physical explanation of the $\Omega$ dependence of $a_3$ at high $T$ and $D_1 = 0$ is as follows. The barrier height is $E_{bar} = D_2^2/4$ and the barrier frequency obeys $\nu_{bar}^2 = 2 D_2$. Thus, if unstable modes were from barrier tops alone, the relation is $E(\nu) = \nu^4/16$, $a_3=4$. However, the entire unstable region contributes to the unstable DOS. The minimum $D_2$ that contributes at a given $\nu$ is indeed the one with $\nu_{bar} = \nu$, but for $D_2$ above the minimum value, $\nu$ occurs at points that have lower energy than the barrier top. The range of energies contributing to $E(\nu)$ is determined by the upper $D_2$ cutoff, $\Omega^2/4$. Decreasing $\Omega$ confines the range closer to the barrier top, making $a_3$ closer to $4$, as we have found. A simple estimate is less obvious for low $T$. However, it seems safe to say that SP with barrier heights, $D_2^2/4$, much greater than $T$ will not contribute. The $D_2$ average will have an effective cutoff proportional to $T^{1/2}$, pre-empting the $\Omega^2/4$ cutoff and giving a result independent of $\Omega$. This is the behavior seen at low $T$ for the coefficient $a_3$ in Fig.~\ref{acoff}. All in all, the simple argument above gives useful insight into Fig.~\ref{acoff} for $a_3$. It should be noted that in liquids, $a_3=2$ has been associated with a ``uniformly rough" potential energy landscape \cite{doi:10.1126/science.267.5206.1935,keyes1997instantaneous}, and $a_3=4$ with a ``nonuniformly rough" landscape \cite{doi:10.1063/1.479810,vijay2}. It is not clear to us if this idea is applicable to the SPM. Let us also repeat that the validity of the SPM for small values of $\Omega$ and large values of $T$ is very questionable. Thus, it does not come as a surprise that the physical argument above is ``violated'' by the $\Omega=3,4$ curves in the bottom panel of Fig.\ref{acoff}, where, above $T \approx 50 T_g$, $a_3$ for $\Omega=4$ becomes larger than that for $\Omega=3$.\\

At this point, we would like to focus on two specific aspects of the unstable DOS: (I) the low frequency linear slope, indicated as $a_1(T)$, and (II) the position of the maximum in $g_u(\nu)$.

\subsection{The linear slope and its temperature dependence}
Despite the many possible derivations of the linear in frequency behavior for the density of states of liquids, most of them are not able to make a robust prediction for the temperature dependence of its coefficient, $g_u(\nu)=a_1(T) \nu+\dots$\\
In \cite{doi:10.1063/1.468407}, Keyes derived the following expression:
\begin{equation}\label{p1}
    a_1(T)\propto \left(\frac{\alpha}{3z}-f_u(T)\right),
\end{equation}
where $\alpha$ is the average number of coordinates with downward curvature at the barrier top and $z$ the number of atoms used to partition the liquid into small cooperative systems (see details in \cite{doi:10.1063/1.468407}). 

The ratio $\alpha/3z$ is the maximum value for $f_u(T)$. Given that the fraction of unstable modes always grows with temperature in liquids,  Eq.\eqref{p1} predicts that the linear coefficient $a_1(T)$ decreases with temperature, usually following a power-law scaling. This prediction turned out to be accurate for simple LJ liquids \cite{keyes1997instantaneous}, but not in general. Indeed, the linear coefficient of many liquids (\textit{e.g.}, \cite{doi:10.1063/1.479810,doi:10.1063/1.5127821,doi:10.1073/pnas.2022303118}) increases with temperature. In \cite{doi:10.1063/1.468407,doi:10.1063/1.479810}, Li and Keyes recognized that to have an unstable mode, the system must attain at least the energy of an inflection point and gave the more general expression:
\begin{align}\label{p2}
    a_1(T)\propto &\left(f_u^{max}-f_u(T)\right)\,\langle e^{-\beta E(\omega=0)} \rangle\nonumber \\ \approx  &\left(f_u^{max}-f_u(T)\right)\, e^{-\beta \langle E(\omega=0)\rangle}, 
\end{align}
where $f_u^{max}$ is the maximum possible value of $f_u(T)$, and $E(\omega=0)$ is the  inflection point energy. In the last step, fluctuations are neglected in order to simplify the final expression. This approximation is exact for a Gaussian distribution. On the other hand, for a generic distribution, the Jensen's inequality $\langle e^{-\beta E(\omega)} \rangle \geq e^{-\beta \langle E(\omega)\rangle}$ holds.
\begin{figure}
    \centering
    \includegraphics[width=\linewidth]{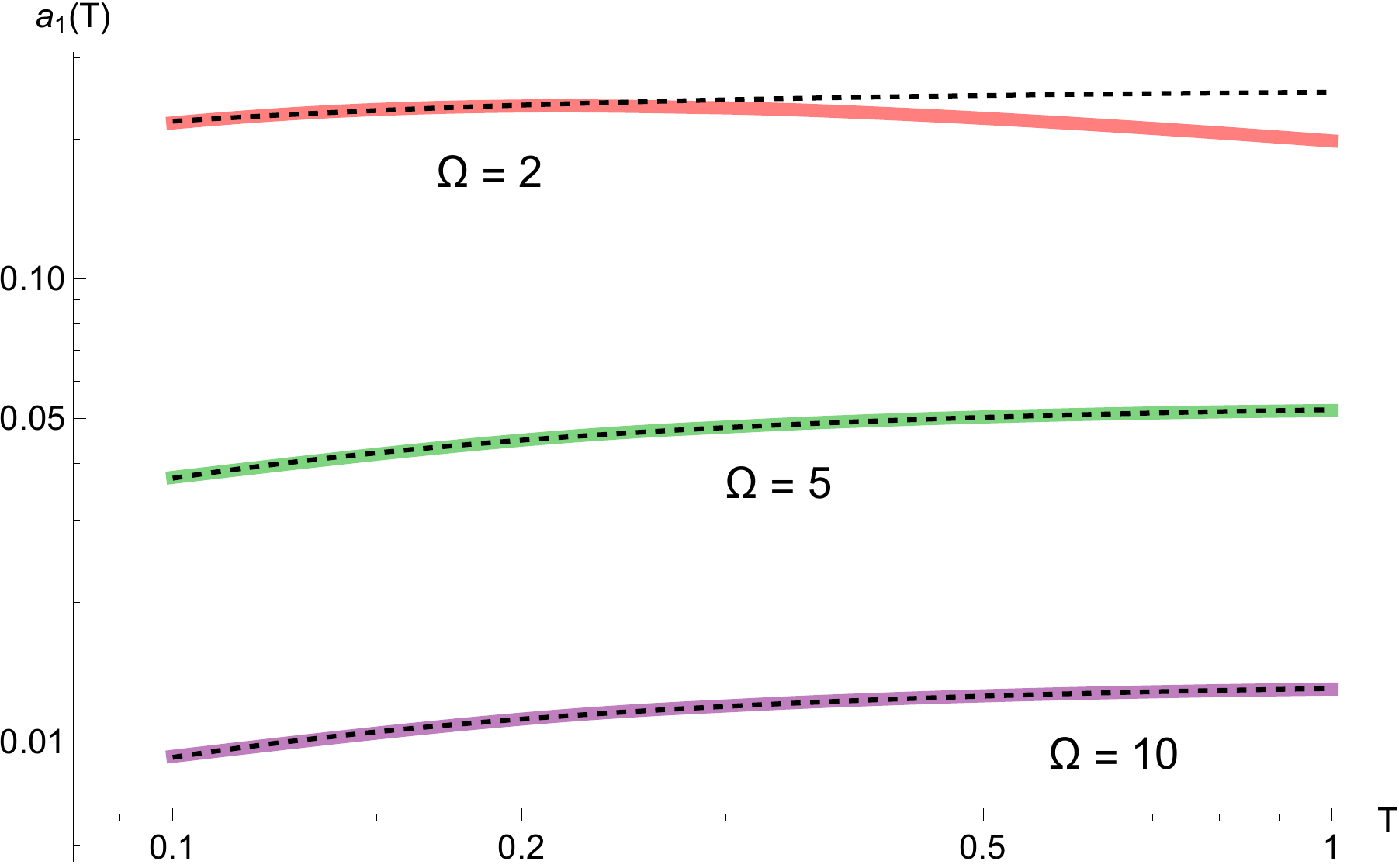}

    \vspace{0.5cm}
    
    \includegraphics[width=\linewidth]{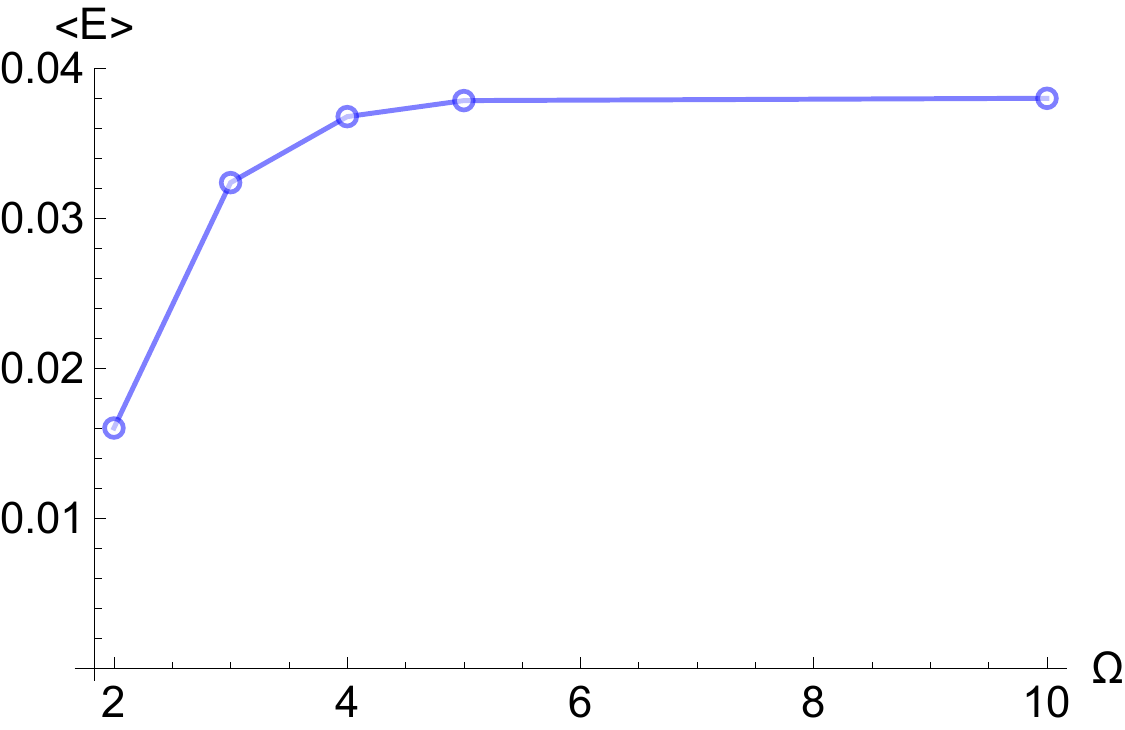}
    \caption{Top: fitting the numerical data for the low-frequency slope $a_1(T)$ with Eq.\eqref{p2}. The black dashed lines are the results of the fits. Bottom: the value of $b_3 = E(\omega=0)/k_B$ as a function of $\Omega$ obtained from the fits.}
    \label{fig:new}
\end{figure}
Now, Eq.~\ref{p2} suggests a richer dynamics. Since $f_u$ increases with $T$, $a_1(T)$ is the product of $(f_u^{max}-f_u(T))$, a decreasing function, and the Arrhenius exponential, an increasing function. The qualitative predictions are clear. For high barriers, $f_u$ is small, $(f_u^{max}-f_u(T))$ is approximately constant, $E(\omega=0)$ is large, and $a_1(T)$ is a monotonically increasing function. For low barriers, $E(\omega=0)$ is small, the exponential is approximately constant, $f_u$ is significant, and $a_1(T)$ has the monotonic decrease of $(f_u^{max}-f_u(T))$. 
The top panel of Fig.\ref{fig:new} shows some corresponding numerical results and fit. LJ liquids follow this scenario because of the well known, extremely low barrier heights. Away from those limiting cases, upon increasing $T$, a rise in $a_1(T)$ followed by a fall is predicted.  

Because the parameter $\Omega$ controls the height of the barrier and the inflection point in the SPM, studying the $\Omega$-dependence of $a_1(T)$ provides a test of the above ideas about its barrier height dependence.

The liquid theory results results were obtained using the potential energy landscape picture of liquids. At a given temperature, Boltzmann averaging selects which local minima are most visited. The unstable fraction, $f_u$, increases monotonically with $T$ to a high-$T$ plateau, because more unstable regions are sampled with increasing $T$. The theory uses the assumption that the landscape environment of all local minima is similar.

However, the assumptions in the previous paragraph do not hold for the SPM. The minima to be sampled are determined by the distribution of $D_1$ and $D_2$, and Boltzmann averaging only affects the distribution of the coordinate within a given minimum. With increasing $T$, we find that $f_u$ initially rises, but then falls. The fall occurs because, once $T$ is high enough that the unstable region of the soft potential is freely accessible, further increases make available more of the stable range of the coordinate at large $\pm x$. Finally, the environment of the different minima are not similar in the SPM; deeper minima have higher barriers.

Thus, we cannot expect the liquid theory for $a_1(T)$ to be quantitative for the SPM. Nevertheless, we will show that it provides a correct qualitative description.  

First, let us look at the analytical formula for the slope of the unaveraged (unstable) DOS, Eq.(\ref{tt}). For simplicity, we set $D_1=0$, and obtain
\begin{equation}
    G_u(\nu^2) = \frac{1}{\sqrt{12}W N_x}\frac{1}{\sqrt{2D_2-\nu^2/W}}\exp{\left(-\frac{V(x_\nu^+)}{T}\right)}.\label{lili}
\end{equation}

Compared with Eq.(\ref{p2}), we clearly see that the unaveraged slope has an Arrhenius exponential dependence on temperature. Moreover, the value of $V(x_\nu^+)$, which corresponds to the value of potential at the inflection point, increases as we increase the height of the barrier.

Turning to numerical fits, the decrease of $f_u(T)$ at higher $T$ for some $\Omega$ is problematic, causing $(f_u^{max}-f_u(T))$ to be an increasing function and making a fit to Eq.~\ref{p2} where $a_1(T)$ is decreasing to be unphysical. Thus we fit the numerical data for selected $\Omega$ over the intervals where $a_1(T)$ is increasing to the function
\begin{equation}\label{pheno1}
    a_1(T)= b_1 e^{-b_3/T},
\end{equation}
varying $b_1$ and  $b_3$. For $\Omega$ going from 2 to 5, $b_3 = E(\omega=0)/k_B$ rises monotonically from $0.016$ to $0.036$ (bottom panel of Fig.\ref{fig:new}), as expected qualitatively. This behavior is also in agreement with the analytical result presented above in Eq.\eqref{lili}.

In sum, despite the differences between liquids and the SPM, the liquid theory for $a_1(T,\Omega)$ is quite useful for the SPM. Conversely, and more importantly, the physical significance of the coefficient of the linear term in the unstable DOS for liquids is further clarified.

\begin{figure}
    \centering
    \includegraphics[width=\linewidth]{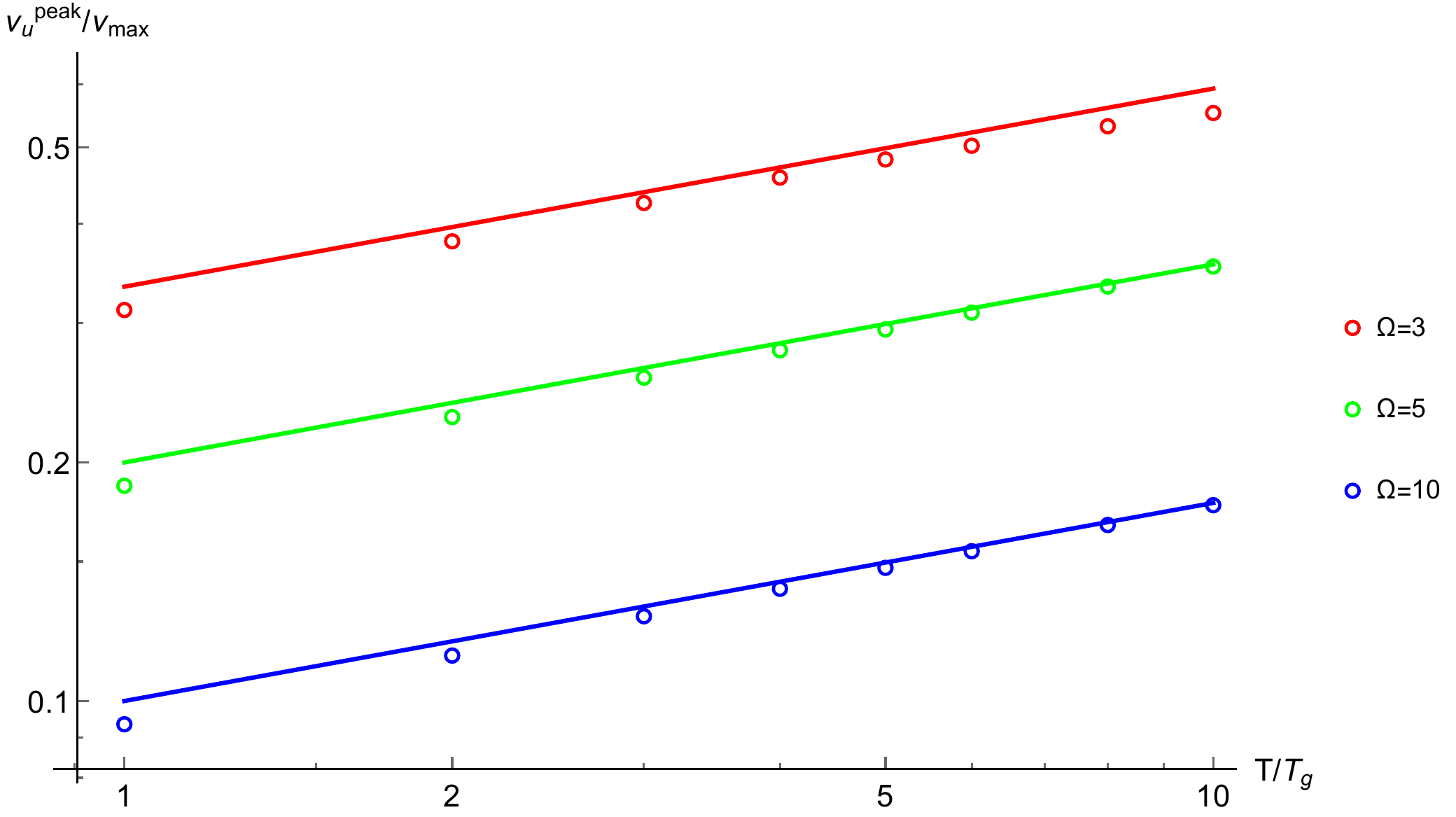}
    
    \caption{The position of the peak in the DOS of unstable INM. The data correspond to those in the top panel of Fig.\ref{figdos}. The line shows the asymptotic behavior $\sim 1/\Omega (T/T_g)^{1/4}$. The three figures correspond to $\Omega=3$, $\Omega=5$, and $\Omega=10$, respectively.}
    \label{fig:peaks}
\end{figure}
\subsection{The position of the maximum in the unstable density of states}
Another interesting observation that has been made in several instances is that the density of unstable INM displays a peak at a certain value of the imaginary frequency $\nu$. The position of this peak moves to higher $\nu$ with increasing temperature \cite{doi:10.1063/1.5127821,doi:10.1063/1.3701564,madan1993unstable}, and it has been observed also in the energy density of states\footnote{Caution must be taken in comparing directly the density of states $g(E)$ extracted from experiments and the density of states of instantaneous normal modes.} from direct experiments using neutron-scattering techniques \cite{stamper2022experimental}. 

The same peak is observed within the soft-potential model both for the stable and unstable parts of the density of states. In Fig.\ref{fig:peaks}, we show the behavior of the peak in the unstable part of the INM DOS as a function of temperature. In agreement  with the studies cited above, the position of the peak grows with temperature. Once normalized by the maximum frequency $\nu_{\text{max}}$ (see Eq.\eqref{maxi}), it follows an asymptotic trend:
\begin{equation}
    \nu_u^{\text{peak}}\approx \frac{\nu_{\text{max}}}{\Omega}\,\left(\frac{T}{T_g}\right)^{1/4}\,,
\end{equation}
which is confirmed numerically for large enough values of the frequency cutoff $\Omega$. This phenomenological expression starts failing when the cutoff frequency becomes small, $\Omega \approx 3$. Finally, we notice that the position of the peak grows by decreasing the cutoff frequency $\Omega$. This implies that the linear regime $g_u(\nu)\propto a_1(T) \nu$ extends towards larger and larger (imaginary) frequencies by decreasing $\Omega$. On the contrary, the peak in the stable part of the INM DOS is approximately constant in temperature (see Fig.\ref{fig3}), and it depends only on the parameters of the model such as the cutoff frequency $\Omega$. Since $\Omega$ controls the maximum barrier height, these observations are relevant to the ``landscape" interpretation of the DOS also. 

The asymmetry between the two behaviors is not surprising. Indeed, the stable and unstable parts of the INM DOS are symmetric only at low frequency where the spectrum is dominated by saddle points rather than maxima or minima. At frequencies in which maxima and minima become important, then, the asymmetry emerges as already shown in Fig.\ref{fig3}.
\begin{figure}[h]
    \centering
    \includegraphics[width=0.8\linewidth]{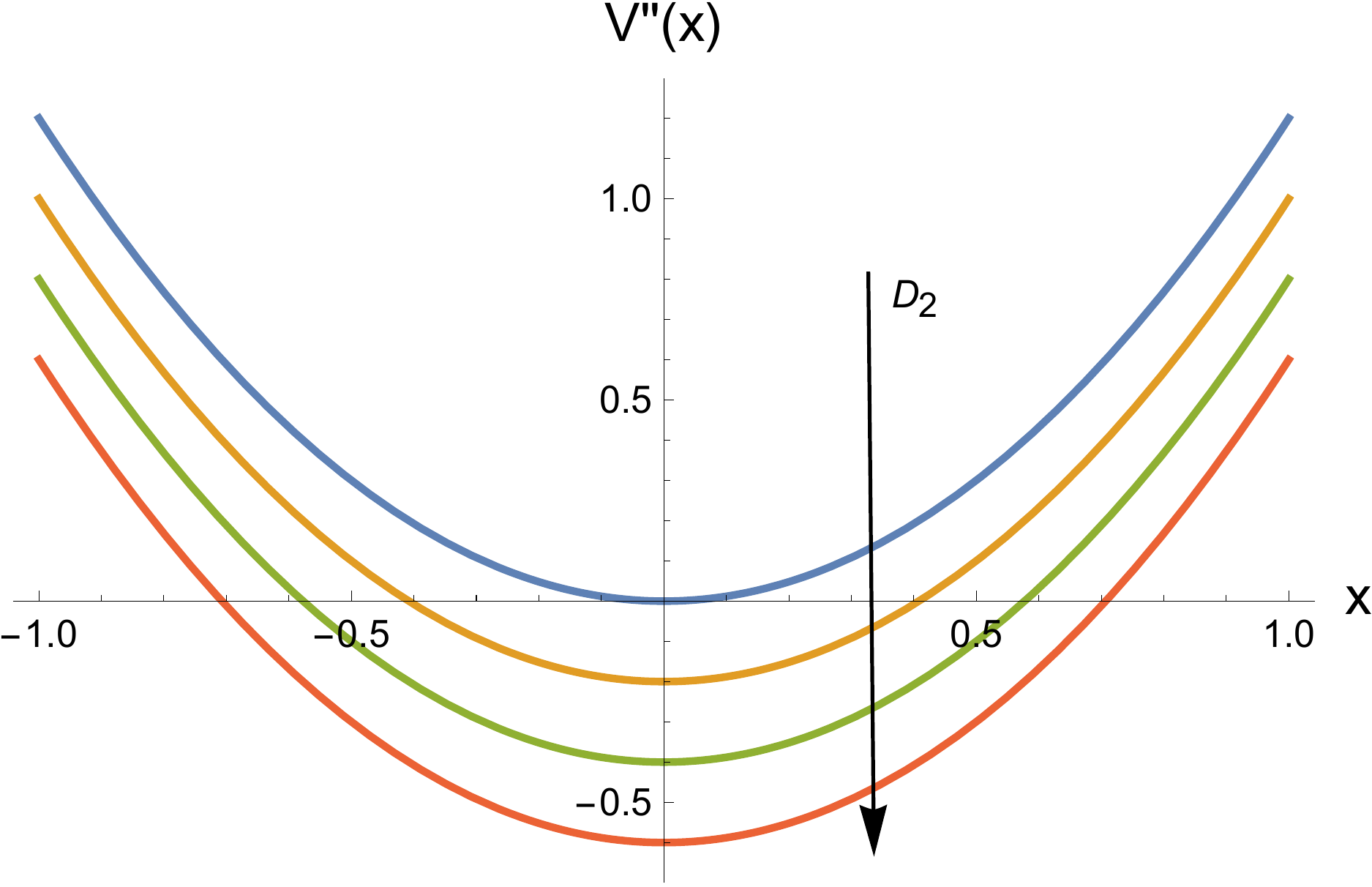}

    \vspace{0.2cm}
    
     \includegraphics[width=0.8\linewidth]{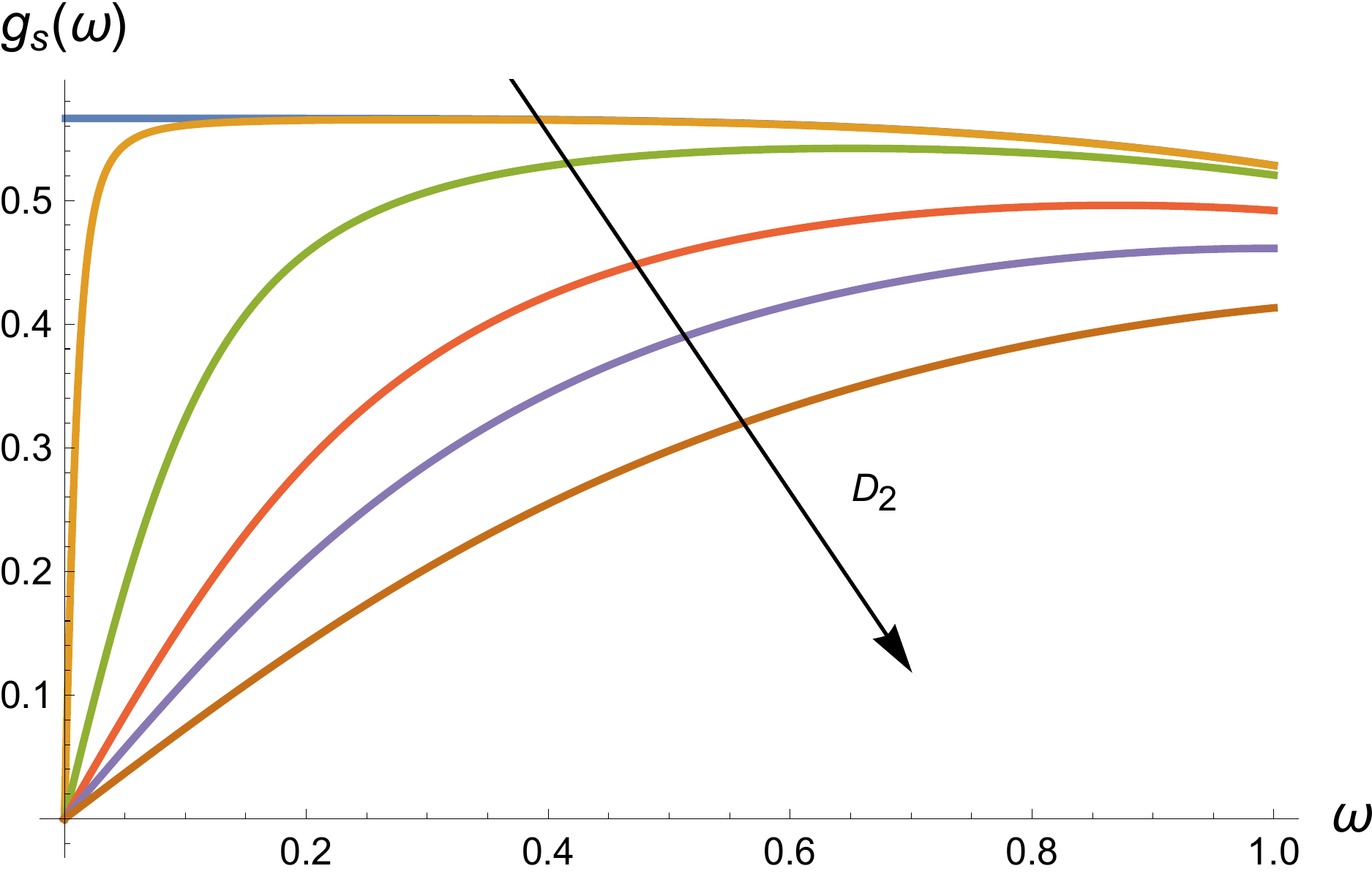}

     \vspace{0.2cm}
     
      \includegraphics[width=0.8\linewidth]{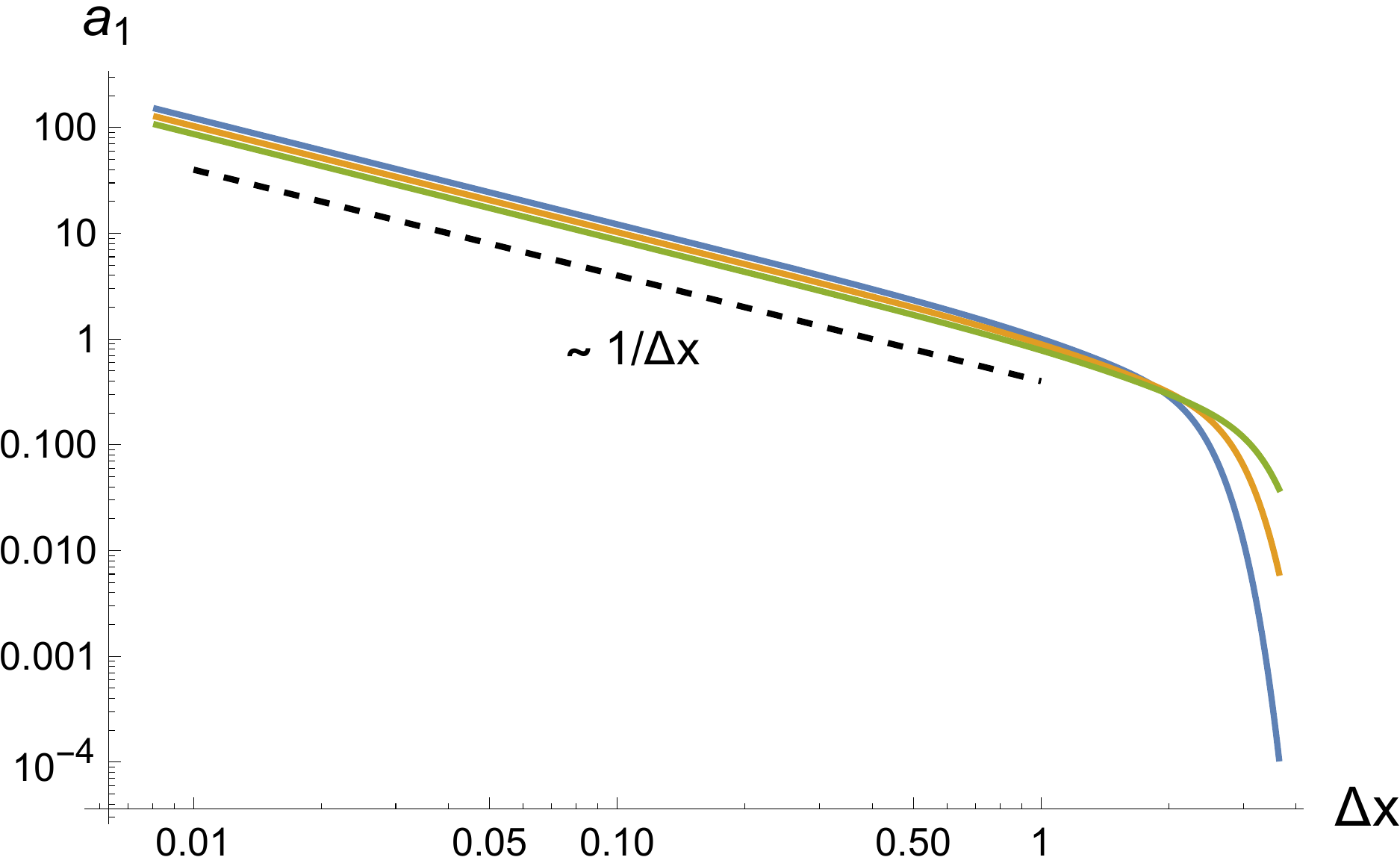}
    \caption{Top: The potential curvature as a function of $x$ for different and positive values of $D_2$. Center: The corresponding density of states for stable modes at low frequency. Bottom: The linear slope of the density of states as a function of the size of the negative curvature region.}
    \label{wo}
\end{figure}
\subsection{The fundamental ingredients for a linear in frequency DOS}
In the previous sections, we have performed an extensive analysis within the SPM model to understand the properties of the stable and unstable vibrational modes. Among the various findings, we found that using the SPM one can easily derive the universal linear in frequency scaling of the density of states at low frequency, ubiquitously observed in liquids. Here, we would like to identify which are the fundamental ingredients for this behavior so distinct from the Debye law of solids. In order to do so, we will neglect the effects of disorder which was implemented through a random distribution of the SPM parameters. As we will see explicitly, disorder is indeed not fundamental to derive such a universal scaling. Ignoring the possible high-frequency behavior, which as we describe above would be singular without the averaging procedure, let us focus on the low-frequency regime and on a single potential of the form:
\begin{equation}
    V(x)=- D_2 x^2+x^4,
\end{equation}
which corresponds to the SPM general case with $W=1$ and $D_1=0$. This potential is symmetric and for $D_2 
\leq 0$ does not contain any region with local negative curvature. On the contrary, by increasing the value of $D_2>0$, negative curvature regions appear and become larger and larger, as shown in the top panel of Fig.\ref{wo}. Using the tools described in the previous section, we can compute the density of states of the stable INM $g_s(\omega)$ as a function of the frequency by varying the parameter $D_2$. When $D_2=0$, the potential does not present any negative curvature region. The corresponding VDOS at low frequency is constant. This is simply the density of states for a perfectly quartic potential. Now, by increasing the value of $D_2$ along the positive axes, a region with negative curvature appears in the potential. The size of this region is given by:
\begin{equation}
    \Delta x=2 \,\frac{\sqrt{D_2}}{\sqrt{6}}.
\end{equation}
As shown in the central panel of Fig.\ref{wo}, as soon as a negative curvature region appears, the density of states becomes linear in the low frequency region. Additionally, in the bottom panel of Fig.\ref{wo}, we find that the linear slope of the VDOS at low frequency is inversely proportional to the size of the negative curvature. This seems to be consistent with our findings within the full SPM, where the slope decreases with the parameter $\Omega$ (see Fig.\ref{figdos}). Indeed, the larger $\Omega$, the larger the region with negative curvature.

In conclusions, using this simplified toy model, we can clearly demonstrate that the presence of a negative curvature region is the fundamental ingredient behind the universal linear scaling of the VDOS in liquids. In a similar spirit \cite{doi:10.1073/pnas.2022303118},  the linear scaling was  derived just assuming the presence of a single unstable mode with imaginary frequency. These findings agree with the argument~\cite{keyes-2005} that, if negative INM eigenvalues exist, the eigenvalue DOS $g(\epsilon)$ will be nonzero at $\epsilon=0$. Then, transforming to $g(\nu)$ with $\epsilon=\nu^2$, the Jacobian gives the linear scaling. Reversing the argument, Fig. 1 of~\cite{PureTR97} clearly shows the onset of low-frequency curvature as the unstable modes vanish with decreasing $T$.

\begin{figure}
    \centering
    \includegraphics[width=\linewidth]{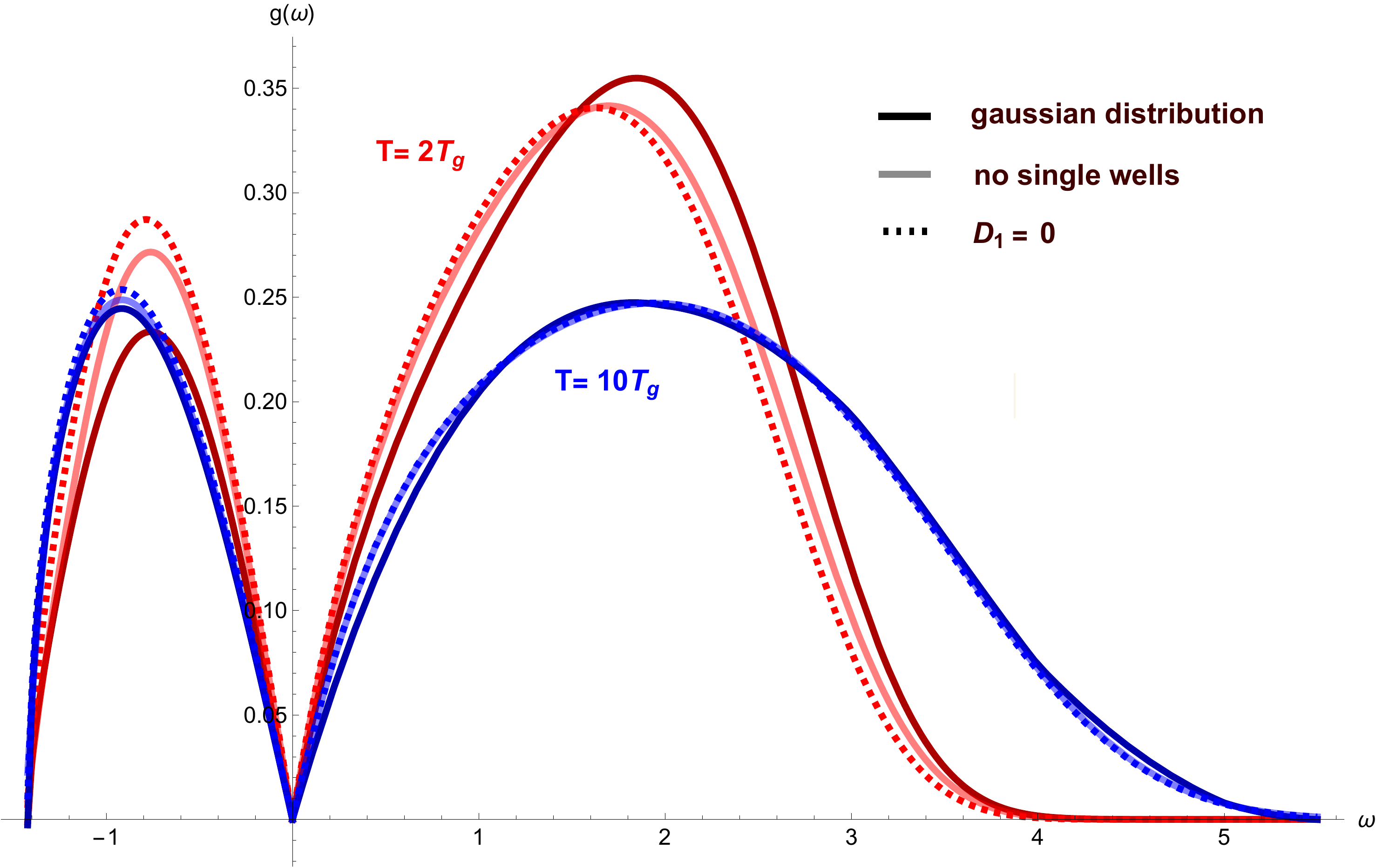}
    \caption{A study of the impact of $D_1$ on the INM density of states. This example is performed for $\Omega=2$. Darker colors indicate the gaussian distribution in \eqref{distr}. Lighter colors indicate a constant distribution cutoff at a maximum value of $D_1$ such that no single wells appear. The dashed lines are for $D_1=0$. Blue and red lines refer to the two temperatures $T=10 T_g$ and $T=2 T_g$ respectively.}
    \label{fignew}
\end{figure}
\subsection{The role of the asymmetry term in the soft potential model for liquids}
As explicitly shown in the previous section, the fundamental ingredient behind the appearance of the unstable modes and of a universal linear in frequency behavior for the INM VDOS is simply the presence and exploration of regions with negative curvature in the potential landscape. Nevertheless, because of our choice of the distribution for the asymmetry parameter $D_1$ in \eqref{distr}, our potential landscape contains also asymmetric single well potential for large values of $D_1$. First, let us notice that, because of the exponential distribution $p(D_1) \propto \exp(-\alpha D_1^2)$, those configurations are highly suppressed. Nevertheless, it is relevant to ask which are the effects of such configurations, and ultimately how the presence of single wells affect our results.

In order to check this point, we have performed the computations of the INM VDOS using two additional different distributions for the asymmetry parameter $D_1$. First, instead of \eqref{distr}, we have considered a constant distribution for $D_1$ with a cutoff $D_1^{max}$ such that single wells do not appear. Second, and even more drastically, we have simply set $D_1=0$ and considered a fully symmetric double-well potential. The results for the INM VDOS for all these different scenarios are shown in Fig.\ref{fignew} for two characteristic values of the temperature. We observe that the different assumptions on the asymmetry of the potential affect only minimally our results, in particular in the low frequency region. Some differences do appear in the high frequency region, close to the maximum of the VDOS and in the large frequency tail. Nevertheless, at large temperature, even these minimal differences are completely washed out. As expected, removing the anharmonic single wells which appear for large values of $D_1$ increases slightly the fraction of unstable modes.

The results of this short analysis confirm that the role of the asymmetry parameter in the SPM plays a very minor role in the description of liquid dynamics. This is comforting as it also proves that the choice of the linear term $\propto D_1$ in the SPM potential rather than a cubic one would have minimal effects on our main conclusions. In the future, it would be interesting to extend our description to consider also harmonic modes in the SPM, corresponding to negative $D_2$ or large $D_1$ values and give a more comprehensive picture of the full potential landscape for liquids. We leave this study for future work.

\section{Heat capacity, a statistical mechanics analysis}
After discussing the vibrational and dynamical properties of liquids within the soft potential model, we turn to the thermodynamic properties and in particular the behavior of the heat capacity. Given the difficulties related with a mode analysis of the heat capacity, we resort to a different, and in a way more standard, description in terms of familiar statistical mechanics concepts.

Let us consider the  unaveraged partition function for a single particle in one dimension
\begin{equation}
    Z_x = \int_{-\infty}^{\infty}\exp{\left(-\frac{V(x)}{T}\right)}dx\label{part},
\end{equation}
which is given in terms of the potential energy $V(x)$.
Here, we treat the parameters of the potential, $D_1$ and $D_2$, as fixed and first express all physical quantities as a function of them.
Using Eq.\eqref{part}, we can calculate the unaveraged energy:
\begin{equation}
    U_0=-\frac{\partial \log Z_x}{\partial \beta},
\end{equation}
where $\beta\equiv 1/k_B T$. As already mentioned, for simplicity, we set $k_B=1$. Following simple mathematical steps, we obtain the following expression
\begin{equation}
    U_0=\,\frac{\Bar{V}_x}{Z_x}\,,\qquad \Bar{V}_x\equiv\int_{-\infty}^{\infty}V(x)\,\exp{\left(-\frac{V(x)}{T}\right)}dx.
\end{equation}
Then, we can extract the corresponding heat capacity as
\begin{equation}
    c_0 = \left( \frac{\partial U_0}{\partial T}  \right)_V\,.
\end{equation}
Importantly, let us notice that this expression for the heat capacity takes into account only the potential energy of the 1-dimensional single particle. As a matter of fact, the total heat capacity of a single mode is then given by
\begin{equation}
    c_V=c_0 + \frac{1}{2},
\end{equation}
where the $1/2$ factor comes from the kinetic energy contribution which follows from the equipartition theorem.

\begin{figure}
    \centering
    \includegraphics[width=\linewidth]{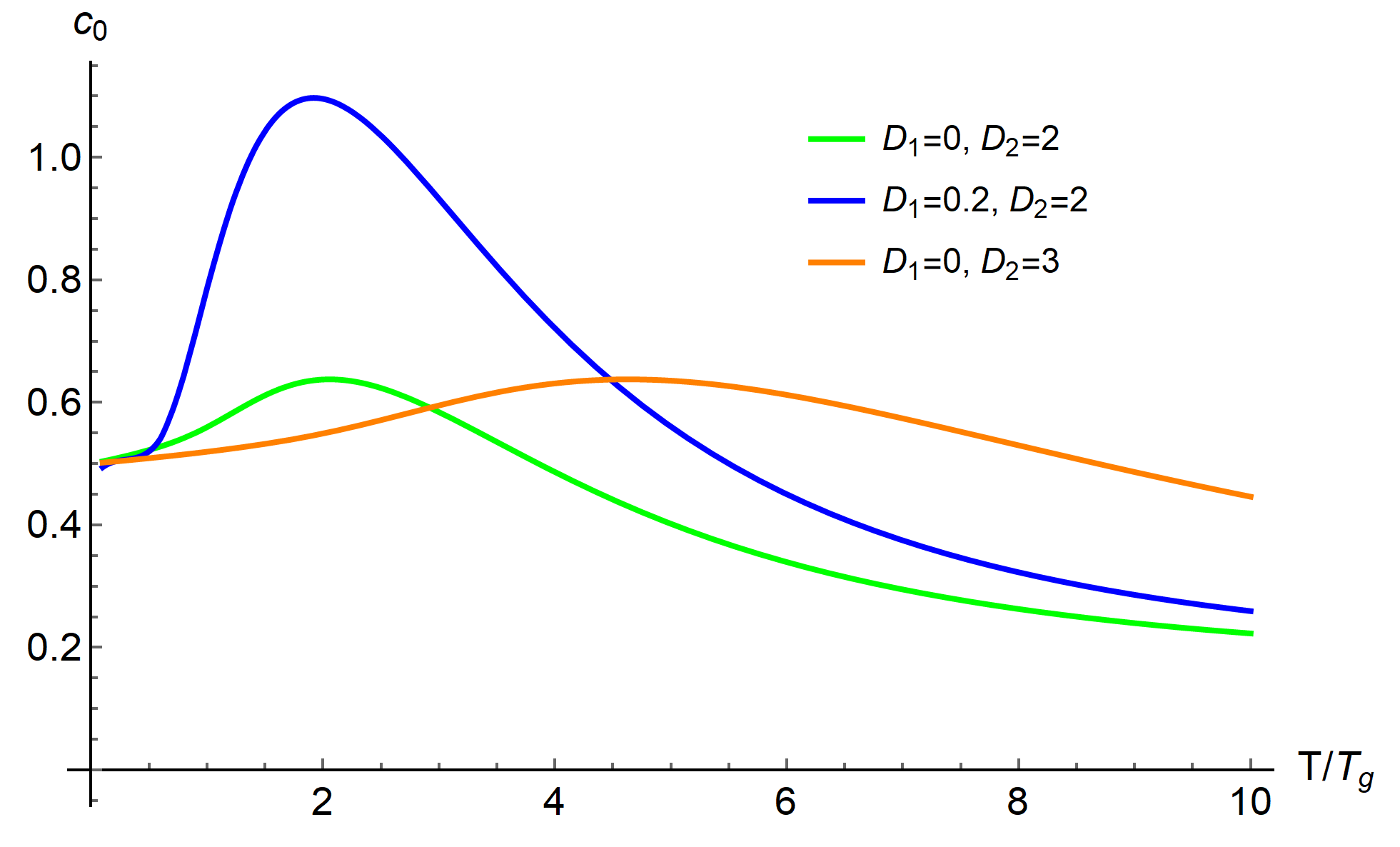}
    \caption{The unaveraged heat capacity (per mode) for different values of $D_1,D_2$, with $W=1$, $T_g=0.1$. For simplicity, we show only the contribution from the potential energy.}
    \label{unave}
\end{figure}

Using the framework explained above, we can obtain numerically the single mode unaveraged heat capacity for different values of $D_1$ and $D_2$. The result is shown in Fig.\ref{unave}, where for simplicity only the potential energy contribution is shown. Interestingly, the unaveraged heat capacity displays a peculiar non-monotonic behavior with a pronounced peak which strongly depends on the value of the parameters in the potential. This peak could be rationalized by computing the unaveraged heat capacity analytically in the limit of small and large temperature. Notice also that the unaveraged heat capacity always interpolates (after adding 1/2 for the kinetic part) between the zero temperature value $c_V=1$ to the infinite temperature value $c_V=3/4$. This can be understood by looking at the potential contribution $c_0$. At small temperature, the particle explore mainly the basin of the potential in which the potential is well approximated by a quadratic shape. There, the value of heat capacity per mode $c_V=1=1/2+1/2$ is just the result of the equipartition theorem for a quadratic potential. At large temperature, the particle explores the far away region of the potential which is dominated by the quartic term. Calculating the partition function and then the heat capacity for the quartic potential, one immediately obtains $c_V=1/2+1/4=3/4$, where $1/4$ is the heat capacity contributed by the quartic potential.  

The unaveraged heat capacity $c_V=c_0+1/2$ per particle in 1-dimension is not the actual heat capacity of our system, since we have not taken into account yet its random nature, and in particular the statistical distribution of the parameters $D_1$ and $D_2$ in the soft potential model. In order to find the heat capacity, we need to average over the potential parameters as follows
\begin{align}
    \langle c_V \rangle=&\frac{1}{2}+ \langle c_0\rangle_{D_1,D_2}\nonumber\\=&1/2+\int\int c_0\, p_1(D_1) p_2 (D_2) dD_1 dD_2,
\end{align}
where the factor of $1/2$ is still the contribution from the kinetic energy. Finally, by assuming the independence of the motion in the $3$-dimensional space, we can simply write that the 3D heat capacity per particle is given by
\begin{equation}
    C=3 \langle c_V \rangle.
\end{equation}
Note that in this notation, crystals would have heat capacity per particle $C=3$, and the ideal monoatomic gas would have $C=3/2$.

\begin{figure}
    \centering
     \includegraphics[width=\linewidth]{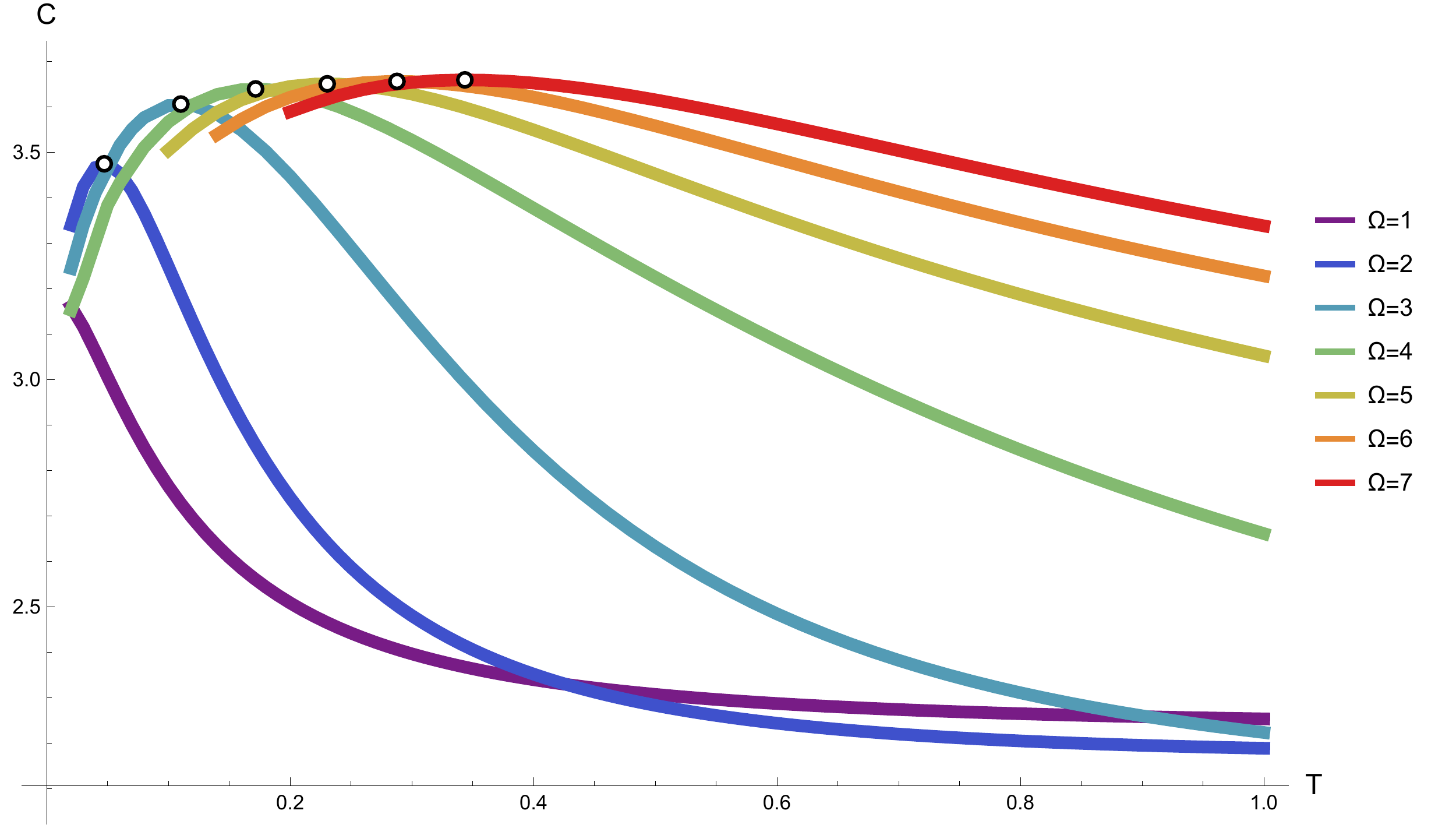} 
     
     \vspace{0.5cm}
     
     \includegraphics[width=\linewidth]{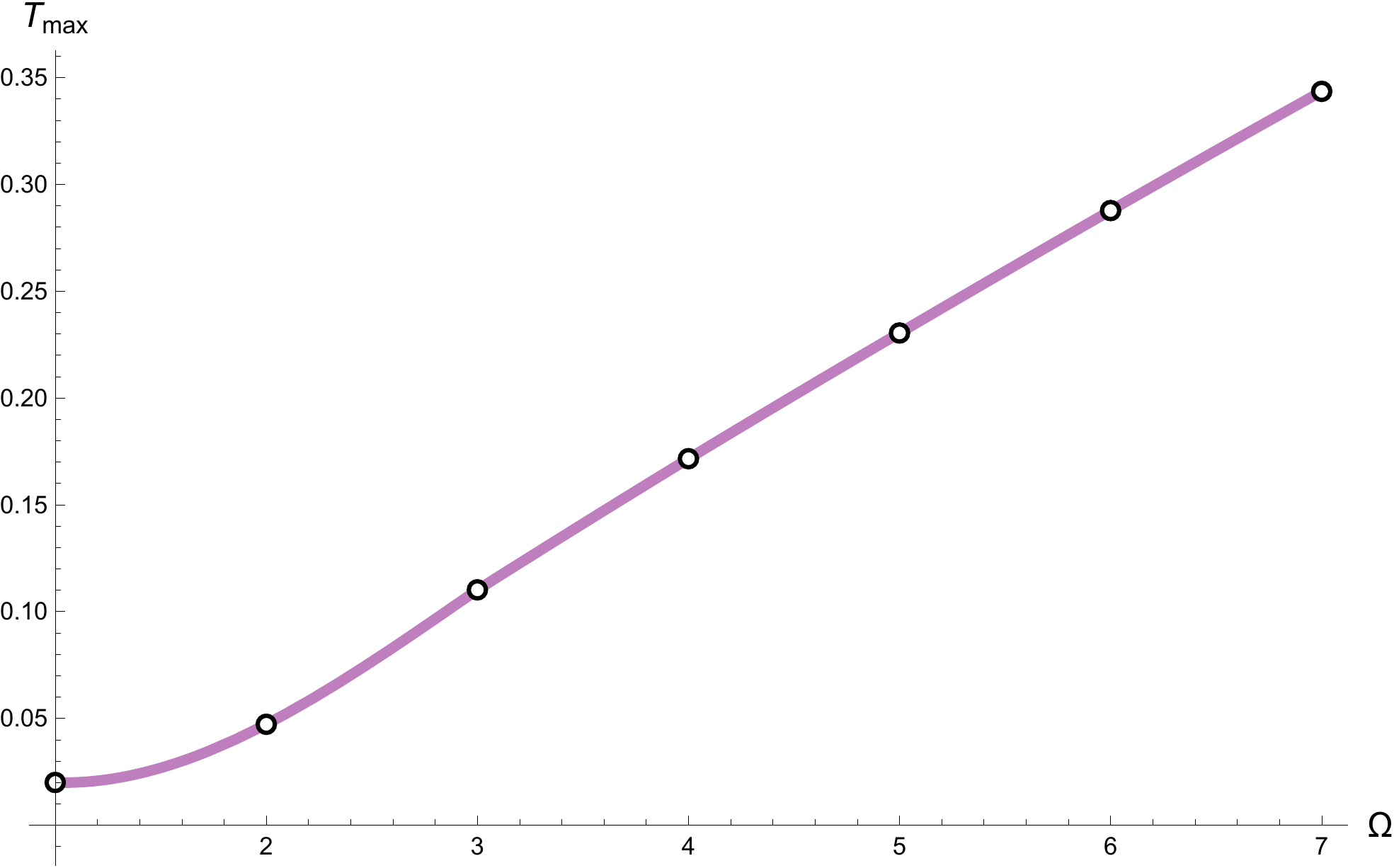}
    \caption{Top: The heat capacity as a function of the reduced temperature for different values of $\Omega$, with $W=1$, $T_g=0.1$. The black circles indicate the position of the maxima. Bottom: The position of the maxima in the heat capacity as a function of the parameter $\Omega$. }
    \label{heatfig}
\end{figure}

The numerical results are shown in Figure \ref{heatfig}. In the obtained heat capacity, we observe a clear non-monotonic behavior as a function of temperature. The heat capacity first grows with $T$ and then, after reaching a maximum at $T=T_{max}$, decreases at large temperatures. The location of the maximum strongly depends on the value of the parameter $\Omega$ which controls the distribution of the energy barriers. For small values of $\Omega$ the heat capacity decreases monotonically with $T$, or show a peak at very low temperature, below, or at least close to, the glass transition temperature $T_g$. By increasing the value of $\Omega$, the maximum shifts towards larger temperatures as shown in the bottom panel of Figure \ref{heatfig}. The dependence of $T_{max}$ on $\Omega$ is approximately linear. 

In the large temperature limit, we also observe that all curves approach the universal value $C=3(1/2+1/4)=9/4$. This can be rationalized as follows. The SPM does not have a gas phase, so $C$ does not reach the monoatomic gas-like value $C=3/2$ (free particles) at high $T$. Rather, in that limit the system senses only the $|x|^4$ potential at large $\pm x$, and attains the $C=9/4$ potential energy described above. Notice also that, due to the strong anharmonicities, the heat capacity can take values larger than the harmonic result $C=3$. This is more evident for large $\Omega$.

Additionally, we observe that a larger $\Omega$ implies a slower temperature decay of the heat capacity. A larger $\Omega$ indicates a larger sample of deep double-wells with higher barriers, and implies a larger averaged barrier energy. We have used the SPM to understand INM of liquids, but how much a model designed for glasses can describe a liquid over a broad temperature range is an open question. Nevertheless, this seems, at least qualitatively, in agreement with what observed in \cite{PhysRevE.104.014103} as a function of the LJ barrier energy. On the other hand, let us also emphasize that the previous analysis for the temperature dependence of the fraction of unstable modes (see Fig.\ref{fig:4}), showed non-liquid-like behavior in the limit of large temperature (and not too large $\Omega$).

Finally, let us comment on the peak which we observe in the top panel of Fig.\ref{heatfig}, and whose dependence on $\Omega$ is shown in the corresponding bottom panel. This feature is usually not observed in the liquid phase. Nevertheless, as we attempt to adapt the SPM to liquids, the system must be liquid above $T_g$. Therefore, the monotonically decreasing heat capacity found above the peak is in agreement with the behavior of real liquids. In addition to that, we observe that such a peak moves towards higher temperature by increasing the value of $\Omega$. This is consistent with the interpretation that the peak occurs because at intermediate $T$, increasing $T$ lets the system be on the barrier top, which make the energy to rise more strongly. Following this intuition, the peak position would increase with increasing $\Omega$ (higher barrier), since a higher barrier requires higher $T$ to visit. This is exactly what we observe in the bottom panel of Fig.\ref{heatfig}. It would be interesting to do further analysis in this direction to corroborate whether this simple argument is valid.

\section{Outlook}
In this work, we have revisited the anharmonic soft potential model as an effective model for liquids dynamics, and in particular to describe the vibrational density of states and the heat capacity of liquids. We found that the SPM correctly reproduces the linear frequency scaling in the VDOS and a heat capacity which decreases with temperature, which both are hallmark features of classical liquids. In doing so, we have corrected some minor mistakes appearing in the computation of the VDOS in \cite{PhysRevE.55.6917} and expanded the analysis therein to the thermodynamic properties. Moreover, we have discussed and tested with the SPM several approximate expressions for the VDOS, its frequency dependence and its temperature dependence, which have appeared in the past literature. Finally, using a simplified toy model, we have shown that the linear in frequency scaling of the VDOS, ubiquitously observed in liquids, is a direct consequence of the appearance of potential regions with negative curvature and, as such, tightly correlated to the existence of unstable INM.

Much of INM theory for liquids is based on the potential energy landscape viewpoint. The SPM is a highly simplified landscape with topography precisely known and controlled by the parameters $D_1, D_2$ and $\Omega$. Thus by varying them we were able to verify several ideas about the relation of the DOS to the topography proposed for liquids.

Our computation of the heat capacity is based on the standard statistical mechanics concept of partition function and it does not rely on any normal mode analysis, as usually done for the Debye model in solids. Despite recent attempts \cite{PhysRevE.104.014103}, it is not obvious if a normal mode derivation of the heat capacity of liquids is possible and how to perform it. For a suggestion based upon combining INM and Stillinger's inherent structure theory \cite{stillinger2015energy}, see Appendix \ref{appb}.

Given the popular explanation for the heat capacity of liquids based on the disappearance of transverse shear waves, \textit{i.e.}, gapped momentum states \cite{BAGGIOLI20201}, it would be interesting to explore whether there is a relation between such a scenario and that based on unstable modes. The equation for gapped shear waves necessarily involves modes with purely imaginary frequency, which might be thought as unstable INM. At the same time, a correlation between the temperature dependence of the fraction of unstable modes, governed by changing access to the barrier regions on the landscape, and the heat capacity has been observed in liquids simulations \cite{madan1993unstable}. Thus there are several hints that INM might play a fundamental role for the heat capacity of liquids, and that the system gaining access to the barrier in the SPM might be the cause of the peak in $C(T)$.

\begin{acknowledgments}
We would like to thank H.~Liang, A.~Zaccone, J.~Douglas, C.~Yang, Y.~Yu, X.~Fan, D.~Yu, C.~Jiang, S.~Jin and Y.~Feng for fruitful discussions and related collaborations on the topic of liquids.
M.B. acknowledges the support of the Shanghai Municipal Science and Technology Major Project (Grant No.2019SHZDZX01) and the sponsorship from the Yangyang Development Fund. 
\end{acknowledgments}

\appendix

\section{From liquids to glasses and back}\label{secfrom}
In this work, we have extensively used the soft potential model to describe the vibrational dynamics and thermodynamic properties of liquids. Nevertheless, as stressed in the introduction, the SPM has been originally introduced to describe the vibrational dynamics of glasses (see \cite{doi:10.1142/9781800612587_0008} or chapter $9$ in \cite{esquinazi2013tunneling} for a nice review of the topic). In particular, the SPM was proposed as an extension of the famous tunneling two-level systems (TLS) theory \cite{doi:10.1142/9781800612587_0004} to explain the boson peak anomaly in amorphous systems. The SPM encompasses the idea that glasses possess a large number of quasi-localized soft modes with a inherently anharmonic dynamics which strongly interact with acoustic extended phononic modes around the boson peak frequency. In the SPM model for glasses, the vibrational density of (stable) states is found to obey a quartic law:
\begin{equation}
    g(\omega)\propto \omega^4,
\end{equation}
which turns into a faster than Debye contribution to the heat capacity:
\begin{equation}
    C(T)\propto T^5,
\end{equation}
giving rise to the infamous boson peak anomaly.\\

It must be emphasized that this paper is entirely classical, while the Debye $C(T) \propto T^3$ for crystals requires quantum mechanics. No direct comparison of our results with a system that is in the quantum regime, or a theory using quantum mechanics, \textit{e.g.}, Debye, is meaningful. We simply mention the $T^3$ law for completeness.

How does all of this relate to what discussed in our work in the context of liquids? Liquids do not have a density of states quartic in the frequency, they do not have a faster than Debye heat capacity and they do have a considerable amount of unstable normal modes. Which is the main difference between the two scenarios? How can we recover the glassy dynamics from our results?

\begin{figure}[h]
    \centering
    \includegraphics[width=0.8\linewidth]{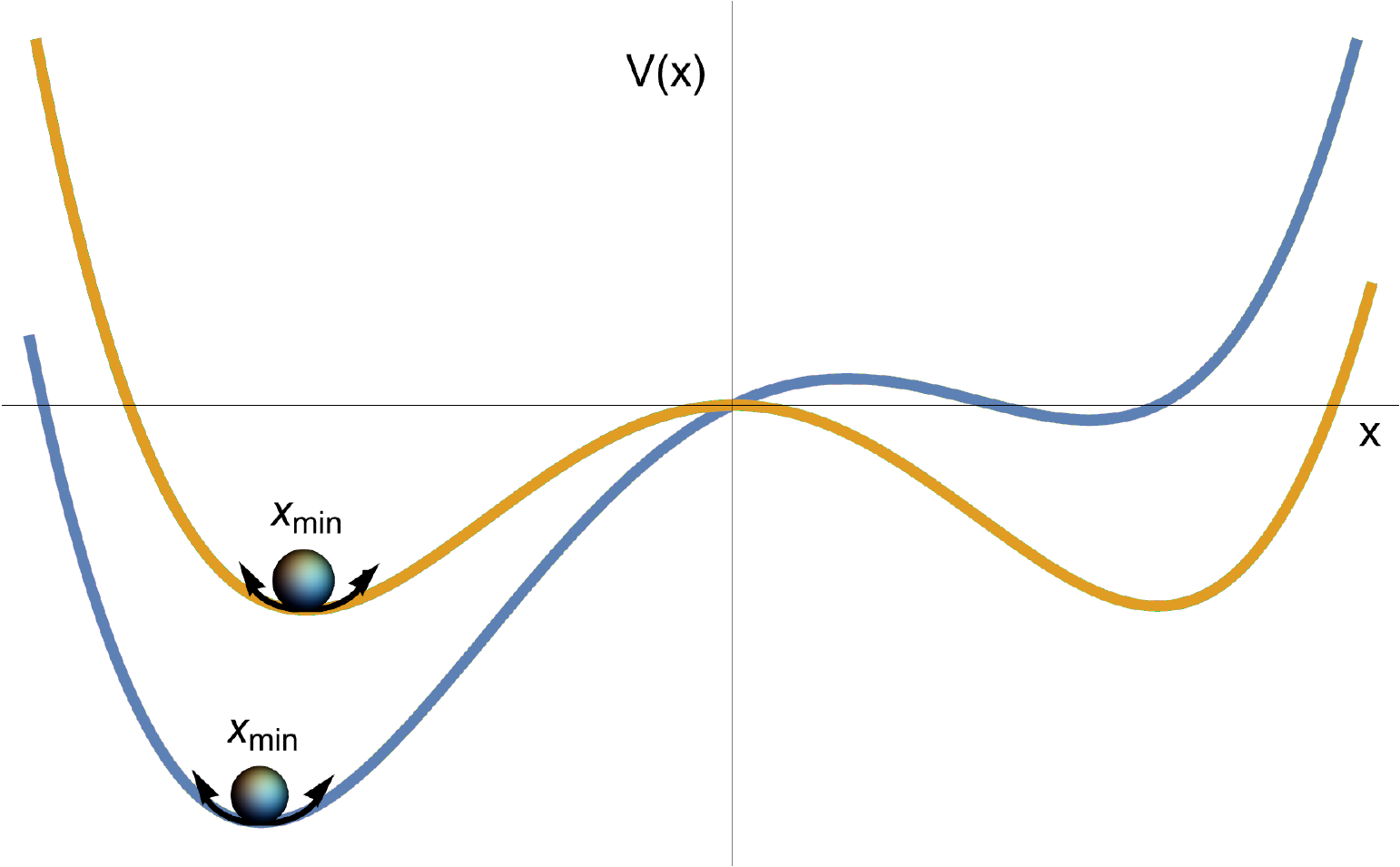}

    \vspace{0.5cm}
    
    \includegraphics[width=0.8\linewidth]{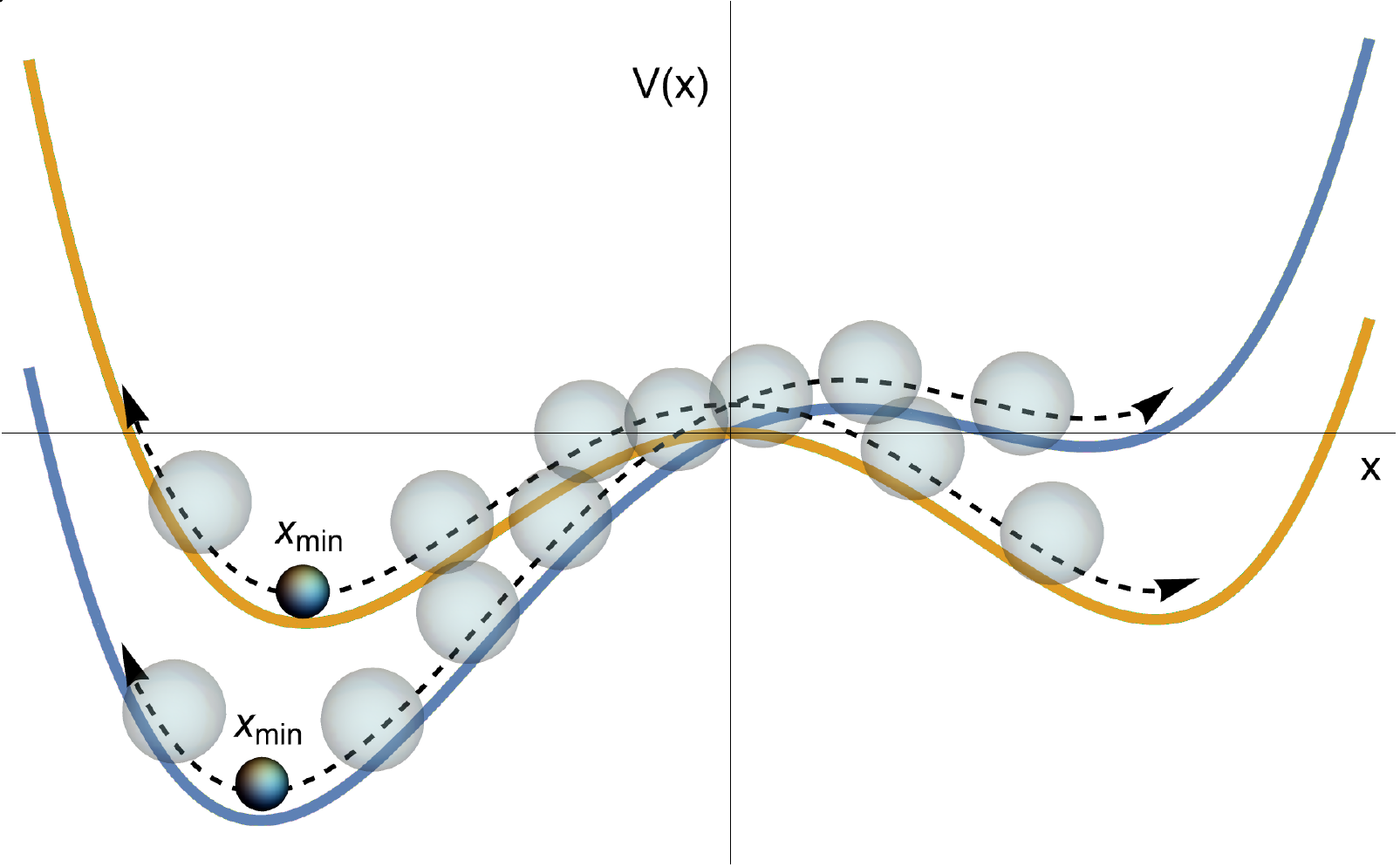}
    \caption{Particle dynamics in a glassy system is localized near the minimum of the different replica of the anharmonic potential (top). In liquids, the particle explores the whole range of the anharmonic potential, including saddle points and regions with negative curvature (bottom).}
    \label{fig:app1}
\end{figure}

In order to understand this point, we need to go back to the original formula for the DOS:
\begin{equation}\label{equa}
    g(\omega)\propto \omega \int  \delta\left(\frac{d^2 V}{dx^2}-\omega^2\right)\,e^{-V(x)/T}\,p(D_1)\,p(D_2)\,dx\,dD_1\,dD_2.
\end{equation}
This expression contains two distinctive different averages. First, we average over the parameters of the anharmonic potential $D_1,D_2$. Second, we average over the spatial direction $x$ using a Boltzmann distribution which characterizes how likely is for a particle to explore a certain point $x$ in the  "landscape" $V(x)$. Both in the context of glasses and liquids, a uniform distribution for $D_2$ and a Gaussian distribution $p(D_1)\propto \exp(-\alpha D_1^2)$ are assumed. 

The major difference is in the treatment of the spatial coordinate $x$. In glasses, the dynamics is mostly localized at the minimum of the potential, $x=x_{\text{min}}$ and the dynamics is dominated by mostly anharmonic soft modes, $\omega \ll W$. In other words, the integration over $x$ in Eq.\eqref{equa} is localized at the minima of the potential and:
\begin{equation}
    \exp\left(-V(x)/T\right)\quad \longrightarrow \quad \delta\left(x-x_{\text{min}}\right)\,.
\end{equation}
This approximation gives immediately a quartic density of states as already anticipated.

In the cases of liquids, the dynamics in the spatial direction is profoundly different. The height of the potential, which is proportional to $W/T$, is not large enough to constrain the dynamics of the particle around the minimum $x=x_{\text{min}}$. On the contrary, the particle explores the full range of the anharmonic potential (see Fig.\ref{fig:app1}), including the saddle points and the regions with negative curvature. As showed in the main text, this gives a fundamentally different result for the DOS. It gives a DOS which is linear in the frequency at small frequency, and also induces the presence of an increasingly number of unstable modes with imaginary frequency.

More in general, by tuning the parameters in the opportune way, we do expect that our results which are relevant to liquids should turn into the original results for the SPM in glasses. In general, by computing the density of states for stable modes, we do expect a form of the type
\begin{equation}
    g_s(\omega)=d_1\,\omega + d_2\,\omega^4 ,
\end{equation}
in the limit of small frequency. In addition, we expect that in the glassy limit, one would recover $d_1 \rightarrow 0$, and a quartic scaling for the DOS.

There are two important factors that control the crossover to the glassy dynamics. More precisely, one wants the height of the potential to be large compared to the temperature, such that jumping the energy barrier around the minima becomes highly improbable. In addition, one would expect a  larger cutoff frequency $\Omega$ which in turn leads to a wider distribution of $D_2$. Larger $D_2$ means deeper double wells, and increases the likelihood that the particle remains localized around the minima of the potential.

\begin{figure}
    \centering
    \includegraphics[width=0.8\linewidth]{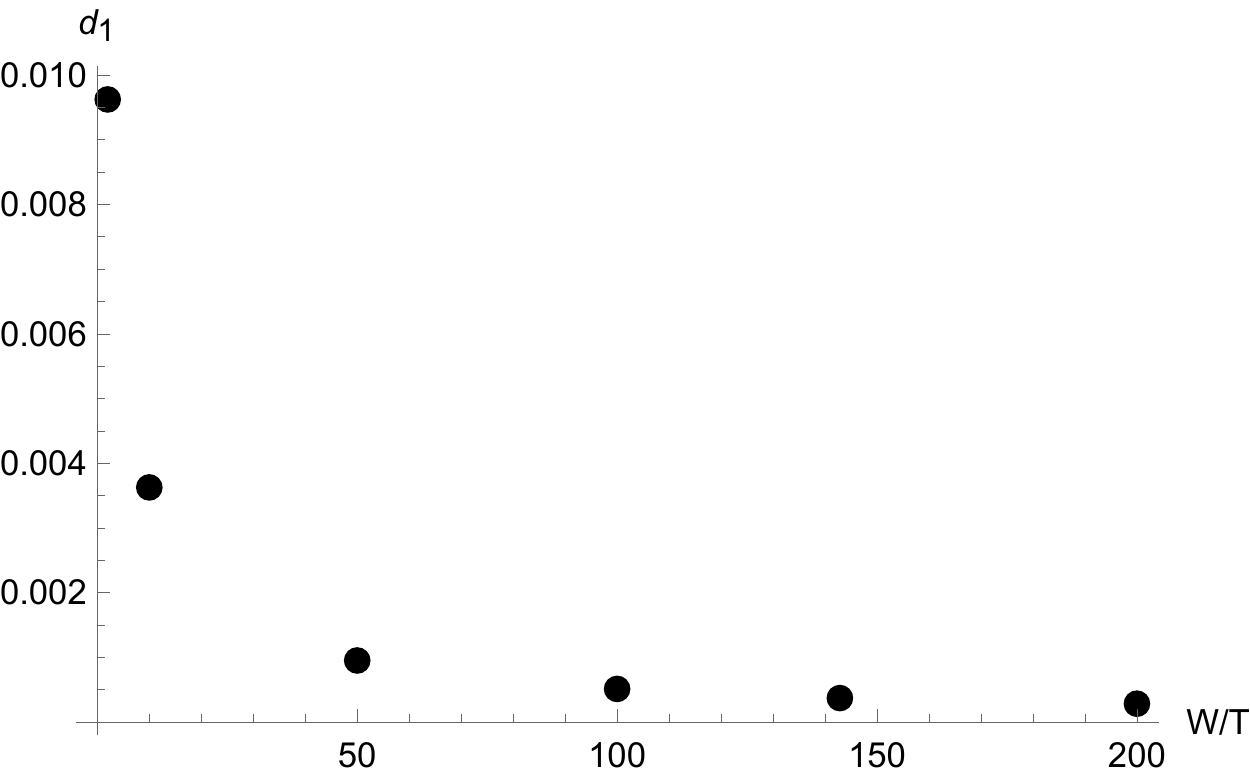}

    \vspace{0.2cm}
    
    \includegraphics[width=0.8\linewidth]{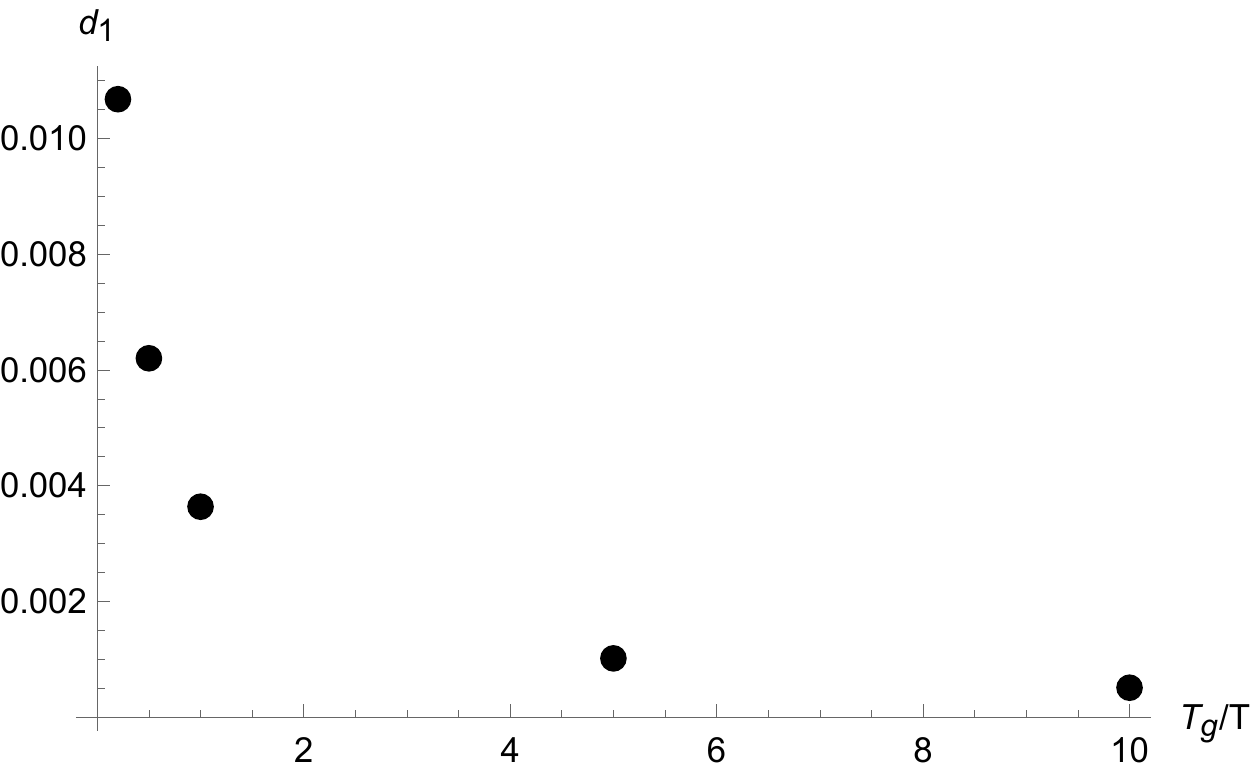}

    \vspace{0.2cm}
    
    \includegraphics[width=0.8\linewidth]{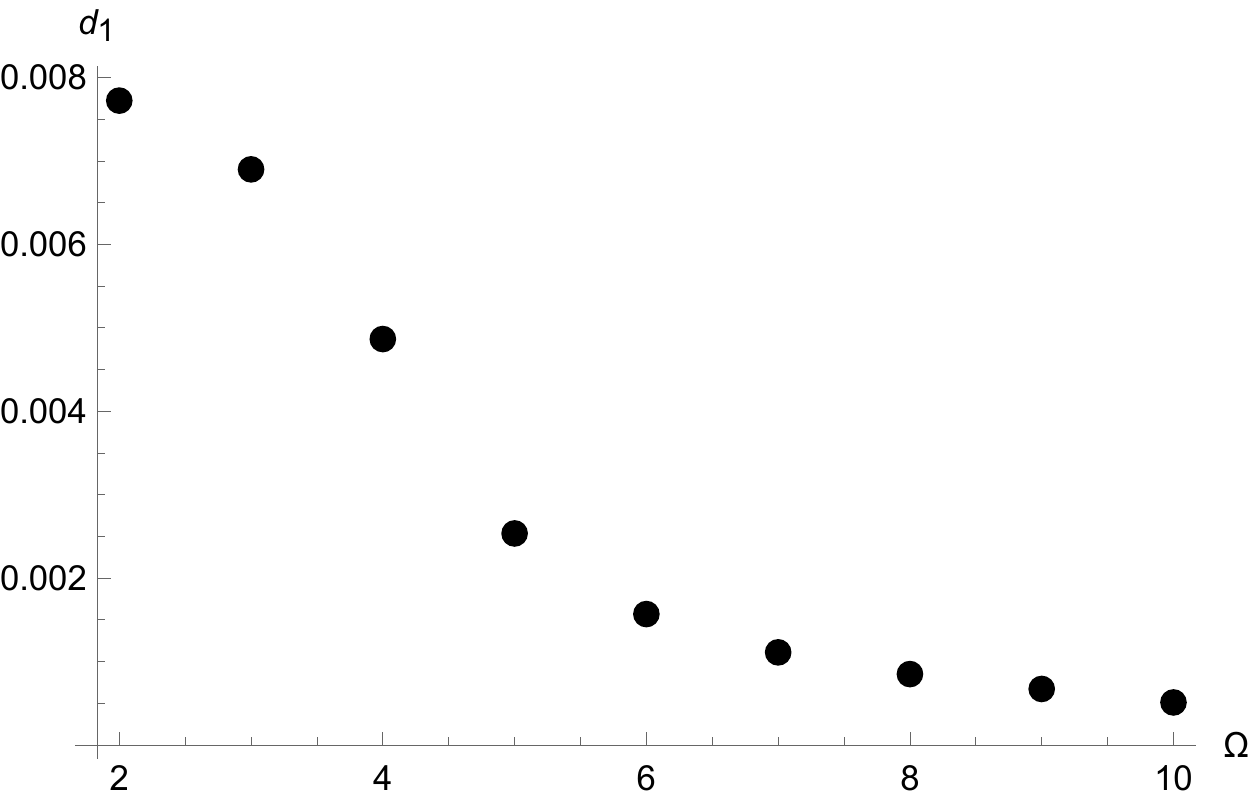}
    \caption{The transition to a glassy phase as the vanishing of the linear in frequency coefficient $d_1$. Top: as a function of the reduced barrier height $W/T$. Here, $W/T_g=1$ and $\Omega=10$. Center: as a function of the glass transition temperature $T_g$. Here, $W=10$ and $T=0.1$. Bottom: as a function of the cutoff frequency $\Omega$. Here, $W=\Omega=10$ and $T=0.1$.}
    \label{fig:tr}
\end{figure}

All these features are observed numerically in Fig.\ref{fig:tr}. There, we see that by increasing $W$, by decreasing the temperature $T$ (with respect to the glass transition temperature $T_g$), and by increasing the width of the $D_2$ distribution, the linear coefficient in the stable DOS decreases towards zero and. As a consequence, the low-frequency behavior is then dominated by a quartic term as in glasses, which becomes exact only in the ``perfectly glassy limit", in which the dynamics is completely localized at the minimum of the potential $x=x_{\text{min}}$. 
This amounts to yet another demonstration that the linear scaling follows from the presence of unstable modes.

Finally, let us notice that very recently similar arguments about the relations between the liquid VDOS and that of quasi-localised modes in glasses have appeared in \cite{moriel2023boson}. The qualitative picture is similar to that presented in this section, where quenched liquid-like modes localized at the minima of the potential give the glassy quasi-localized dynamics.
\section{Does liquid=solid+gas?}\label{appb}
In this appendix, we would like to comment on a recently proposed  model for the heat capacity of solids, liquids and gases \cite{moon2022microscopic}. The main idea is that we can somehow describe a liquid as a mixture of solid-like and gas-like degrees of freedom (see also the very recent Ref.\cite{moon2023atomic}). In the classical limit, each solid-like degree of freedom contributes to the energy as $k_B T$, while gas-like ones contribute only with a kinetic term and therefore give $k_B T/2$. As such, the total energy of a liquid can be written as:
\begin{equation}
    E=k_BT\,\int g_{\text{solid}}(\omega)\,d\omega\,+\, \frac{k_BT}{2}\,\int g_{\text{gas}}(\omega)\,d\omega\,\,.
\end{equation}
Given that the total number of degrees of freedom is $3N$:
\begin{equation}
    3N\,=\,\int g_{\text{solid}}(\omega)\,d\omega\,+\, \,\int g_{\text{gas}}(\omega)\,d\omega\,
\end{equation}
one can re-write the expression for the energy as:
\begin{equation}
    E= 3 Nk_B T \left(1-\mathcal{G}\right)+ \frac{3}{2}N k_B T\,\mathcal{G},
\end{equation}
where $\mathcal{G}$ is a parameter which the define the ``\textit{gasness}'' of the liquid. In particular, in solids we do expect that $\mathcal{G}=0$, while in gases $\mathcal{G}=1$. In other words, a parameter is defined which interpolates between $0$ at the solidification temperature to $1$ at the evaporation temperature. In \cite{moon2022microscopic}, the authors proposed two different phenomenological gasness parameters given by:
\begin{equation}\label{g1}
    \text{IP}_1(T)\equiv 2 N_u\,,\quad \text{IP}_2(T)\equiv \frac{N_u}{1-N_u}\,.
\end{equation}
Both parameters are defined using the assumption that the fraction of unstable, or in their language imaginary modes, is zero in the solid phase. Moreover, it is argued that at large temperature, in the gas phase, one would expect a homogeneous distribution of maxima and minima in the potential and therefore that $N_u=N_s=1/2 N$. Given these two assumptions, both parameters interpolates between $0$ in the solid phase and $1$ in the solid phase. 

\begin{figure}
    \centering
    \includegraphics[width=\linewidth]{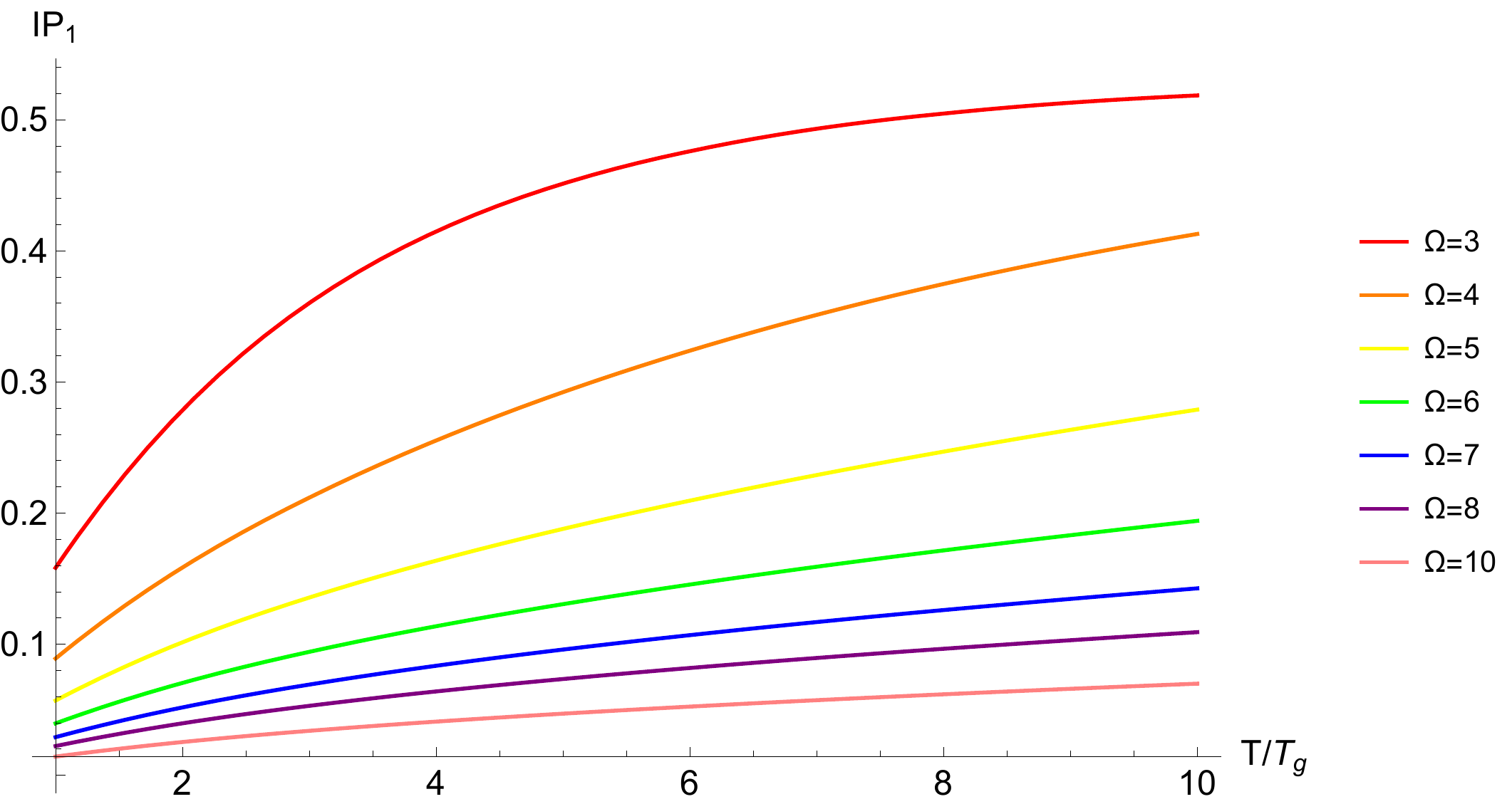}
    
    \vspace{0.5cm}
    
    \includegraphics[width=\linewidth]{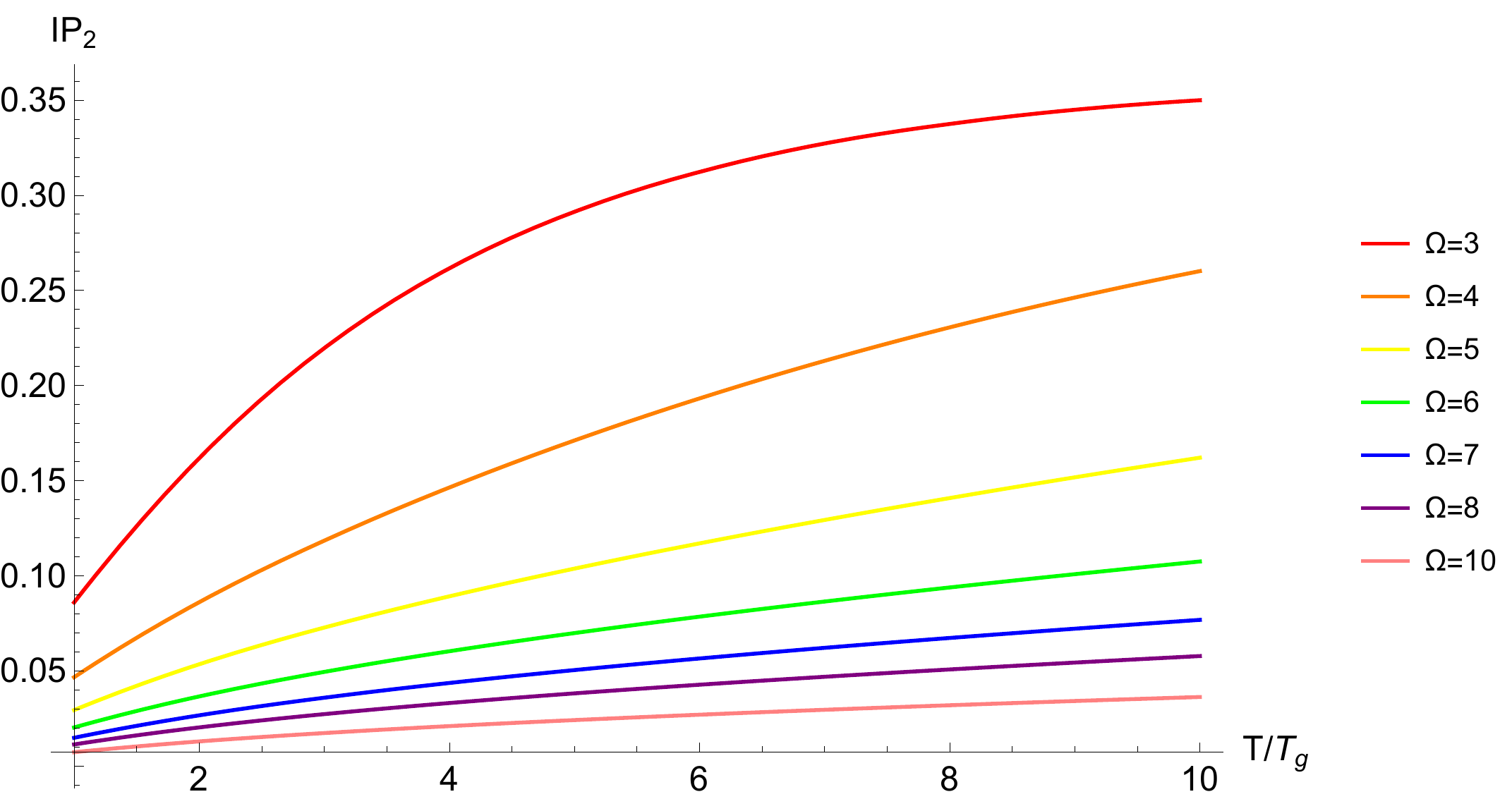}
    \caption{Top: The gasness parameter $\text{IP}_{1}$ as a function of temperature for different values of $\Omega$. Bottom: The gasness parameter $\text{IP}_{2}$ as a function of temperature for different values of $\Omega$.}
    \label{gass}
\end{figure}

In Fig.\ref{gass}, we show the values of the two phenomenological gasness parameters in Eq.\eqref{g1} as a function of the temperature for different values of the cutoff frequency $\Omega$. We observe that both parameters, indeed, grow monotonically with temperature.

After defining the gasness parameters, the heat capacity can be easily obtained from the energy and it is given by:
\begin{align}
    C_{1,2}=&\left(1-\text{IP}_{1,2}\right)3N_{tot}\, k_B \,+\,\text{IP}_{1,2}\,\left(\frac{3}{2}N_{tot} \,k_B \right)\nonumber\\
    &-\frac{d \text{IP}_{1,2}}{dT}\,\left(\frac{3}{2}N_{tot} \,k_B T\right)\,.\label{fifi}
\end{align}

\begin{figure}
    \centering
    \includegraphics[width=1\linewidth]{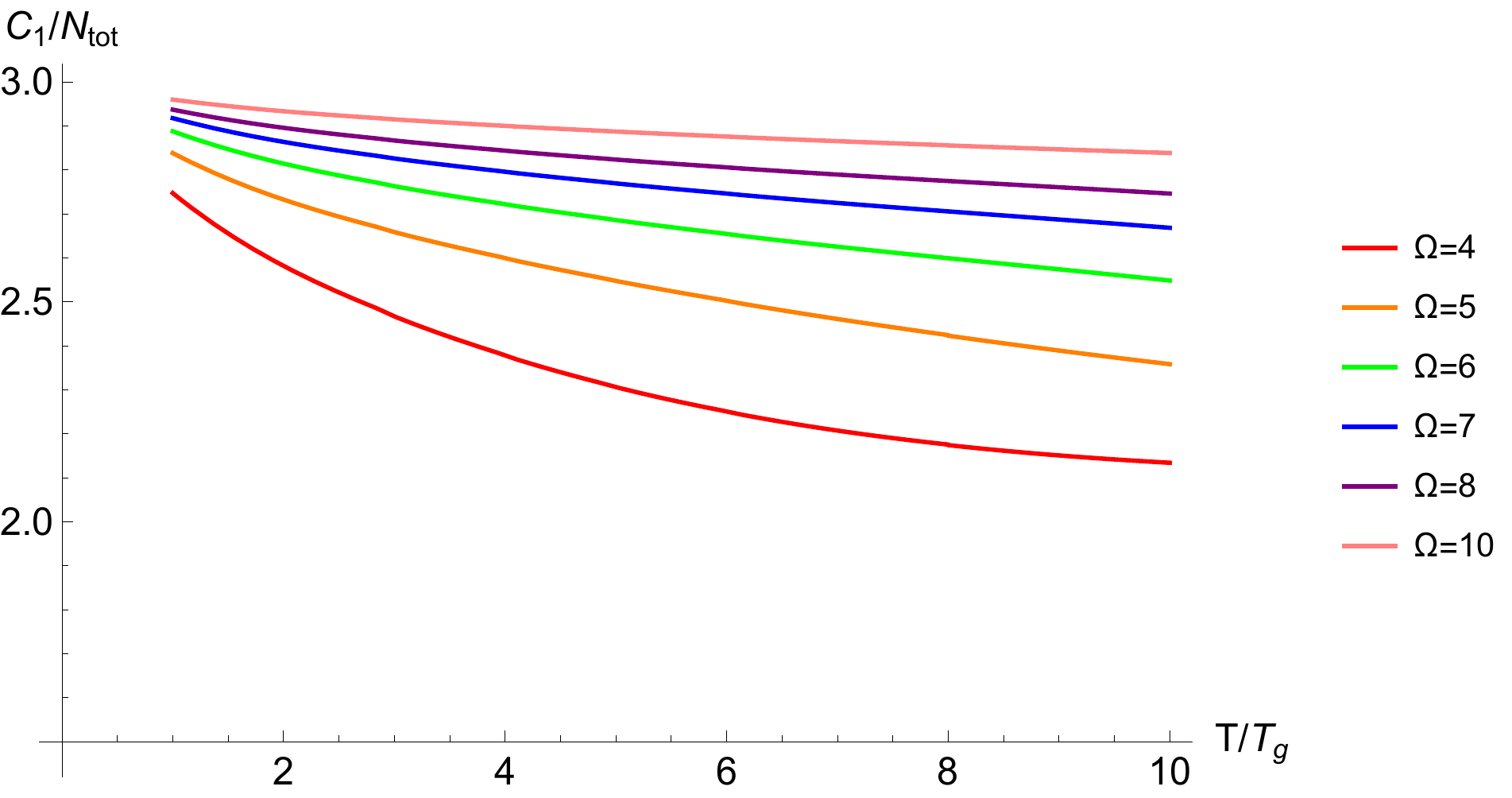}

    \vspace{0.5cm}
    
    \includegraphics[width=1\linewidth]{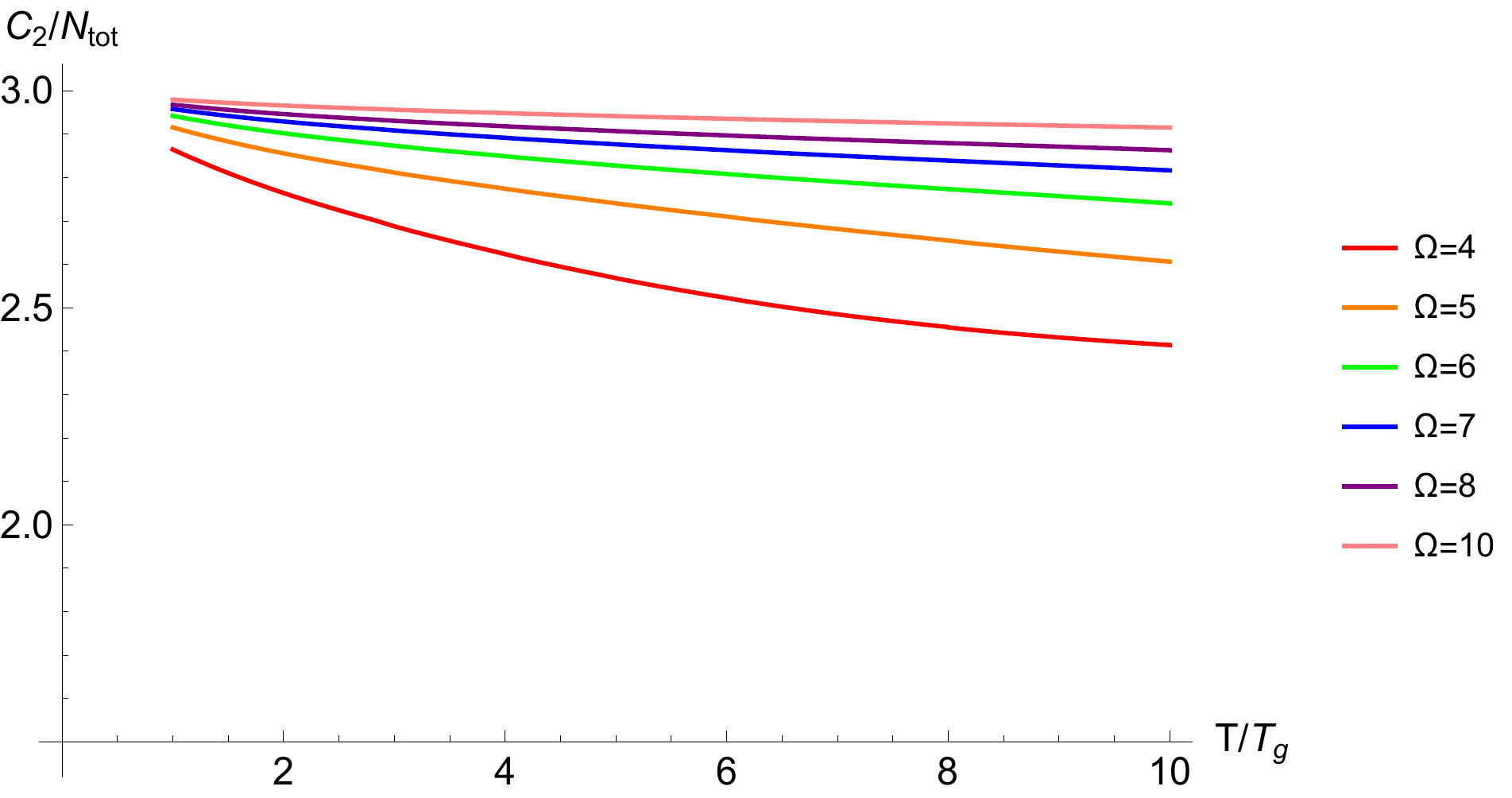}
    \caption{The heat capacity as a function of temperature computing using the phenomenological formula in Eq.\eqref{fifi}.}
    \label{heatfig2}
\end{figure}

Since the gasness parameters can be determined using the SPM model, we can also analyze the heat capacity using the phenomenological formula presented above, Eq.\eqref{fifi}. The results are shown as a function of the reduced temperature and for different values of the frequency cutoff $\Omega$ in Fig.\ref{heatfig2}. The top and bottom panels correspond respectively to the two different phenomenological choices in Eq.\eqref{g1}.

Several points for discussion are in order. (I) As expected, the heat capacity decreases monotonically with temperature from a $T=0$ value of $C=3 N k_B T$. This is in a way trivial since it is built into the phenomenological formula. Indeed, this statement is equivalent to impose that the gasness parameter grows with temperature. This is guaranteed from the expressions in Eq.\eqref{g1}. (II) The heat capacity does not reach the monoatomic gas-like value at large temperature, $C=3/2 N k_B T$, since the SPM does not describe a gas. 

In addition, the large temperature behavior depends crucially on the parameters of the systems, \textit{e.g.}, the cutoff frequency $\Omega$. This is again not surprising. The simple reason is that the gasness parameters defined in Eq.\eqref{g1} do not asymptote to a constant unit value at large temperature (see Fig.\ref{gass}). (III) The most interesting feature in the heat capacity computed using Eq.\eqref{g1} is its shape as a function of the cutoff frequency $\Omega$. The latter controls both the absolute value of the heat capacity and also the speed of its decay as a function of $T$. A larger $\Omega$ corresponds to a larger and flatter heat capacity. Qualitatively, this is in agreement with the results obtained in the previous section using purely statistical mechanics arguments. Despite the shape of the heat capacity from the two procedure is very different, some of the qualitative aspects seem universal. \\

More generally, we do not agree that a liquid has any gas-like properties. Our view of the heat capacity of liquids and normal modes is as follows.
Normal modes do not give the total energy of a system. They give additive contributions to the energy at which they are calculated. Stillinger's Inherent Structure (IS) theory \cite{stillinger2015energy} is well suited to implement this idea. The IS are the local minima of the multidimensional potential energy landscape. The energy is rigorously expressed as the average IS energy at a given $T$, plus the ``vibrational" contribution computed within the basins of attraction of those IS. There have been many calculations of the vibrational energy using harmonic normal modes, and some non-mode-based calculations of the anharmonic contribution \cite{JCFHSfest}.

The primary anharmonicity is the finite extent of the basins. Once $T$ is high enough, a mode will freely sample the configurations within the basin along the finite extent of its normal coordinate. Its energy will reach a constant value, and it  will have zero potential energy contribution to $C(T)$. The mode will be gas-like in its contribution in this sense (\textit{cfr.} \cite{moon2023atomic}), but it has nothing to do with ``gasness".

We suggest that the modes sufficiently excited to freely sample the basin are, to first approximation, the unstable modes. Thus we also would have them contributing $k_B/2$ to $C(T)$ instead of $k_B$, giving a decrease of $C(T)$ with increasing $T$, in agreement with \cite{moon2022microscopic}, for a completely different reason.

In addition, the IS energy is known to initially increase sharply with $T$, and reach a plateau at high $T$ \cite{JCFHSfest}. Thus, its contribution to $C(T)$ decreases with increasing $T$. In sum, the decreasing $C(T)$ in liquids arises from: (I) the IS energy approaching a constant following a sharp rise, and (II) some vibrational modes become capable of freely sampling their finite normal coordinate, and further $T$ increases cannot increase their energy.  In a future paper we will give a detailed theory based on anharmonic modes in the basins, in particular, INM and SPM modes.


\begin{thebibliography}{58}%
\makeatletter
\providecommand \@ifxundefined [1]{%
 \@ifx{#1\undefined}
}%
\providecommand \@ifnum [1]{%
 \ifnum #1\expandafter \@firstoftwo
 \else \expandafter \@secondoftwo
 \fi
}%
\providecommand \@ifx [1]{%
 \ifx #1\expandafter \@firstoftwo
 \else \expandafter \@secondoftwo
 \fi
}%
\providecommand \natexlab [1]{#1}%
\providecommand \enquote  [1]{``#1''}%
\providecommand \bibnamefont  [1]{#1}%
\providecommand \bibfnamefont [1]{#1}%
\providecommand \citenamefont [1]{#1}%
\providecommand \href@noop [0]{\@secondoftwo}%
\providecommand \href [0]{\begingroup \@sanitize@url \@href}%
\providecommand \@href[1]{\@@startlink{#1}\@@href}%
\providecommand \@@href[1]{\endgroup#1\@@endlink}%
\providecommand \@sanitize@url [0]{\catcode `\\12\catcode `\$12\catcode
  `\&12\catcode `\#12\catcode `\^12\catcode `\_12\catcode `\%12\relax}%
\providecommand \@@startlink[1]{}%
\providecommand \@@endlink[0]{}%
\providecommand \url  [0]{\begingroup\@sanitize@url \@url }%
\providecommand \@url [1]{\endgroup\@href {#1}{\urlprefix }}%
\providecommand \urlprefix  [0]{URL }%
\providecommand \Eprint [0]{\href }%
\providecommand \doibase [0]{http://dx.doi.org/}%
\providecommand \selectlanguage [0]{\@gobble}%
\providecommand \bibinfo  [0]{\@secondoftwo}%
\providecommand \bibfield  [0]{\@secondoftwo}%
\providecommand \translation [1]{[#1]}%
\providecommand \BibitemOpen [0]{}%
\providecommand \bibitemStop [0]{}%
\providecommand \bibitemNoStop [0]{.\EOS\space}%
\providecommand \EOS [0]{\spacefactor3000\relax}%
\providecommand \BibitemShut  [1]{\csname bibitem#1\endcsname}%
\let\auto@bib@innerbib\@empty
\bibitem [{\citenamefont {Trachenko}(2023)}]{trachenko_2023}%
  \BibitemOpen
  \bibfield  {author} {\bibinfo {author} {\bibfnamefont {K.}~\bibnamefont
  {Trachenko}},\ }\href {\doibase 10.1017/9781009355483} {\emph {\bibinfo
  {title} {Theory of Liquids: From Excitations to Thermodynamics}}}\ (\bibinfo
  {publisher} {Cambridge University Press},\ \bibinfo {year}
  {2023})\BibitemShut {NoStop}%
\bibitem [{\citenamefont {Moon}\ \emph {et~al.}(2022)\citenamefont {Moon},
  \citenamefont {Th{\'e}baud}, \citenamefont {Lindsay},\ and\ \citenamefont
  {Egami}}]{moon2022microscopic}%
  \BibitemOpen
  \bibfield  {author} {\bibinfo {author} {\bibfnamefont {J.}~\bibnamefont
  {Moon}}, \bibinfo {author} {\bibfnamefont {S.}~\bibnamefont {Th{\'e}baud}},
  \bibinfo {author} {\bibfnamefont {L.}~\bibnamefont {Lindsay}}, \ and\
  \bibinfo {author} {\bibfnamefont {T.}~\bibnamefont {Egami}},\ }\bibfield
  {title} {\enquote {\bibinfo {title} {Microscopic view of heat capacity of
  matter: solid, liquid, and gas},}\ }\href@noop {} {\bibfield  {journal}
  {\bibinfo  {journal} {arXiv preprint arXiv:2210.06218}\ } (\bibinfo {year}
  {2022})}\BibitemShut {NoStop}%
\bibitem [{\citenamefont {Zwanzig}(1983)}]{doi:10.1063/1.446338}%
  \BibitemOpen
  \bibfield  {author} {\bibinfo {author} {\bibfnamefont {R.}~\bibnamefont
  {Zwanzig}},\ }\bibfield  {title} {\enquote {\bibinfo {title} {On the relation
  between self‐diffusion and viscosity of liquids},}\ }\href {\doibase
  10.1063/1.446338} {\bibfield  {journal} {\bibinfo  {journal} {The Journal of
  Chemical Physics}\ }\textbf {\bibinfo {volume} {79}},\ \bibinfo {pages}
  {4507--4508} (\bibinfo {year} {1983})}\BibitemShut {NoStop}%
\bibitem [{\citenamefont {Stillinger}\ and\ \citenamefont
  {Weber}(1983)}]{PhysRevA.28.2408}%
  \BibitemOpen
  \bibfield  {author} {\bibinfo {author} {\bibfnamefont {F.~H.}\ \bibnamefont
  {Stillinger}}\ and\ \bibinfo {author} {\bibfnamefont {T.~A.}\ \bibnamefont
  {Weber}},\ }\bibfield  {title} {\enquote {\bibinfo {title} {Dynamics of
  structural transitions in liquids},}\ }\href {\doibase
  10.1103/PhysRevA.28.2408} {\bibfield  {journal} {\bibinfo  {journal} {Phys.
  Rev. A}\ }\textbf {\bibinfo {volume} {28}},\ \bibinfo {pages} {2408--2416}
  (\bibinfo {year} {1983})}\BibitemShut {NoStop}%
\bibitem [{\citenamefont {Weber}\ and\ \citenamefont
  {Stillinger}(1984)}]{doi:10.1063/1.447498}%
  \BibitemOpen
  \bibfield  {author} {\bibinfo {author} {\bibfnamefont {T.~A.}\ \bibnamefont
  {Weber}}\ and\ \bibinfo {author} {\bibfnamefont {F.~H.}\ \bibnamefont
  {Stillinger}},\ }\bibfield  {title} {\enquote {\bibinfo {title} {Inherent
  structures and distribution functions for liquids that freeze into bcc
  crystals},}\ }\href {\doibase 10.1063/1.447498} {\bibfield  {journal}
  {\bibinfo  {journal} {The Journal of Chemical Physics}\ }\textbf {\bibinfo
  {volume} {81}},\ \bibinfo {pages} {5089--5094} (\bibinfo {year}
  {1984})}\BibitemShut {NoStop}%
\bibitem [{\citenamefont {Stillinger}(2015)}]{stillinger2015energy}%
  \BibitemOpen
  \bibfield  {author} {\bibinfo {author} {\bibfnamefont {F.}~\bibnamefont
  {Stillinger}},\ }\href {https://books.google.com/books?id=qKC4CgAAQBAJ}
  {\emph {\bibinfo {title} {Energy Landscapes, Inherent Structures, and
  Condensed-Matter Phenomena}}}\ (\bibinfo  {publisher} {Princeton University
  Press},\ \bibinfo {year} {2015})\BibitemShut {NoStop}%
\bibitem [{\citenamefont {Chowdhary}\ and\ \citenamefont
  {Keyes}(2004)}]{JCFHSfest}%
  \BibitemOpen
  \bibfield  {author} {\bibinfo {author} {\bibfnamefont {J.}~\bibnamefont
  {Chowdhary}}\ and\ \bibinfo {author} {\bibfnamefont {T.}~\bibnamefont
  {Keyes}},\ }\bibfield  {title} {\enquote {\bibinfo {title} {Thermodynamics
  and dynamics for a model potential energy landscape},}\ }\href {\doibase
  10.1021/jp047615t} {\bibfield  {journal} {\bibinfo  {journal} {The Journal of
  Physical Chemistry B}\ }\textbf {\bibinfo {volume} {108}},\ \bibinfo {pages}
  {19786--19798} (\bibinfo {year} {2004})}\BibitemShut {NoStop}%
\bibitem [{\citenamefont {Seeley}\ and\ \citenamefont
  {Keyes}(1989)}]{doi:10.1063/1.457664}%
  \BibitemOpen
  \bibfield  {author} {\bibinfo {author} {\bibfnamefont {G.}~\bibnamefont
  {Seeley}}\ and\ \bibinfo {author} {\bibfnamefont {T.}~\bibnamefont {Keyes}},\
  }\bibfield  {title} {\enquote {\bibinfo {title} {Normal‐mode analysis of
  liquid‐state dynamics},}\ }\href {\doibase 10.1063/1.457664} {\bibfield
  {journal} {\bibinfo  {journal} {The Journal of Chemical Physics}\ }\textbf
  {\bibinfo {volume} {91}},\ \bibinfo {pages} {5581--5586} (\bibinfo {year}
  {1989})}\BibitemShut {NoStop}%
\bibitem [{\citenamefont {Keyes}(1997)}]{doi:10.1021/jp963706h}%
  \BibitemOpen
  \bibfield  {author} {\bibinfo {author} {\bibfnamefont {T.}~\bibnamefont
  {Keyes}},\ }\bibfield  {title} {\enquote {\bibinfo {title} {Instantaneous
  normal mode approach to liquid state dynamics},}\ }\href {\doibase
  10.1021/jp963706h} {\bibfield  {journal} {\bibinfo  {journal} {The Journal of
  Physical Chemistry A}\ }\textbf {\bibinfo {volume} {101}},\ \bibinfo {pages}
  {2921--2930} (\bibinfo {year} {1997})}\BibitemShut {NoStop}%
\bibitem [{\citenamefont {Stratt}(1995)}]{doi:10.1021/ar00053a001}%
  \BibitemOpen
  \bibfield  {author} {\bibinfo {author} {\bibfnamefont {R.~M.}\ \bibnamefont
  {Stratt}},\ }\bibfield  {title} {\enquote {\bibinfo {title} {The
  instantaneous normal modes of liquids},}\ }\href {\doibase
  10.1021/ar00053a001} {\bibfield  {journal} {\bibinfo  {journal} {Accounts of
  Chemical Research}\ }\textbf {\bibinfo {volume} {28}},\ \bibinfo {pages}
  {201--207} (\bibinfo {year} {1995})}\BibitemShut {NoStop}%
\bibitem [{\citenamefont {Keyes}(1994)}]{doi:10.1063/1.468407}%
  \BibitemOpen
  \bibfield  {author} {\bibinfo {author} {\bibfnamefont {T.}~\bibnamefont
  {Keyes}},\ }\bibfield  {title} {\enquote {\bibinfo {title} {Unstable modes in
  supercooled and normal liquids: Density of states, energy barriers, and
  self‐diffusion},}\ }\href {\doibase 10.1063/1.468407} {\bibfield  {journal}
  {\bibinfo  {journal} {The Journal of Chemical Physics}\ }\textbf {\bibinfo
  {volume} {101}},\ \bibinfo {pages} {5081--5092} (\bibinfo {year}
  {1994})}\BibitemShut {NoStop}%
\bibitem [{\citenamefont {Madan}, \citenamefont {Keyes},\ and\ \citenamefont
  {Seeley}(1990)}]{doi:10.1063/1.458192}%
  \BibitemOpen
  \bibfield  {author} {\bibinfo {author} {\bibfnamefont {B.}~\bibnamefont
  {Madan}}, \bibinfo {author} {\bibfnamefont {T.}~\bibnamefont {Keyes}}, \ and\
  \bibinfo {author} {\bibfnamefont {G.}~\bibnamefont {Seeley}},\ }\bibfield
  {title} {\enquote {\bibinfo {title} {Diffusion in supercooled liquids via
  normal mode analysis},}\ }\href {\doibase 10.1063/1.458192} {\bibfield
  {journal} {\bibinfo  {journal} {The Journal of Chemical Physics}\ }\textbf
  {\bibinfo {volume} {92}},\ \bibinfo {pages} {7565--7569} (\bibinfo {year}
  {1990})}\BibitemShut {NoStop}%
\bibitem [{\citenamefont {Keyes}(1995)}]{doi:10.1063/1.469947}%
  \BibitemOpen
  \bibfield  {author} {\bibinfo {author} {\bibfnamefont {T.}~\bibnamefont
  {Keyes}},\ }\bibfield  {title} {\enquote {\bibinfo {title} {Normal mode
  theory of diffusion in liquids for a broad temperature range},}\ }\href
  {\doibase 10.1063/1.469947} {\bibfield  {journal} {\bibinfo  {journal} {The
  Journal of Chemical Physics}\ }\textbf {\bibinfo {volume} {103}},\ \bibinfo
  {pages} {9810--9812} (\bibinfo {year} {1995})}\BibitemShut {NoStop}%
\bibitem [{\citenamefont {Clapa}, \citenamefont {Kottos},\ and\ \citenamefont
  {Starr}(2012)}]{doi:10.1063/1.3701564}%
  \BibitemOpen
  \bibfield  {author} {\bibinfo {author} {\bibfnamefont {V.~I.}\ \bibnamefont
  {Clapa}}, \bibinfo {author} {\bibfnamefont {T.}~\bibnamefont {Kottos}}, \
  and\ \bibinfo {author} {\bibfnamefont {F.~W.}\ \bibnamefont {Starr}},\
  }\bibfield  {title} {\enquote {\bibinfo {title} {Localization transition of
  instantaneous normal modes and liquid diffusion},}\ }\href {\doibase
  10.1063/1.3701564} {\bibfield  {journal} {\bibinfo  {journal} {The Journal of
  Chemical Physics}\ }\textbf {\bibinfo {volume} {136}},\ \bibinfo {pages}
  {144504} (\bibinfo {year} {2012})}\BibitemShut {NoStop}%
\bibitem [{\citenamefont {Gezelter}, \citenamefont {Rabani},\ and\
  \citenamefont {Berne}(1999)}]{doi:10.1063/1.478211}%
  \BibitemOpen
  \bibfield  {author} {\bibinfo {author} {\bibfnamefont {J.~D.}\ \bibnamefont
  {Gezelter}}, \bibinfo {author} {\bibfnamefont {E.}~\bibnamefont {Rabani}}, \
  and\ \bibinfo {author} {\bibfnamefont {B.~J.}\ \bibnamefont {Berne}},\
  }\bibfield  {title} {\enquote {\bibinfo {title} {Calculating the hopping rate
  for diffusion in molecular liquids: Cs2},}\ }\href {\doibase
  10.1063/1.478211} {\bibfield  {journal} {\bibinfo  {journal} {The Journal of
  Chemical Physics}\ }\textbf {\bibinfo {volume} {110}},\ \bibinfo {pages}
  {3444--3452} (\bibinfo {year} {1999})}\BibitemShut {NoStop}%
\bibitem [{\citenamefont {Li}, \citenamefont {Keyes},\ and\ \citenamefont
  {Sciortino}(1998)}]{doi:10.1063/1.475376}%
  \BibitemOpen
  \bibfield  {author} {\bibinfo {author} {\bibfnamefont {W.-X.}\ \bibnamefont
  {Li}}, \bibinfo {author} {\bibfnamefont {T.}~\bibnamefont {Keyes}}, \ and\
  \bibinfo {author} {\bibfnamefont {F.}~\bibnamefont {Sciortino}},\ }\bibfield
  {title} {\enquote {\bibinfo {title} {Three-flavor instantaneous normal mode
  formalism: Diffusion, harmonicity, and the potential energy landscape of
  liquid cs2},}\ }\href {\doibase 10.1063/1.475376} {\bibfield  {journal}
  {\bibinfo  {journal} {The Journal of Chemical Physics}\ }\textbf {\bibinfo
  {volume} {108}},\ \bibinfo {pages} {252--260} (\bibinfo {year}
  {1998})}\BibitemShut {NoStop}%
\bibitem [{\citenamefont {Chowdhary}\ and\ \citenamefont
  {Keyes}(2002)}]{PhysRevE.65.026125}%
  \BibitemOpen
  \bibfield  {author} {\bibinfo {author} {\bibfnamefont {J.}~\bibnamefont
  {Chowdhary}}\ and\ \bibinfo {author} {\bibfnamefont {T.}~\bibnamefont
  {Keyes}},\ }\bibfield  {title} {\enquote {\bibinfo {title} {Conjugate
  gradient filtering of instantaneous normal modes, saddles on the energy
  landscape, and diffusion in liquids},}\ }\href {\doibase
  10.1103/PhysRevE.65.026125} {\bibfield  {journal} {\bibinfo  {journal} {Phys.
  Rev. E}\ }\textbf {\bibinfo {volume} {65}},\ \bibinfo {pages} {026125}
  (\bibinfo {year} {2002})}\BibitemShut {NoStop}%
\bibitem [{\citenamefont {Stamper}\ \emph {et~al.}(2022)\citenamefont
  {Stamper}, \citenamefont {Cortie}, \citenamefont {Yue}, \citenamefont
  {Wang},\ and\ \citenamefont {Yu}}]{stamper2022experimental}%
  \BibitemOpen
  \bibfield  {author} {\bibinfo {author} {\bibfnamefont {C.}~\bibnamefont
  {Stamper}}, \bibinfo {author} {\bibfnamefont {D.}~\bibnamefont {Cortie}},
  \bibinfo {author} {\bibfnamefont {Z.}~\bibnamefont {Yue}}, \bibinfo {author}
  {\bibfnamefont {X.}~\bibnamefont {Wang}}, \ and\ \bibinfo {author}
  {\bibfnamefont {D.}~\bibnamefont {Yu}},\ }\bibfield  {title} {\enquote
  {\bibinfo {title} {Experimental confirmation of the universal law for the
  vibrational density of states of liquids},}\ }\href@noop {} {\bibfield
  {journal} {\bibinfo  {journal} {The Journal of Physical Chemistry Letters}\
  }\textbf {\bibinfo {volume} {13}},\ \bibinfo {pages} {3105--3111} (\bibinfo
  {year} {2022})}\BibitemShut {NoStop}%
\bibitem [{\citenamefont {Phillips}\ \emph {et~al.}(1989)\citenamefont
  {Phillips}, \citenamefont {Buchenau}, \citenamefont {N\"ucker}, \citenamefont
  {Dianoux},\ and\ \citenamefont {Petry}}]{PhysRevLett.63.2381}%
  \BibitemOpen
  \bibfield  {author} {\bibinfo {author} {\bibfnamefont {W.~A.}\ \bibnamefont
  {Phillips}}, \bibinfo {author} {\bibfnamefont {U.}~\bibnamefont {Buchenau}},
  \bibinfo {author} {\bibfnamefont {N.}~\bibnamefont {N\"ucker}}, \bibinfo
  {author} {\bibfnamefont {A.-J.}\ \bibnamefont {Dianoux}}, \ and\ \bibinfo
  {author} {\bibfnamefont {W.}~\bibnamefont {Petry}},\ }\bibfield  {title}
  {\enquote {\bibinfo {title} {Dynamics of glassy and liquid selenium},}\
  }\href {\doibase 10.1103/PhysRevLett.63.2381} {\bibfield  {journal} {\bibinfo
   {journal} {Phys. Rev. Lett.}\ }\textbf {\bibinfo {volume} {63}},\ \bibinfo
  {pages} {2381--2384} (\bibinfo {year} {1989})}\BibitemShut {NoStop}%
\bibitem [{\citenamefont {Buchenau}(1993)}]{BUCHENAU1993275}%
  \BibitemOpen
  \bibfield  {author} {\bibinfo {author} {\bibfnamefont {U.}~\bibnamefont
  {Buchenau}},\ }\bibfield  {title} {\enquote {\bibinfo {title} {Soft modes in
  undercooled liquids},}\ }\href {\doibase
  https://doi.org/10.1016/0022-2860(93)80144-K} {\bibfield  {journal} {\bibinfo
   {journal} {Journal of Molecular Structure}\ }\textbf {\bibinfo {volume}
  {296}},\ \bibinfo {pages} {275--283} (\bibinfo {year} {1993})}\BibitemShut
  {NoStop}%
\bibitem [{\citenamefont {Jin}\ \emph {et~al.}(2023)\citenamefont {Jin},
  \citenamefont {Fan}, \citenamefont {Stamper}, \citenamefont {Mole},
  \citenamefont {Yu}, \citenamefont {Hong}, \citenamefont {Yu},\ and\
  \citenamefont {Baggioli}}]{jin2023dissecting}%
  \BibitemOpen
  \bibfield  {author} {\bibinfo {author} {\bibfnamefont {S.}~\bibnamefont
  {Jin}}, \bibinfo {author} {\bibfnamefont {X.}~\bibnamefont {Fan}}, \bibinfo
  {author} {\bibfnamefont {C.}~\bibnamefont {Stamper}}, \bibinfo {author}
  {\bibfnamefont {R.~A.}\ \bibnamefont {Mole}}, \bibinfo {author}
  {\bibfnamefont {Y.}~\bibnamefont {Yu}}, \bibinfo {author} {\bibfnamefont
  {L.}~\bibnamefont {Hong}}, \bibinfo {author} {\bibfnamefont {D.}~\bibnamefont
  {Yu}}, \ and\ \bibinfo {author} {\bibfnamefont {M.}~\bibnamefont
  {Baggioli}},\ }\href@noop {} {\enquote {\bibinfo {title} {Dissecting the
  experimental vibrational density of states of liquids using instantaneous
  normal mode theory},}\ } (\bibinfo {year} {2023}),\ \Eprint
  {http://arxiv.org/abs/2304.14609} {arXiv:2304.14609 [cond-mat.soft]}
  \BibitemShut {NoStop}%
\bibitem [{\citenamefont {Xu}\ and\ \citenamefont
  {Stratt}(1989)}]{doi:10.1063/1.457564}%
  \BibitemOpen
  \bibfield  {author} {\bibinfo {author} {\bibfnamefont {B.}~\bibnamefont
  {Xu}}\ and\ \bibinfo {author} {\bibfnamefont {R.~M.}\ \bibnamefont
  {Stratt}},\ }\bibfield  {title} {\enquote {\bibinfo {title} {Liquid theory
  for band structure in a liquid},}\ }\href {\doibase 10.1063/1.457564}
  {\bibfield  {journal} {\bibinfo  {journal} {The Journal of Chemical Physics}\
  }\textbf {\bibinfo {volume} {91}},\ \bibinfo {pages} {5613--5627} (\bibinfo
  {year} {1989})}\BibitemShut {NoStop}%
\bibitem [{\citenamefont {Wu}\ and\ \citenamefont
  {Loring}(1993)}]{doi:10.1063/1.465563}%
  \BibitemOpen
  \bibfield  {author} {\bibinfo {author} {\bibfnamefont {T.}~\bibnamefont
  {Wu}}\ and\ \bibinfo {author} {\bibfnamefont {R.~F.}\ \bibnamefont
  {Loring}},\ }\bibfield  {title} {\enquote {\bibinfo {title} {Collective
  motions in liquids with a normal mode approach},}\ }\href {\doibase
  10.1063/1.465563} {\bibfield  {journal} {\bibinfo  {journal} {The Journal of
  Chemical Physics}\ }\textbf {\bibinfo {volume} {99}},\ \bibinfo {pages}
  {8936--8947} (\bibinfo {year} {1993})}\BibitemShut {NoStop}%
\bibitem [{\citenamefont {Wan}\ and\ \citenamefont
  {Stratt}(1994)}]{doi:10.1063/1.467178}%
  \BibitemOpen
  \bibfield  {author} {\bibinfo {author} {\bibfnamefont {Y.}~\bibnamefont
  {Wan}}\ and\ \bibinfo {author} {\bibfnamefont {R.~M.}\ \bibnamefont
  {Stratt}},\ }\bibfield  {title} {\enquote {\bibinfo {title} {Liquid theory
  for the instantaneous normal modes of a liquid},}\ }\href {\doibase
  10.1063/1.467178} {\bibfield  {journal} {\bibinfo  {journal} {The Journal of
  Chemical Physics}\ }\textbf {\bibinfo {volume} {100}},\ \bibinfo {pages}
  {5123--5138} (\bibinfo {year} {1994})}\BibitemShut {NoStop}%
\bibitem [{\citenamefont {Zaccone}\ and\ \citenamefont
  {Baggioli}(2021)}]{doi:10.1073/pnas.2022303118}%
  \BibitemOpen
  \bibfield  {author} {\bibinfo {author} {\bibfnamefont {A.}~\bibnamefont
  {Zaccone}}\ and\ \bibinfo {author} {\bibfnamefont {M.}~\bibnamefont
  {Baggioli}},\ }\bibfield  {title} {\enquote {\bibinfo {title} {Universal law
  for the vibrational density of states of liquids},}\ }\href {\doibase
  10.1073/pnas.2022303118} {\bibfield  {journal} {\bibinfo  {journal}
  {Proceedings of the National Academy of Sciences}\ }\textbf {\bibinfo
  {volume} {118}},\ \bibinfo {pages} {e2022303118} (\bibinfo {year}
  {2021})}\BibitemShut {NoStop}%
\bibitem [{\citenamefont {Schirmacher}, \citenamefont {Bryk},\ and\
  \citenamefont {Ruocco}(2022)}]{doi:10.1073/pnas.2119288119}%
  \BibitemOpen
  \bibfield  {author} {\bibinfo {author} {\bibfnamefont {W.}~\bibnamefont
  {Schirmacher}}, \bibinfo {author} {\bibfnamefont {T.}~\bibnamefont {Bryk}}, \
  and\ \bibinfo {author} {\bibfnamefont {G.}~\bibnamefont {Ruocco}},\
  }\bibfield  {title} {\enquote {\bibinfo {title} {Modeling the instantaneous
  normal mode spectra of liquids as that of unstable elastic media},}\ }\href
  {\doibase 10.1073/pnas.2119288119} {\bibfield  {journal} {\bibinfo  {journal}
  {Proceedings of the National Academy of Sciences}\ }\textbf {\bibinfo
  {volume} {119}},\ \bibinfo {pages} {e2119288119} (\bibinfo {year}
  {2022})}\BibitemShut {NoStop}%
\bibitem [{\citenamefont {Li}\ and\ \citenamefont
  {Keyes}(1999)}]{doi:10.1063/1.479810}%
  \BibitemOpen
  \bibfield  {author} {\bibinfo {author} {\bibfnamefont {W.-X.}\ \bibnamefont
  {Li}}\ and\ \bibinfo {author} {\bibfnamefont {T.}~\bibnamefont {Keyes}},\
  }\bibfield  {title} {\enquote {\bibinfo {title} {Instantaneous normal mode
  theory of diffusion and the potential energy landscape: Application to
  supercooled liquid cs2},}\ }\href {\doibase 10.1063/1.479810} {\bibfield
  {journal} {\bibinfo  {journal} {The Journal of Chemical Physics}\ }\textbf
  {\bibinfo {volume} {111}},\ \bibinfo {pages} {5503--5513} (\bibinfo {year}
  {1999})}\BibitemShut {NoStop}%
\bibitem [{\citenamefont {Z\"urcher}\ and\ \citenamefont
  {Keyes}(1997)}]{PhysRevE.55.6917}%
  \BibitemOpen
  \bibfield  {author} {\bibinfo {author} {\bibfnamefont {U.}~\bibnamefont
  {Z\"urcher}}\ and\ \bibinfo {author} {\bibfnamefont {T.}~\bibnamefont
  {Keyes}},\ }\bibfield  {title} {\enquote {\bibinfo {title} {Anharmonic
  potentials in supercooled liquids: The soft-potential model},}\ }\href
  {\doibase 10.1103/PhysRevE.55.6917} {\bibfield  {journal} {\bibinfo
  {journal} {Phys. Rev. E}\ }\textbf {\bibinfo {volume} {55}},\ \bibinfo
  {pages} {6917--6927} (\bibinfo {year} {1997})}\BibitemShut {NoStop}%
\bibitem [{\citenamefont {Wu}\ and\ \citenamefont
  {Loring}(1992)}]{doi:10.1063/1.463375}%
  \BibitemOpen
  \bibfield  {author} {\bibinfo {author} {\bibfnamefont {T.}~\bibnamefont
  {Wu}}\ and\ \bibinfo {author} {\bibfnamefont {R.~F.}\ \bibnamefont
  {Loring}},\ }\bibfield  {title} {\enquote {\bibinfo {title} {Phonons in
  liquids: A random walk approach},}\ }\href {\doibase 10.1063/1.463375}
  {\bibfield  {journal} {\bibinfo  {journal} {The Journal of Chemical Physics}\
  }\textbf {\bibinfo {volume} {97}},\ \bibinfo {pages} {8568--8575} (\bibinfo
  {year} {1992})}\BibitemShut {NoStop}%
\bibitem [{\citenamefont {Keyes}, \citenamefont {Vijayadamodar},\ and\
  \citenamefont {Zurcher}(1997)}]{keyes1997instantaneous}%
  \BibitemOpen
  \bibfield  {author} {\bibinfo {author} {\bibfnamefont {T.}~\bibnamefont
  {Keyes}}, \bibinfo {author} {\bibfnamefont {G.}~\bibnamefont
  {Vijayadamodar}}, \ and\ \bibinfo {author} {\bibfnamefont {U.}~\bibnamefont
  {Zurcher}},\ }\bibfield  {title} {\enquote {\bibinfo {title} {An
  instantaneous normal mode description of relaxation in supercooled
  liquids},}\ }\href@noop {} {\bibfield  {journal} {\bibinfo  {journal} {The
  Journal of chemical physics}\ }\textbf {\bibinfo {volume} {106}},\ \bibinfo
  {pages} {4651--4657} (\bibinfo {year} {1997})}\BibitemShut {NoStop}%
\bibitem [{\citenamefont {Kittel}(2004)}]{kittel2004introduction}%
  \BibitemOpen
  \bibfield  {author} {\bibinfo {author} {\bibfnamefont {C.}~\bibnamefont
  {Kittel}},\ }\href {https://books.google.co.kr/books?id=kym4QgAACAAJ} {\emph
  {\bibinfo {title} {Introduction to Solid State Physics}}}\ (\bibinfo
  {publisher} {Wiley},\ \bibinfo {year} {2004})\BibitemShut {NoStop}%
\bibitem [{\citenamefont {Togo}\ and\ \citenamefont
  {Tanaka}(2015)}]{TOGO20151}%
  \BibitemOpen
  \bibfield  {author} {\bibinfo {author} {\bibfnamefont {A.}~\bibnamefont
  {Togo}}\ and\ \bibinfo {author} {\bibfnamefont {I.}~\bibnamefont {Tanaka}},\
  }\bibfield  {title} {\enquote {\bibinfo {title} {First principles phonon
  calculations in materials science},}\ }\href {\doibase
  https://doi.org/10.1016/j.scriptamat.2015.07.021} {\bibfield  {journal}
  {\bibinfo  {journal} {Scripta Materialia}\ }\textbf {\bibinfo {volume}
  {108}},\ \bibinfo {pages} {1--5} (\bibinfo {year} {2015})}\BibitemShut
  {NoStop}%
\bibitem [{\citenamefont {Bolmatov}, \citenamefont {Brazhkin},\ and\
  \citenamefont {Trachenko}(2012)}]{bolmatov2012phonon}%
  \BibitemOpen
  \bibfield  {author} {\bibinfo {author} {\bibfnamefont {D.}~\bibnamefont
  {Bolmatov}}, \bibinfo {author} {\bibfnamefont {V.}~\bibnamefont {Brazhkin}},
  \ and\ \bibinfo {author} {\bibfnamefont {K.}~\bibnamefont {Trachenko}},\
  }\bibfield  {title} {\enquote {\bibinfo {title} {The phonon theory of liquid
  thermodynamics},}\ }\href@noop {} {\bibfield  {journal} {\bibinfo  {journal}
  {Scientific reports}\ }\textbf {\bibinfo {volume} {2}},\ \bibinfo {pages}
  {1--6} (\bibinfo {year} {2012})}\BibitemShut {NoStop}%
\bibitem [{\citenamefont {Bryk}, \citenamefont {Scopigno},\ and\ \citenamefont
  {Ruocco}(2015)}]{bryk2015heat}%
  \BibitemOpen
  \bibfield  {author} {\bibinfo {author} {\bibfnamefont {T.}~\bibnamefont
  {Bryk}}, \bibinfo {author} {\bibfnamefont {T.}~\bibnamefont {Scopigno}}, \
  and\ \bibinfo {author} {\bibfnamefont {G.}~\bibnamefont {Ruocco}},\
  }\bibfield  {title} {\enquote {\bibinfo {title} {Heat capacity of liquids: A
  hydrodynamic approach},}\ }\href@noop {} {\bibfield  {journal} {\bibinfo
  {journal} {arXiv preprint arXiv:1504.01224}\ } (\bibinfo {year}
  {2015})}\BibitemShut {NoStop}%
\bibitem [{\citenamefont {Baggioli}\ and\ \citenamefont
  {Zaccone}(2021)}]{PhysRevE.104.014103}%
  \BibitemOpen
  \bibfield  {author} {\bibinfo {author} {\bibfnamefont {M.}~\bibnamefont
  {Baggioli}}\ and\ \bibinfo {author} {\bibfnamefont {A.}~\bibnamefont
  {Zaccone}},\ }\bibfield  {title} {\enquote {\bibinfo {title} {Explaining the
  specific heat of liquids based on instantaneous normal modes},}\ }\href
  {\doibase 10.1103/PhysRevE.104.014103} {\bibfield  {journal} {\bibinfo
  {journal} {Phys. Rev. E}\ }\textbf {\bibinfo {volume} {104}},\ \bibinfo
  {pages} {014103} (\bibinfo {year} {2021})}\BibitemShut {NoStop}%
\bibitem [{\citenamefont {Madan}\ and\ \citenamefont
  {Keyes}(1993)}]{madan1993unstable}%
  \BibitemOpen
  \bibfield  {author} {\bibinfo {author} {\bibfnamefont {B.}~\bibnamefont
  {Madan}}\ and\ \bibinfo {author} {\bibfnamefont {T.}~\bibnamefont {Keyes}},\
  }\bibfield  {title} {\enquote {\bibinfo {title} {Unstable modes in liquids
  density of states, potential energy, and heat capacity},}\ }\href@noop {}
  {\bibfield  {journal} {\bibinfo  {journal} {The Journal of chemical physics}\
  }\textbf {\bibinfo {volume} {98}},\ \bibinfo {pages} {3342--3350} (\bibinfo
  {year} {1993})}\BibitemShut {NoStop}%
\bibitem [{\citenamefont {Rosenfeld}\ and\ \citenamefont
  {Tarazona}(1998)}]{doi:10.1080/00268979809483145}%
  \BibitemOpen
  \bibfield  {author} {\bibinfo {author} {\bibfnamefont {Y.}~\bibnamefont
  {Rosenfeld}}\ and\ \bibinfo {author} {\bibfnamefont {P.}~\bibnamefont
  {Tarazona}},\ }\bibfield  {title} {\enquote {\bibinfo {title} {Density
  functional theory and the asymptotic high density expansion of the free
  energy of classical solids and fluids},}\ }\href {\doibase
  10.1080/00268979809483145} {\bibfield  {journal} {\bibinfo  {journal}
  {Molecular Physics}\ }\textbf {\bibinfo {volume} {95}},\ \bibinfo {pages}
  {141--150} (\bibinfo {year} {1998})}\BibitemShut {NoStop}%
\bibitem [{\citenamefont {Wallace}(1998)}]{PhysRevE.57.1717}%
  \BibitemOpen
  \bibfield  {author} {\bibinfo {author} {\bibfnamefont {D.~C.}\ \bibnamefont
  {Wallace}},\ }\bibfield  {title} {\enquote {\bibinfo {title} {Liquid dynamics
  theory of high-temperature specific heat},}\ }\href {\doibase
  10.1103/PhysRevE.57.1717} {\bibfield  {journal} {\bibinfo  {journal} {Phys.
  Rev. E}\ }\textbf {\bibinfo {volume} {57}},\ \bibinfo {pages} {1717--1722}
  (\bibinfo {year} {1998})}\BibitemShut {NoStop}%
\bibitem [{\citenamefont {Ramos}(2022)}]{doi:10.1142/q0371}%
  \BibitemOpen
  \bibfield  {author} {\bibinfo {author} {\bibfnamefont {M.~A.}\ \bibnamefont
  {Ramos}},\ }\href {\doibase 10.1142/q0371} {\emph {\bibinfo {title}
  {Low-Temperature Thermal and Vibrational Properties of Disordered Solids}}}\
  (\bibinfo  {publisher} {WORLD SCIENTIFIC (EUROPE)},\ \bibinfo {year}
  {2022})\BibitemShut {NoStop}%
\bibitem [{\citenamefont {U.Buchenau}()}]{doi:10.1142/9781800612587_0008}%
  \BibitemOpen
  \bibfield  {author} {\bibinfo {author} {\bibnamefont {U.Buchenau}},\
  }\enquote {\bibinfo {title} {The soft-potential model and its extensions},}\
  in\ \href {\doibase 10.1142/9781800612587_0008} {\emph {\bibinfo {booktitle}
  {Low-Temperature Thermal and Vibrational Properties of Disordered Solids}}},\
  Chap.~\bibinfo {chapter} {8}, pp.\ \bibinfo {pages} {299--330}\BibitemShut
  {NoStop}%
\bibitem [{\citenamefont {Moriel}, \citenamefont {Lerner},\ and\ \citenamefont
  {Bouchbinder}(2023)}]{moriel2023boson}%
  \BibitemOpen
  \bibfield  {author} {\bibinfo {author} {\bibfnamefont {A.}~\bibnamefont
  {Moriel}}, \bibinfo {author} {\bibfnamefont {E.}~\bibnamefont {Lerner}}, \
  and\ \bibinfo {author} {\bibfnamefont {E.}~\bibnamefont {Bouchbinder}},\
  }\href@noop {} {\enquote {\bibinfo {title} {The boson peak in the vibrational
  spectra of glasses},}\ } (\bibinfo {year} {2023}),\ \Eprint
  {http://arxiv.org/abs/2304.03661} {arXiv:2304.03661 [cond-mat.dis-nn]}
  \BibitemShut {NoStop}%
\bibitem [{\citenamefont {Karpov}\ and\ \citenamefont
  {Klinger}(1983)}]{karpov1983theory}%
  \BibitemOpen
  \bibfield  {author} {\bibinfo {author} {\bibfnamefont {V.}~\bibnamefont
  {Karpov}}\ and\ \bibinfo {author} {\bibfnamefont {I.}~\bibnamefont
  {Klinger}},\ }\bibfield  {title} {\enquote {\bibinfo {title} {Theory of the
  low-temperature anomalies in the thermal properties of amorphous
  structures},}\ }\href@noop {} {\bibfield  {journal} {\bibinfo  {journal} {Zh.
  Eksp. Teor. Fiz}\ }\textbf {\bibinfo {volume} {84}},\ \bibinfo {pages} {775}
  (\bibinfo {year} {1983})}\BibitemShut {NoStop}%
\bibitem [{\citenamefont {Parshin}(1993)}]{Parshin_1993}%
  \BibitemOpen
  \bibfield  {author} {\bibinfo {author} {\bibfnamefont {D.~A.}\ \bibnamefont
  {Parshin}},\ }\bibfield  {title} {\enquote {\bibinfo {title} {Soft potential
  model and universal properties of glasses},}\ }\href {\doibase
  10.1088/0031-8949/1993/T49A/030} {\bibfield  {journal} {\bibinfo  {journal}
  {Physica Scripta}\ }\textbf {\bibinfo {volume} {1993}},\ \bibinfo {pages}
  {180} (\bibinfo {year} {1993})}\BibitemShut {NoStop}%
\bibitem [{\citenamefont {Buchenau}\ \emph {et~al.}(1991)\citenamefont
  {Buchenau}, \citenamefont {Galperin}, \citenamefont {Gurevich},\ and\
  \citenamefont {Schober}}]{PhysRevB.43.5039}%
  \BibitemOpen
  \bibfield  {author} {\bibinfo {author} {\bibfnamefont {U.}~\bibnamefont
  {Buchenau}}, \bibinfo {author} {\bibfnamefont {Y.~M.}\ \bibnamefont
  {Galperin}}, \bibinfo {author} {\bibfnamefont {V.~L.}\ \bibnamefont
  {Gurevich}}, \ and\ \bibinfo {author} {\bibfnamefont {H.~R.}\ \bibnamefont
  {Schober}},\ }\bibfield  {title} {\enquote {\bibinfo {title} {Anharmonic
  potentials and vibrational localization in glasses},}\ }\href {\doibase
  10.1103/PhysRevB.43.5039} {\bibfield  {journal} {\bibinfo  {journal} {Phys.
  Rev. B}\ }\textbf {\bibinfo {volume} {43}},\ \bibinfo {pages} {5039--5045}
  (\bibinfo {year} {1991})}\BibitemShut {NoStop}%
\bibitem [{\citenamefont {Buchenau}\ \emph {et~al.}(1992)\citenamefont
  {Buchenau}, \citenamefont {Galperin}, \citenamefont {Gurevich}, \citenamefont
  {Parshin}, \citenamefont {Ramos},\ and\ \citenamefont
  {Schober}}]{PhysRevB.46.2798}%
  \BibitemOpen
  \bibfield  {author} {\bibinfo {author} {\bibfnamefont {U.}~\bibnamefont
  {Buchenau}}, \bibinfo {author} {\bibfnamefont {Y.~M.}\ \bibnamefont
  {Galperin}}, \bibinfo {author} {\bibfnamefont {V.~L.}\ \bibnamefont
  {Gurevich}}, \bibinfo {author} {\bibfnamefont {D.~A.}\ \bibnamefont
  {Parshin}}, \bibinfo {author} {\bibfnamefont {M.~A.}\ \bibnamefont {Ramos}},
  \ and\ \bibinfo {author} {\bibfnamefont {H.~R.}\ \bibnamefont {Schober}},\
  }\bibfield  {title} {\enquote {\bibinfo {title} {Interaction of soft modes
  and sound waves in glasses},}\ }\href {\doibase 10.1103/PhysRevB.46.2798}
  {\bibfield  {journal} {\bibinfo  {journal} {Phys. Rev. B}\ }\textbf {\bibinfo
  {volume} {46}},\ \bibinfo {pages} {2798--2808} (\bibinfo {year}
  {1992})}\BibitemShut {NoStop}%
\bibitem [{\citenamefont {Gil}\ \emph {et~al.}(1993)\citenamefont {Gil},
  \citenamefont {Ramos}, \citenamefont {Bringer},\ and\ \citenamefont
  {Buchenau}}]{PhysRevLett.70.182}%
  \BibitemOpen
  \bibfield  {author} {\bibinfo {author} {\bibfnamefont {L.}~\bibnamefont
  {Gil}}, \bibinfo {author} {\bibfnamefont {M.~A.}\ \bibnamefont {Ramos}},
  \bibinfo {author} {\bibfnamefont {A.}~\bibnamefont {Bringer}}, \ and\
  \bibinfo {author} {\bibfnamefont {U.}~\bibnamefont {Buchenau}},\ }\bibfield
  {title} {\enquote {\bibinfo {title} {Low-temperature specific heat and
  thermal conductivity of glasses},}\ }\href {\doibase
  10.1103/PhysRevLett.70.182} {\bibfield  {journal} {\bibinfo  {journal} {Phys.
  Rev. Lett.}\ }\textbf {\bibinfo {volume} {70}},\ \bibinfo {pages} {182--185}
  (\bibinfo {year} {1993})}\BibitemShut {NoStop}%
\bibitem [{\citenamefont {Buchenau}(1994)}]{BUCHENAU1994391}%
  \BibitemOpen
  \bibfield  {author} {\bibinfo {author} {\bibfnamefont {U.}~\bibnamefont
  {Buchenau}},\ }\bibfield  {title} {\enquote {\bibinfo {title} {Anharmonic
  effects in undercooled liquids},}\ }\href {\doibase
  https://doi.org/10.1016/0022-3093(94)90462-6} {\bibfield  {journal} {\bibinfo
   {journal} {Journal of Non-Crystalline Solids}\ }\textbf {\bibinfo {volume}
  {172-174}},\ \bibinfo {pages} {391--395} (\bibinfo {year} {1994})},\ \bibinfo
  {note} {relaxations in Complex Systems}\BibitemShut {NoStop}%
\bibitem [{\citenamefont {Frenkel}(1946)}]{FrenkelBook46}%
  \BibitemOpen
  \bibfield  {author} {\bibinfo {author} {\bibfnamefont {J.}~\bibnamefont
  {Frenkel}},\ }\href@noop {} {\emph {\bibinfo {title} {Kinetic Theory of
  liquids}}}\ (\bibinfo  {publisher} {Oxford University Press, Oxford, UK},\
  \bibinfo {year} {1946})\BibitemShut {NoStop}%
\bibitem [{\citenamefont {Ferrari}, \citenamefont {Phillips},\ and\
  \citenamefont {Russo}(1987)}]{ferrari1987heat}%
  \BibitemOpen
  \bibfield  {author} {\bibinfo {author} {\bibfnamefont {L.}~\bibnamefont
  {Ferrari}}, \bibinfo {author} {\bibfnamefont {W.}~\bibnamefont {Phillips}}, \
  and\ \bibinfo {author} {\bibfnamefont {G.}~\bibnamefont {Russo}},\ }\bibfield
   {title} {\enquote {\bibinfo {title} {Heat capacity at the glass
  transition},}\ }\href@noop {} {\bibfield  {journal} {\bibinfo  {journal}
  {Europhysics Letters}\ }\textbf {\bibinfo {volume} {3}},\ \bibinfo {pages}
  {611} (\bibinfo {year} {1987})}\BibitemShut {NoStop}%
\bibitem [{\citenamefont
  {Stillinger}(1995)}]{doi:10.1126/science.267.5206.1935}%
  \BibitemOpen
  \bibfield  {author} {\bibinfo {author} {\bibfnamefont {F.~H.}\ \bibnamefont
  {Stillinger}},\ }\bibfield  {title} {\enquote {\bibinfo {title} {A
  topographic view of supercooled liquids and glass formation},}\ }\href
  {\doibase 10.1126/science.267.5206.1935} {\bibfield  {journal} {\bibinfo
  {journal} {Science}\ }\textbf {\bibinfo {volume} {267}},\ \bibinfo {pages}
  {1935--1939} (\bibinfo {year} {1995})},\ \Eprint
  {http://arxiv.org/abs/https://www.science.org/doi/pdf/10.1126/science.267.5206.1935}
  {https://www.science.org/doi/pdf/10.1126/science.267.5206.1935} \BibitemShut
  {NoStop}%
\bibitem [{\citenamefont {Vijayadamodar}\ and\ \citenamefont
  {Nitzan}(1995)}]{vijay2}%
  \BibitemOpen
  \bibfield  {author} {\bibinfo {author} {\bibfnamefont {G.~V.}\ \bibnamefont
  {Vijayadamodar}}\ and\ \bibinfo {author} {\bibfnamefont {A.}~\bibnamefont
  {Nitzan}},\ }\bibfield  {title} {\enquote {\bibinfo {title} {{On the
  application of instantaneous normal mode analysis to long time dynamics of
  liquids}},}\ }\href {\doibase 10.1063/1.469693} {\bibfield  {journal}
  {\bibinfo  {journal} {The Journal of Chemical Physics}\ }\textbf {\bibinfo
  {volume} {103}},\ \bibinfo {pages} {2169--2177} (\bibinfo {year} {1995})},\
  \Eprint
  {http://arxiv.org/abs/https://pubs.aip.org/aip/jcp/article-pdf/103/6/2169/8105088/2169\_1\_online.pdf}
  {https://pubs.aip.org/aip/jcp/article-pdf/103/6/2169/8105088/2169\_1\_online.pdf}
  \BibitemShut {NoStop}%
\bibitem [{\citenamefont {Zhang}, \citenamefont {Douglas},\ and\ \citenamefont
  {Starr}(2019)}]{doi:10.1063/1.5127821}%
  \BibitemOpen
  \bibfield  {author} {\bibinfo {author} {\bibfnamefont {W.}~\bibnamefont
  {Zhang}}, \bibinfo {author} {\bibfnamefont {J.~F.}\ \bibnamefont {Douglas}},
  \ and\ \bibinfo {author} {\bibfnamefont {F.~W.}\ \bibnamefont {Starr}},\
  }\bibfield  {title} {\enquote {\bibinfo {title} {What does the instantaneous
  normal mode spectrum tell us about dynamical heterogeneity in glass-forming
  fluids?}}\ }\href {\doibase 10.1063/1.5127821} {\bibfield  {journal}
  {\bibinfo  {journal} {The Journal of Chemical Physics}\ }\textbf {\bibinfo
  {volume} {151}},\ \bibinfo {pages} {184904} (\bibinfo {year}
  {2019})}\BibitemShut {NoStop}%
\bibitem [{\citenamefont {Keyes}(2005)}]{keyes-2005}%
  \BibitemOpen
  \bibfield  {author} {\bibinfo {author} {\bibfnamefont {T.}~\bibnamefont
  {Keyes}},\ }\enquote {\bibinfo {title} {Normal mode analysis: Theory and
  applications to biological and chemical systems},}\ \ (\bibinfo  {publisher}
  {Chapman and Hall/CRC},\ \bibinfo {year} {2005})\ Chap.\ \bibinfo {chapter}
  {The Relation Between Unstable Instantaneous Normal Modes and
  Diffusion}\BibitemShut {NoStop}%
\bibitem [{\citenamefont {Li}\ and\ \citenamefont {Keyes}(1997)}]{PureTR97}%
  \BibitemOpen
  \bibfield  {author} {\bibinfo {author} {\bibfnamefont {W.-X.}\ \bibnamefont
  {Li}}\ and\ \bibinfo {author} {\bibfnamefont {T.}~\bibnamefont {Keyes}},\
  }\bibfield  {title} {\enquote {\bibinfo {title} {Pure translation
  instantaneous normal modes: Imaginary frequency contributions vanish at the
  glass transition in cs2},}\ }\href@noop {} {\bibfield  {journal} {\bibinfo
  {journal} {The Journal of chemical physics}\ }\textbf {\bibinfo {volume}
  {107}},\ \bibinfo {pages} {7275--7277} (\bibinfo {year} {1997})}\BibitemShut
  {NoStop}%
\bibitem [{\citenamefont {Baggioli}\ \emph {et~al.}(2020)\citenamefont
  {Baggioli}, \citenamefont {Vasin}, \citenamefont {Brazhkin},\ and\
  \citenamefont {Trachenko}}]{BAGGIOLI20201}%
  \BibitemOpen
  \bibfield  {author} {\bibinfo {author} {\bibfnamefont {M.}~\bibnamefont
  {Baggioli}}, \bibinfo {author} {\bibfnamefont {M.}~\bibnamefont {Vasin}},
  \bibinfo {author} {\bibfnamefont {V.}~\bibnamefont {Brazhkin}}, \ and\
  \bibinfo {author} {\bibfnamefont {K.}~\bibnamefont {Trachenko}},\ }\bibfield
  {title} {\enquote {\bibinfo {title} {Gapped momentum states},}\ }\href
  {\doibase https://doi.org/10.1016/j.physrep.2020.04.002} {\bibfield
  {journal} {\bibinfo  {journal} {Physics Reports}\ }\textbf {\bibinfo {volume}
  {865}},\ \bibinfo {pages} {1--44} (\bibinfo {year} {2020})},\ \bibinfo {note}
  {gapped momentum states}\BibitemShut {NoStop}%
\bibitem [{\citenamefont {Esquinazi}(2013)}]{esquinazi2013tunneling}%
  \BibitemOpen
  \bibfield  {author} {\bibinfo {author} {\bibfnamefont {P.}~\bibnamefont
  {Esquinazi}},\ }\href {https://books.google.co.th/books?id=37ToCAAAQBAJ}
  {\emph {\bibinfo {title} {Tunneling Systems in Amorphous and Crystalline
  Solids}}}\ (\bibinfo  {publisher} {Springer Berlin Heidelberg},\ \bibinfo
  {year} {2013})\BibitemShut {NoStop}%
\bibitem [{\citenamefont {Yu}\ and\ \citenamefont
  {Carruzzo}()}]{doi:10.1142/9781800612587_0004}%
  \BibitemOpen
  \bibfield  {author} {\bibinfo {author} {\bibfnamefont {C.~C.}\ \bibnamefont
  {Yu}}\ and\ \bibinfo {author} {\bibfnamefont {H.~M.}\ \bibnamefont
  {Carruzzo}},\ }\enquote {\bibinfo {title} {Two-level systems and the
  tunneling model: A critical view},}\ in\ \href {\doibase
  10.1142/9781800612587_0004} {\emph {\bibinfo {booktitle} {Low-Temperature
  Thermal and Vibrational Properties of Disordered Solids}}},\ Chap.~\bibinfo
  {chapter} {4}, pp.\ \bibinfo {pages} {113--139}\BibitemShut {NoStop}%
\bibitem [{\citenamefont {Moon}, \citenamefont {Lindsay},\ and\ \citenamefont
  {Egami}(2023)}]{moon2023atomic}%
  \BibitemOpen
  \bibfield  {author} {\bibinfo {author} {\bibfnamefont {J.}~\bibnamefont
  {Moon}}, \bibinfo {author} {\bibfnamefont {L.}~\bibnamefont {Lindsay}}, \
  and\ \bibinfo {author} {\bibfnamefont {T.}~\bibnamefont {Egami}},\
  }\href@noop {} {\enquote {\bibinfo {title} {Atomic dynamics in fluids: Normal
  mode analysis revisited},}\ } (\bibinfo {year} {2023}),\ \Eprint
  {http://arxiv.org/abs/2304.09778} {arXiv:2304.09778 [cond-mat.dis-nn]}
  \BibitemShut {NoStop}%
\end{thebibliography}
\end{document}